\documentclass[noshowpacs,prx,aps,superscriptaddress,floatfix,twocolumn,nofootinbib]{revtex4-2}
\usepackage[utf8]{inputenc}
\usepackage{lineno,mathtools,url}
\usepackage{amsmath}
\usepackage{amssymb,mathrsfs}
\usepackage{graphicx}
\usepackage[colorlinks=true]{hyperref}
\usepackage[dvipsnames]{xcolor}
\usepackage{dsfont}
\usepackage{mathbbol} 
\usepackage[normalem]{ulem}

\newcommand{\beq}{\begin{eqnarray}}
\newcommand{\eeq}{\end{eqnarray}}
\newcommand{\bem}{\begin{pmatrix}}
\newcommand{\eem}{\end{pmatrix}}

\renewcommand{\r}{{\bf r}}

\begin{document}

\title{Boson-fermion universality of mesoscopic entanglement fluctuations in free systems}
\author{Cunzhong Lou}
\email{These authors contribute equally to the work.}
\address{School of Physics, Zhejiang  University,  Hangzhou, Zhejiang  310027, China}
\address{CAS Key Laboratory of Theoretical Physics and Institute of Theoretical Physics, Chinese Academy of Sciences,  Beijing 100190, China}

\author{Chushun Tian}	
\email{These authors contribute equally to the work.}	
\address{CAS Key Laboratory of Theoretical Physics and Institute of Theoretical Physics, Chinese Academy of Sciences,  Beijing 100190, China}

\author{Zhixing Zou}
\address{Department of Physics and Key Laboratory of Low Dimensional Condensed Matter Physics (Department of Education of Fujian Province), Xiamen University, Xiamen, Fujian 361005, China}
\address{CAS Key Laboratory of Theoretical Physics and Institute of Theoretical Physics, Chinese Academy of Sciences,  Beijing 100190, China}

\author{Tao Shi}
\address{CAS Key Laboratory of Theoretical Physics and Institute of Theoretical Physics, Chinese Academy of Sciences,  Beijing 100190, China}

\author{Lih-King Lim}		
\email{lihking@zju.edu.cn}
\address{School of Physics, Zhejiang  University,  Hangzhou, Zhejiang  310027, China}

\date{\today}

\begin{abstract}
Entanglement fluctuations associated with Schr\"{o}dinger evolution of wavefunctions offer a unique perspective on various fundamental issues ranging from quantum thermalization to state preparation in quantum devices. Very recently, a subset of present authors have shown that in a class of free-fermion lattice models and interacting spin chains, entanglement dynamics enters into a new regime at long time, with entanglement probes displaying persistent temporal fluctuations, whose statistics falls into the seemingly disparate paradigm of mesoscopic fluctuations in condensed matter physics. This motivate us to revisit here entanglement dynamics of a canonical bosonic model in many-body physics, i.e., a coupled harmonic oscillator chain. We find that when the system is driven out of equilibrium, the long-time entanglement dynamics exhibits strictly the same statistical behaviors as that of free-fermion models. Specifically, irrespective of entanglement probes and microscopic parameters, the statistical distribution of entanglement fluctuations is flanked by asymmetric tails: sub-Gaussian for upward fluctuations and sub-Gamma for downward; moreover, the variance exhibits a crossover from the scaling $\sim 1/L$ to $\sim L_A^3/L^2$, as the subsystem size $L_A$ increases ($L$ the total system size). This insensitivity to the particle statistics, dubbed boson-fermion universality, is contrary to the common wisdom that statistical phenomena of many-body nature depend strongly on particle statistics. Together with our previous work, the present work indicates rich fluctuation phenomena in entanglement dynamics awaiting in-depth explorations.
\end{abstract}

\maketitle

\section{\label{sec_1} Introduction}
Quantum entanglement is a key to understanding quantum phases of matter. The focus has long been on many-body ground states, since their entanglement properties are believed to be responsible for exotic behaviors of strongly correlated systems \cite{Anderson_1973, Li_PRL_2008, Coleman2015} especially when they are in a topological phase \cite{Kitaev_AP_2003, Chen_2010}. Recent progresses achieved in realizing highly tunable quantum simulators have opened up a door to direct measurements of quantum entanglement \cite{Islam_2015, Satzinger_2021, Semeghini_2021, Joshi_2023,Karamlou_2024}, allowing experimental studies of various subjects of nonequilibrium dynamics, ranging from quantum thermalization \cite{Kaufman_2016} to many-body quantum scars \cite{Bernien_2017,Turner_2018}. This has in turn triggered keen interests in entanglement of many-body excited states.

For excited states, especially highly excited ones, the study of their entanglement properties remains a challenging task. So far, theoretical works have been based on an ensemble of states subjected to certain kinematic constraints \cite{Page_1993, Popescu_2006, Vidmar_2017, Nakagawa_NC_2018, Yu_PRR_2023}. In this direction, Page's study of entanglement entropy \cite{Page_1993} for states randomly drawn from the Haar measure is a benchmark, and is believed to capture entanglement properties at `infinite temperature'. Recently, the study has been extended to the ensemble of random energy eigenstates \cite{Vidmar_2017}. These works show that depending on the ensemble of random states statistical behaviors of entanglement differ significantly \cite{Bianchi_2022}. However, all these studies are kinematic and mute on entanglement dynamics.

When a quantum system undergoes a sudden quench, whereby parameters of system's Hamiltonian are changed instantaneously, many high-lying eigenstates of the post-quench Hamiltonian are excited \cite{Essler_JSM_2016}. Subsequent unitary time evolution generates an infinitely long path in the Hilbert space; each point of this path, labelled by the time parameter $t\in [0,\infty)$, represents an instantaneous system state, $\psi(t)$. One may view this path as a special `ensemble' of states, each of which is a linear superposition of many excited eigenstates. A question then arises: What happens to entanglement fluctuations of this ensemble? Studies of this question go far beyond previous kinematic studies of entanglement fluctuations for two fundamental reasons: First, the variation of the subsystem's entanglement along the unitary time evolution path is genuinely of dynamical origin, i.e., entanglement fluctuations are of out-of-equilibrium nature. Secondly, the path is an exceptional,
i.e., a zero-measure, set of the Hilbert space and must be distinguished from the kinematic
sampling of states in the Hilbert space \cite{Page_1993, Popescu_2006, Vidmar_2017, Nakagawa_NC_2018, Yu_PRR_2023}. The origin of the two fluctuations is fundamentally different.  Indeed, for {\it infinite} systems undergoing quench dynamics, the celebrated result of Calabrese and Cardy \cite{Calabrese_JSM_2005} states that after initial linear growth entanglement saturates at a value proportional to the subsystem size (the so-called volume law), and no temporal entanglement fluctuations arise, in the limit when the ratio of the subsystem to total system size vanishes.

Yet, in experiments and real applications a system is always {\it finite}, or the subsystem-to-total system size ratio is finite. Besides, recent studies of the foundations of statistical mechanics have focused on isolated quantum systems, which are finite \cite{Gogolin16,Rigol16,Borgonovi16}. For finite systems temporal entanglement fluctuations are ubiquitous in real \cite{Kaufman_2016} and numerical \cite{Nakagawa_NC_2018,Rajabpour14,Faiez_2020} experiments on long-time entanglement dynamics. On one hand, for practical applications it is necessary to control these fluctuations. On the other hand, understanding their statistical behaviors especially their out-of-equilibrium properties is essential to the foundations of statistical mechanics \cite{Faiez_2020}. In fact, as early as in 1929 von Neumann discovered the fundamental importance of temporal fluctuations of the so-called observational entropy to the validity of a statistical description of an isolated quantum system \cite{von_Neumann29,von_Neumann10,Lebowitz10}. In parallel to von Neumann's study a fundamental question is what happens to temporal fluctuations of entanglement entropy \cite{Faiez_2020}. Studies of entanglement dynamics have long been focused on early-time behaviors, such as entanglement growth \cite{Calabrese_JSM_2005} and revival \cite{Modak_JSM_2020,Essler_JSM_2016} and creation and transport of entanglement \cite{Plenio04}, and thus do not address this issue.

Not until very recently has the analytical study of long-time entanglement dynamics been initiated \cite{Lih-King_2023}, which indicates rich out-of-equilibrium fluctuation behaviors of entanglement in finite systems. That work crucially relies on the uncovering of an emergent random structure in the evolution of many-body wavefunctions, and combining that structure with the nonasymptotic probability theory \cite{Boucheron_2013}. In that work, for several integrable lattice models, including free fermions, the transverse field Ising model, and the spin-$1/2$ Heisenberg XXZ chain, it was found that entanglement dynamics settles into a fluctuation regime eventually. The fluctuation phenomenon in that regime, though of out-of-equilibrium nature, carries the universality of the celebrated {\it sample-to-sample} fluctuations in mesoscopic electronic and optical systems \cite{Sheng_2006, Akkermans_2007}, and shares the same physical origin, namely, the wave interference. Specifically, the fluctuations of various entanglement probes, such as the entanglement entropy and the R\'{e}nyi entropy of distinct orders, display the same probability distribution which is asymmetric, with a sub-Gamma lower tail and a sub-Gaussian upper tail. This asymmetry implies that downward fluctuations are more favored than the upward. Moreover, the variances of distinct probes display the same scaling behavior that depends only on the subsystem and total system sizes, independent of system's microscopic parameters.

It is worth noting that entanglement fluctuations found in Ref.~\cite{Lih-King_2023} are of many-body nature. In many-body physics fluctuation properties are often sensitive to the particle statistics: bosons, fermions or anyons. In addition, they can be influenced by the coupling of bosonic and fermionic degrees of freedom, that exists typically in supersymmetric systems. Thus an interesting issue is raised. That is, whether it is possible to detect the particle statistics from temporal fluctuations in entanglement dynamics, akin to detecting different types of composite fermions from the electric current noise in fractional quantum Hall systems \cite{Kane03,Stern06,Simon12}.

Here, we address this issue by generalizing the theory of long-time entanglement dynamics developed for free fermions \cite{Lih-King_2023} to a canonical bosonic model, the coupled harmonic oscillator chain. This model is a prototype of many realistic systems, ranging from one-dimensional solids to the Klein-Gordon field in particle physics. Entanglement properties of this model have been studied by many authors. However, the focus has been the kinematic entanglement structure \cite{Peschel99,Plenio02,Botero04,Plenio05,Eisert_RMP_2010} or the early-time behavior of entanglement dynamics \cite{Modak_JSM_2020,Cotller16,Hackl_PRA_2018,Alba_SPP_2018}. There have been increasing interests in long-time entanglement dynamics of this model \cite{Rajabpour14,Calabrese17}, but so far studies have been restricted to numerical investigations and have paid no attentions to fluctuation phenomena. In this work, we analytically study fluctuation phenomena in long-time entanglement dynamics of this model. We find, surprisingly, that the statistics of out-of-equilibrium entanglement fluctuations of a harmonic oscillator chain is strictly the same as that in free-fermion models.  Our analytical predictions are confirmed by extensive numerical experiments. Thus we establish the boson-fermion universality of mesoscopic fluctuations in long-time entanglement dynamics of free systems.

The rest of this paper is organized in the following way: In Sec.~\ref{sec_2} we introduce the bosonic model, the quench protocol, and entanglement dynamics. We also briefly review some basic concepts and dynamical properties. The problem is formulated and a scope of the work is described. In Secs.~\ref{sec_3}-\ref{sec_3_3} we develop the statistical theory for temporal fluctuations in entanglement dynamics. We begin with showing that the long-time entanglement dynamics is statistically equivalent to some emergent mesoscopic sample-to-sample fluctuations of entanglement (Sec.~\ref{sec_3}). Then we combine the statistical equivalence and the concentration-of-measure theory to show that for different entanglement probes their statistical distributions hold the same, and display the boson-fermion universality (Sec.~\ref{sec_3_2}). We further study the variance of entanglement fluctuations and find its scaling behaviors, and show that those scaling behaviors display the boson-fermion universality also (Sec.~\ref{sec_3_3}). Finally, in Sec.~\ref{Sec:num} extensive exact numerics are given, which verify various analytical predictions. We end the work with conclusions and discussions of outlook in Sec.~\ref{sec_4}. In Appendixes \ref{App_A}-\ref{sec:derivative_matrix_function} we give additional technical details of analytical derivations and more numerical results.

\section{\label{sec_2}Basic notions and motivations}

We begin with a description of the model and a brief review of some basic concepts and dynamical properties. An emphasis is placed on the Gaussianity of the total system and subsystem state, since this notion plays important roles in the studies of entanglement (see, e.g., Ref.~\cite{Adesso07} for an exhaustive review) and in developing our fluctuation theory. With these preparations we formulate the problem to be addressed, and describe the scope of the paper.

\subsection{\label{sec_2_1}Model and quench dynamics}
\begin{figure}
\includegraphics[width=1.0\linewidth]{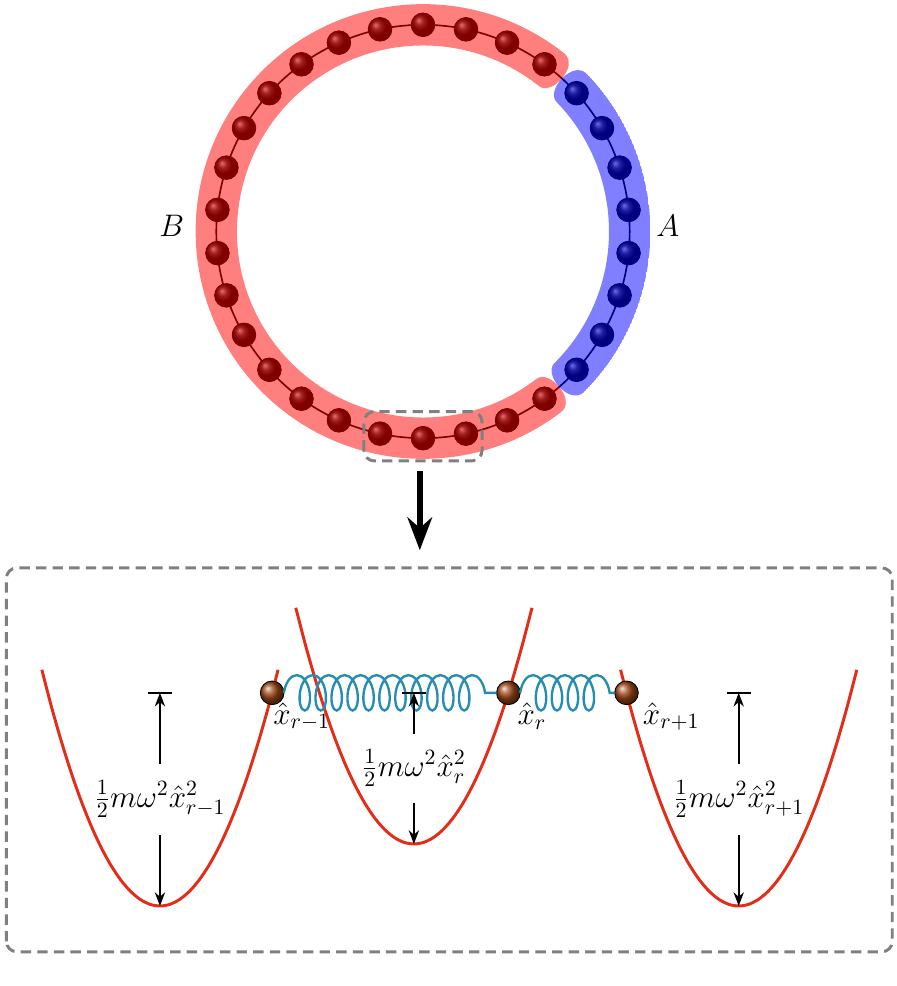}
\caption{\label{setting} Schematic of the system and its bipartition. The system is a chain of $L$ oscillators with nearest neighbors coupled  through harmonic interaction. The subsystem $A$ consists of $L_A$ oscillators while its complement $B$ consists of $L-L_A$ oscillators.}
\end{figure}

We consider a chain of $L$ (assumed to be even and large, but finite throughout) quantum harmonic oscillators, each of which has frequency $\omega$ and mass $m$. They are labelled by $r=1,2,\ldots,L$. Two nearest oscillators are coupled through a harmonic interaction with strength $K'$ (cf.~Fig.~\ref{setting}). The Hamiltonian reads
\begin{eqnarray}
\hat{H}' = \sum\limits_{r=1}^{L} \left( \frac{\hat{p}_r^2}{2m} + \frac{m\omega^2\hat{x}_r^2}{2}   + \frac{K'}{2} (\hat{x}_{r+1} - \hat{x}_r)^2 \right).
\label{eq:75}
\end{eqnarray}
Here the first two terms describe a single oscillator and the last describes the coupling between two nearest oscillators. The periodic boundary condition: $\hat{x}_{L+1}=\hat{x}_1$, $\hat{p}_{L+1}=\hat{p}_1$ is imposed.  The position (displaced from the equilibrium point) and momentum operators $\hat{x}_r$, $\hat{p}_r$ of the $r$th oscillator satisfy the canonical commutation relation $[\hat{x}_r, \hat{p}_s]=i\hbar\delta_{r,s}$. We make the rescaling: $\hat{x}_r\rightarrow \sqrt{m\omega\over \hbar}\,\hat{x}_r$, $\hat{p}_r\rightarrow \hat{p}_r/\sqrt{\hbar m\omega}$, so that $\hat{x}_r$ and $\hat{p}_r$ are both dimensionless, and the commutation relation reduces to $[\hat{x}_r, \hat{p}_s]=i\delta_{r,s}$. Upon this rescaling, the Hamiltonian is transformed to
\begin{eqnarray}
\hat{H}' = {\hbar\omega\over 2}\sum\limits_{r=1}^{L} \left( \hat{p}_r^2 + \hat{x}_r^2   + \frac{K}{\omega^2} (\hat{x}_{r+1} - \hat{x}_r)^2 \right)
\label{eq:109}
\end{eqnarray}
with $K\equiv K'/m$. This Hamiltonian depends on parameters $\omega$ and $K$.

For the convenience below we introduce a vector $\hat{\r}$ formed by $2L$ operators, $\hat{x}$'s and $\hat{p}$'s:
\begin{equation}\label{eq:85}
  \hat{\r} \equiv (\hat{x}_1,\hat{x}_2,\cdots,\hat{x}_{L}, \hat{p}_1,\hat{p}_2,\cdots,\hat{p}_{L})^{\rm T}
\end{equation}
with `T' standing for the transpose. The components of $\hat{\r}$ are labelled by $i\in \{1,\ldots,2L\}$. The first $L$ components, with $1\leq i\leq L$, are position operators and the second $L$ components, with $L+1\leq i\leq 2L$, are conjugated momentum operators. Equivalently, Eq.~(\ref{eq:85}) can be represented as
\begin{equation}\label{eq:79}
  \hat{\r}=\bigoplus_{r=1}^{L}\left(
                               \begin{array}{c}
                                 \hat{x}_r \\
                                 \hat{p}_r \\
                               \end{array}
                             \right).
\end{equation}
So the $2L$ indexes $i$ in the representation Eq.~(\ref{eq:85}) can be reorganized as
\begin{equation}\label{eq:93}
  i\equiv(r\sigma),\quad r=1,2,\ldots,L,\, \sigma=\bar{A},\bar{B}
\end{equation}
with the use of double indexes, $r$ and $\sigma$: The former labels the oscillators, and the latter accommodates the position/momentum degree of freedom carried by an oscillator. With Eq.~(\ref{eq:85}) or Eq.~(\ref{eq:79}), the commutation relation is represented as
\begin{equation}\label{eq:86}
  [\hat{\r},\,\hat{\r}]=iJ\Leftrightarrow [\hat{r}_i,\,\hat{r}_j]=iJ_{ij}.
\end{equation}
Here
\beq
J\equiv\left[\begin{array}{cc}0&\mathbb{I}_L\\ -\mathbb{I}_L&0\end{array}\right]=\left[\begin{array}{cc}0&1\\ -1&0\end{array}\right]\otimes \mathbb{I}_L.
\eeq
is called the symplectic form, whose offdiagonal structure is carried by the $\hat{x}$-$\hat{p}$ sector introduced by the representation
Eq.~(\ref{eq:79}). $\mathbb{I}_L$ is the unit matrix defined in the oscillator space, with the subscript standing for the matrix size.

The chain has $L$ normal modes, each of which carries a wavenumber $k={2\pi n\over L}$, with the integer $n$ satisfying $-({L\over 2}-1)\leq n \leq {L\over 2}$, and a frequency $\omega_k\equiv s_k \omega$ with $s_k=\sqrt{1+{4K\over \omega^2}\sin^2{k\over 2}}$. In the normal mode space, the Hamiltonian (\ref{eq:109}) reads
\begin{eqnarray}
\hat{H}'=\hbar\omega \sum\limits_{k}s_k( \hat{\alpha}^\dagger_k \hat{\alpha}_k + \frac{1}{2}).
\label{eq:106}
\end{eqnarray}
Here $\hat{\alpha}_k = {1\over \sqrt{2}}(\sqrt{s_k}\hat{Q}_k+{i\over \sqrt{s_k}}\,\hat{P}_k)$ and $\hat{\alpha}^\dagger_k= {1\over \sqrt{2}}(\sqrt{s_k}\hat{Q}^\dagger_k-{i\over \sqrt{s_k}}\,\hat{P}^\dagger_k)$ are respectively the annihilation and creation operators of mode $k$. $\hat{Q}_k$ and $\hat{P}_k$ are the Fourier transformation of $\hat{x}_r$ and $\hat{p}_r$, respectively.

Initially, we set the parameter in $\hat{H}'$ to be ($\omega_0,\,K_0$). Moreover, let the system be at the ground state of the ensuing Hamiltonian. This state, denoted as $|\psi_0\rangle$, is the tensor product of the vacuum state of each normal mode:
$|\psi_0\rangle = \bigotimes_k  |0\rangle_k$,
where $\hat{\alpha}_k |0\rangle_k = 0$ for all $k$. At time $t=0$ we implement a global quench by instantaneously changing the parameters,
\begin{eqnarray}
(\omega_0,K_0) \to (\omega,K).
\end{eqnarray}
Upon quench  the state $|\psi_0\rangle$ is driven out of equilibrium, and follows Schr\"odinger equation, i.e.,
\begin{eqnarray}
|\psi(t)\rangle = {\rm e}^{-{i\hat{H}' t\over \hbar}} |\psi_0\rangle={\rm e}^{-i\hat{H} t} |\psi_0\rangle,\quad t>0
\label{eq:87}
\end{eqnarray}
to evolve, where in the second equality the Hamiltonian
\begin{eqnarray}
\hat{H}= {\omega\over 2}\sum\limits_{r=1}^{L} \left( \hat{p}_r^2 + \hat{x}_r^2   + \frac{K}{\omega^2} (\hat{x}_{r+1} - \hat{x}_r)^2 \right)
\label{eq:80}
\end{eqnarray}
since the overall factor $\hbar$ in Eq.~(\ref{eq:109}) is cancelled out by the denominator of the first exponent of Eq.~(\ref{eq:87}). It is the post-quench Hamiltonian $\hat{H}$ given by Eq.~(\ref{eq:80}) that we will adopt below. Note that strictly speaking it has the unit of the frequency rather than the energy, but this plays no roles in subsequent studies since $\hat{H}t$ is dimensionless.

\subsection{Gaussianity of evolving state}
\label{sec:Gaussianity}

A basic property of the Schr\"odinger evolution (\ref{eq:87}), which is crucial to our work, is that at any instant $t>0$ the pure state $\psi(t)$ is Gaussian. This Gaussianity inherits from that of the pre-quench state $\psi_0$. To understand this property we need to introduce the first and second moments of $\hat{\r}$, defined as $\langle \psi(t)|\hat{\bf r}|\psi(t)\rangle$ and $\langle\psi(t)|\hat{\bf r}\hat{\bf r}|\psi(t)\rangle\equiv G(t)$, respectively, whose components are correspondingly $\langle \psi(t)|\hat{r}_i|\psi(t)\rangle$ and $G_{ij}(t)=\langle \psi(t)|\hat{r}_i\hat{r}_j|\psi(t)\rangle$. It is easy to show that the former vanishes and the second is
\begin{equation}\label{eq:77}
  G(t)=\gamma(t)+iJ/2.
\end{equation}
The real part, $\gamma(t)=\{\gamma_{ij}(t)\}$, is the so-called covariance matrix \cite{Adesso07,Eisert_RMP_2010,Hackl_PRA_2018}
\begin{eqnarray}
&&\gamma(t)\equiv\frac{1}{2}\left\langle\psi(t)\left|\{\hat{\bf r},\, \hat{\bf r}\}\right|\psi(t)\right\rangle\nonumber\\
&\Leftrightarrow&
\gamma_{ij}(t)\equiv {1\over 2}\langle \psi(t)|\{\hat{r}_i,\hat{r}_j\}|\psi(t)\rangle,
\label{eq:81}
\end{eqnarray}
with $\{\cdot,\cdot\}$ being the anticommutator, which evolves with time, whereas the imaginary part is $J/2$, which is independent of time. Then, by straightforward calculations one finds that any instantaneous correlations of $h\in \mathbb{N}$ operators: $\hat{r}_1,\,\cdots,\,\hat{r}_{h}$ are given by
\begin{eqnarray}
\label{eq:83}
  &&\langle \psi(t)|\hat{r}_1\cdots\hat{r}_{h}|\psi(t)\rangle\nonumber\\
  &=&\left\{\begin{array}{ll}
                                                                     \sum_{P} \prod_{n=1}^{m}G_{i_ni'_n}(t), & {\rm for}\,\,h=2m, m\in\mathbb{N}; \\
                                                                     0 &  {\rm for}\,\,h=2m+1, m\in \mathbb{N},
                                                                   \end{array}
  \right.,\quad
\end{eqnarray}
where the sum is over all permutations $P:\,i_1\rightarrow i'_1\rightarrow ...\rightarrow i_m\rightarrow i'_m$ of the sequence: $1\rightarrow 2\rightarrow ...\rightarrow 2m$ satisfying $i_{n+1}-i_n>1$ and $i_n<i'_n$. This gives Wick's theorem and is a defining property of the Gaussian state. It implies that the second moment, i.e., the covariant matrix completely determines correlation properties at state $\psi(t)$.

The Gaussianity of the total system state $\psi(t)$, which is pure, can be transferred to the subsystem state which, in contrast, is mixed. Consider the bipartition of the chain, with $A$ being a subsystem and $B$ being its complement (Fig.~\ref{setting}). The former consists of $L_A$ contiguous oscillators. (We assume $L_A \gg 1$ throughout.) The state of subsystem $A$ is described by the reduced density of matrix $\hat{\rho}_A(t)\equiv {\rm Tr}_B(|\psi(t)\rangle\langle\psi(t)|)$, where the trace is restricted to the complement $B$. To analyze the properties of $\hat{\rho}_A(t)$ we project the vector $\hat{\r}$ onto $A$:
\begin{equation}\label{eq:88}
  \hat{\r}_A\equiv\bigoplus_{r=1}^{L_A}\left(
                               \begin{array}{c}
                                 \hat{x}_r \\
                                 \hat{p}_r \\
                               \end{array}
                             \right)
\end{equation}
for which the commutation relation reads
\begin{equation}\label{eq:89}
  [\hat{\r}_A,\,\hat{\r}_A]=iJ_A,\quad J_A\equiv\left[\begin{array}{cc}0&\mathbb{I}_{L_A}\\ -\mathbb{I}_{L_A}&0\end{array}\right].
\end{equation}
So for subsystem $A$ we can also introduce the first and second moments of $\hat{\r}_A$, read ${\rm Tr}_A(\hat{r}_{Ai} \hat{\rho}_A(t))$  and ${\rm Tr}_A(\hat{r}_{Ai}\hat{r}_{Aj} \hat{\rho}_A(t))\equiv (G_A(t))_{ij}$, respectively, with the trace restricted to $A$. One can readily see that these two moments are identical to $\langle \psi(t)|\hat{r}_i|\psi(t)\rangle$ and $G_{ij}(t)$, respectively, with the indexes $i,j$ restricted to $A$. As such Eq.~(\ref{eq:77}) is transferred to
\begin{equation}\label{eq:90}
  G_A(t)=\gamma_A(t)+i{J_A/2}
\end{equation}
with
\beq
\gamma_A(t)=\frac{1}{2}{\rm Tr}_A\left(\{\hat{\bf r}_A,\, \hat{\bf r}_A\}\hat{\rho}_A(t)\right).
\label{eq:91}
\eeq
The matrices: $G_A,\,\gamma_A,\,J_A$ are respectively a submatrix of $G,\,\gamma,\,J$, with their entries restricted to $A$. So, upon restricting all indexes to $A$, Eq.~(\ref{eq:83}) gives
\begin{eqnarray}
\label{eq:92}
  &&{\rm Tr}_A\left(\hat{r}_1\cdots\hat{r}_{h}\hat{\rho}_A(t)\right)\nonumber\\
  &=&\left\{\begin{array}{ll}
                                                                     \sum_{P} \prod_{n=1}^{m}(G_A(t))_{i_ni'_n}, & {\rm for}\,\,h=2m, m\in\mathbb{N}; \\
                                                                     0 &  {\rm for}\,\,h=2m+1, m\in \mathbb{N}.
                                                                   \end{array}
  \right.\quad
\end{eqnarray}
This shows that $\hat{\rho}_A(t)$ is a Gaussian mixed state.

With Eqs.~(\ref{eq:89})-(\ref{eq:92}), one can find $\hat{\rho}_A(t)$ explicitly. To this end we note that Eqs.~(\ref{eq:89})-(\ref{eq:92}) are invariant under the symplectic transformation:
\begin{equation}\label{eq:94}
  \hat{\r}_A \rightarrow \hat{\r}_A'\equiv S\hat{\r}_A,\quad S\in {\rm Sp}(2L_A,\mathbb{R}),
\end{equation}
where $S$ is a $2L_A\times 2L_A$ matrix belonged to the real symplectic group ${\rm Sp}(2L_A,\mathbb{R})$. By definition of that group all matrix elements of $S$ are real, and
\begin{equation}\label{eq:95}
  S J_A S^{\rm T}=J_A \Leftrightarrow S^{\rm T} J_A S=J_A
\end{equation}
is satisfied, i.e., the symplectic form is preserved. Furthermore, because $\gamma_A$ is positive definite and real symmetric, by the Williamson theorem \cite{Williamson36} one can find a special element of ${\rm Sp}(2L_A,\mathbb{R})$, denoted as $S_{\rm w}$, and transform $\gamma_A$ to a diagonal form, the so-called Williamson normal form, as
\begin{equation}\label{eq:96}
  S_{\rm w} \gamma_A S_{\rm w}^{\rm T}=\Lambda \otimes \mathbb{I}_2, \quad \Lambda={\rm diag}(\lambda_1,...,\lambda_{L_A}),
\end{equation}
where $\lambda_r$ is positive and is called symplectic eigenvalue, and the unit matrix $\mathbb{I}_2$ is defined in the $\hat{x}$-$\hat{p}$ sector. It is important that this is not a similarity transformation, and thus in general $\lambda_r$'s are not the eigenvalues of $\gamma_A$. Also, note that both $S_{\rm w}$ and the symplectic eigenvalue spectrum $\{\lambda_r\}$ evolve with time. However, because the matrix
\beq
C(t)\equiv iJ_A\gamma_A(t)=- S_{\rm w}^{\rm T} \left( \Lambda \otimes\sigma_2\right) (S_{\rm w}^{\rm T})^{-1},\label{cor_mat}
\eeq
with the Pauli matrix $\sigma_2$ being defined in the $\hat{x}$-$\hat{p}$ sector, it has the same instantaneous eigenvalue spectrum as the matrix: $-\Lambda \otimes\sigma_2$, whose spectrum is $\{\pm\lambda_r(t)\}$. Thus the eigenvalue spectrum of $C_A(t)$ are real and composed of symmetric positive and negative branch: The positive branch gives the symplectic eigenvalue spectrum of $\gamma_A(t)$. Later on we shall see that it is this property that renders $C_A(t)$ playing much the same role as the correlation matrix in entanglement dynamics of free-fermion models \cite{Peschel_JPA_2003, Vidal_PRL_2003, Peschel_JPA_2009}.

In the basis of the so-called Williamson mode:
\begin{equation}\label{eq:98}
  \hat{\r}_{\rm w} \equiv S_{\rm w}\hat{\r}_A\equiv \bigoplus_{r=1}^{L_A}\left(
                               \begin{array}{c}
                                 \hat{x}_{{\rm w}r} \\
                                 \hat{p}_{{\rm w}r} \\
                               \end{array}
                             \right),
\end{equation}
Eq.~(\ref{eq:90}) takes a simple form as
\begin{equation}\label{eq:99}
  G_A(t)= \left[\begin{array}{cc}
                                       \Lambda & {i\over 2}\,\mathbb{I}_{L_A} \\
                                       -{i\over 2}\,\mathbb{I}_{L_A} & \Lambda
                                     \end{array}
  \right].
\end{equation}
Thanks to this expression and that $\hat{\rho}_A(t)$ obeys Wick's theorem, $\hat{\rho}_A(t)$ must take the form as
\begin{equation}\label{eq:100}
  \hat{\rho}_A(t)=\bigotimes_{r=1}^{L_A} (1-{\rm e}^{-\Omega_r}){\rm e}^{-\Omega_r \hat{a}_{{\rm w}r}^\dagger \hat{a}_{{\rm w}r}}.
\end{equation}
Here $\hat{a}_{{\rm w}r}^\dagger$ and $\hat{a}_{{\rm w}r}$ are respectively the creation and annihilation operator of Williamson mode $r$, given by ($\hat{x}_{{\rm w}r},\,\hat{p}_{{\rm w}r}$):
\begin{equation}\label{eq:102}
  \hat{a}_{{\rm w}r}^\dagger={\hat{x}_{{\rm w}r}-i\hat{p}_{{\rm w}r}\over \sqrt{2}},\quad
  \hat{a}_{{\rm w}r}={\hat{x}_{{\rm w}r}+i\hat{p}_{{\rm w}r}\over \sqrt{2}}.
\end{equation}
To determine $\Omega_r(t)$ we substitute Eq.~(\ref{eq:100}) back to Eq.~(\ref{eq:91}), giving
\begin{eqnarray}\label{eq:103}
  \lambda_r(t)&=&{\rm Tr}_A\left(\hat{x}_{{\rm w}r}^2\hat{\rho}_A(t)\right)={\rm Tr}_A\left(\hat{p}_{{\rm w}r}^2\hat{\rho}_A(t)\right)\nonumber\\
&=&{\rm Tr}_A\left(\hat{a}_{{\rm w}r}^\dagger \hat{a}_{{\rm w}r}\hat{\rho}_A(t)\right)+{1\over 2}.
\end{eqnarray}
It implies that $\lambda_r(t)-1/2$ is the average occupation number of Williamson mode $r$. As a simple corollary,
\begin{equation}\label{eq:104}
  \lambda_r(t)\geq 1/2,
\end{equation}
which reflects the Heisenberg uncertainty. Calculating the average occupation number by using Eq.~(\ref{eq:100}), we reduce Eq.~(\ref{eq:103}) to
\begin{eqnarray}\label{eq:105}
  \lambda_r(t)={1\over e^{\Omega_r(t)}-1}+{1\over 2}
  \Rightarrow\Omega_r(t)&=&\ln\frac{\lambda_r(t)+1/2}{\lambda_r(t)-1/2}.\,\,
\end{eqnarray}
Equation (\ref{eq:100}), together with Eqs.~(\ref{eq:98}), (\ref{eq:102}) and (\ref{eq:105}) gives a closed form for the evolving reduced density of matrix. Finally, we remark that (i) the Williamson modes are different from the normal modes in Eq.~(\ref{eq:106}), and (ii) the Gaussianity can be also studied by using the Weyl representation familiar in quantum optics \cite{Shi18,Shi20}.

\subsection{\label{sec_2_2}Entanglement probes}
We focus on the entanglement entropy $S(t)\equiv -{\rm Tr}_A(\hat{\rho}_A(t)\ln \hat{\rho}_A(t))$ and the $n$-th order R\'{e}nyi entropy $S_n(t)\equiv {1\over 1-n}\ln {\rm Tr}_A(\hat{\rho}_A(t))^n,\,n=2,3,...$. These probes of entanglement dynamics are generally denoted as $O(t)$. With the substitution of Eqs.~(\ref{eq:98}), (\ref{eq:100}), (\ref{eq:102}) and (\ref{eq:105}), one finds
\begin{widetext}
\beq
S(t)&=&\sum_{r=1}^{L_A}\left((\lambda_{r}(t)+\frac{1}{2})\ln(\lambda_{r}(t)+\frac{1}{2})
-(\lambda_{r}(t)-\frac{1}{2})\ln(\lambda_{r}(t)-\frac{1}{2})\right)\nonumber\\
&=&{1\over 2}\sum_{r=1}^{L_A}\left(\ln\left((2\lambda_{r}(t)+1)(2\lambda_{r}(t)-1)\right)
+2\lambda_{r}(t)\ln{2\lambda_{r}(t)+1\over 2\lambda_{r}(t)-1}\right)-L_A\ln 2
\label{EE}
\eeq
and
\beq
S_n(t)&=&\frac{1}{n-1} \sum_{r=1}^{L_A} \ln \left((\lambda_{r}(t)+\frac{1}{2})^n-(\lambda_{r}(t)-\frac{1}{2})^n \right)\nonumber\\
&=&\frac{1}{2(n-1)} \sum_{r=1}^{L_A} \ln \left((\lambda_{r}(t)+\frac{1}{2})^n-(\lambda_{r}(t)-\frac{1}{2})^n \right)^2-L_A\,{n\ln 2\over n-1}.
\label{RE}
\eeq
They are completely determined by the evolving symplectic eigenvalue spectrum $\{\lambda_r(t)\}$. These expressions are convenient for numerical simulations \cite{Calabrese17}, but not for developing a fluctuation theory.

For the latter purpose, we notice an important consequence of the Gaussianity of $\hat{\rho}_A(t)$. That is, with the introduction of $C(t)$ defined in Eq.~(\ref{cor_mat}) and its eigenvalue properties described before, we can invoke the Gaussianity to rewrite the second line of Eqs.~(\ref{EE}) and (\ref{RE}) as
\beq\label{eq:108}
O(t)={1\over 4}\operatorname{Tr}_A h(C(t)),
\eeq
where $h(C)$ is a function of $C$:
\beq
\label{eq:107}
h(C)\equiv\left\{
\begin{array}{ll}
\ln\left((2C+\mathbb{I}_{2L_A})(2C-\mathbb{I}_{2L_A})\right)+2C\ln{2C+\mathbb{I}_{2L_A}\over 2C-\mathbb{I}_{2L_A}}-(2\ln 2)\, \mathbb{I}_{2L_A}, & \quad \textrm{for}\, O=S, \\
\frac{1}{n-1} \ln \left( (2C + \mathbb{I}_{2L_A})^n -  (2C - \mathbb{I}_{2L_A})^{n} \right)^2 - {(2\ln 2)n\over n-1}\,\mathbb{I}_{2L_A}, & \quad \textrm{for}\, O=S_n.
\end{array}
\right. \quad
\eeq
\end{widetext}
Note that although the matrices: $2C\pm \mathbb{I}_{2L_A}$ are not positive definite, those under the logarithms are. Therefore, the matrix logarithms above are well defined, which are the so-called {\it principal logarithm} \cite{Higham08}. Equations (\ref{eq:108}) and (\ref{eq:107}) lay a basis for our systematic analytical treatments of the temporal fluctuations in entanglement dynamics.

\subsection{Boson-fermion duality}
\label{sec:fermionic_dual}

In this part we derive a fermion dual of the entanglement dynamics free-boson systems. We should emphasize that
the duality does {\it not} mapped bosonic entanglement dynamics onto a fermionic one, i.e. does {\it not} fermionize bosonic entanglement dynamics. However, it indicates that the two classes of entanglement dynamics share some common aspects.

The Hamiltonian of a generic free-boson system takes the form as
\begin{equation}
    \label{eq:121}
    \hat{H}_b=\frac{1}{2}\,\hat{\bf r}^\mathrm{T}\Omega_b \hat{\bf r},\quad
    \hat{\r}={1\over \sqrt{2}}\bigoplus_{r=1}^{L}\left[\begin{array}{cc}
                                        1 & 1 \\
                                        -i & i
                                      \end{array}
    \right]\left(
                               \begin{array}{c}
                                 \hat{a}_r \\
                                 \hat{a}^\dagger_r \\
                               \end{array}
                             \right).
\end{equation}
Here $\hat{\bf r}$ is defined in Eq.~\eqref{eq:79}, and $\Omega_b$ is a real symmetric and positive-definite matrix of size $2L$. This Hamiltonian has a coupling between the position $\hat{x}$ and momentum $\hat{p}$ operators, a property which is not carried by the Hamiltonian (\ref{eq:80}) of an oscillator chain. By
the Williamson theorem there exists a real symplectic matrix ${\cal G}_S \in {\rm Sp}(2L,\mathbb{R})$ such that
\begin{equation}
    \label{eq:122}
    {\cal G}^\mathrm{T}_S\,\Omega_b\,{\cal G}_S= D \otimes \mathbb{I}_2,\quad
    D=\text{diag}(d_1,d_2,\cdots,d_L),
\end{equation}
where the symplectic eigenvalues $d_r>0$. Passing to the operator bases $\hat{\bf r}'={\cal G}_S^{-1}\hat{\bf r}=\bigoplus_{r=1}^L\begin{pmatrix}
    \hat{x}'_r \\
    \hat{p}'_r
    \end{pmatrix}$, the Hamiltonian reads
\begin{equation}
    \label{eq:123}
    \hat{H}_b=\frac{1}{2}\sum_{r=1}^L d_r(\hat{x}^{\,\prime 2}_r+\hat{p}_r^{\,\prime 2})=\mathop{\sum}\limits_{r=1}^Ld_r\hat{a}_r^{\,\prime \dagger}\hat{a}_r^{\,\prime}.
\end{equation}
This shows that $d_r$ is the frequency of the $r$-th oscillator spectrum in the bases $\hat{\bf r}'$. The evolution of a Gaussian state
$\psi_b(t)$ is completely determined by the instantaneous covariance matrix $\gamma_b(t)\equiv \frac{1}{2}\langle{\psi_b(t)}|\{ \hat{\bf r},\hat{\bf r}\} |{\psi_b(t)}\rangle$, which satisfies
\begin{equation}
    \label{eq:124}
    \gamma_b(t)=S_b(t)S^\mathrm{T}_b(t), \quad\partial_t S_b(t)=J\Omega_bS_b(t).
\end{equation}
As before we restrict $\gamma_b$ to the subsystem A, and the ensuing matrix is denoted as $\gamma_{bA}$. Furthermore, we define $C_b(t) \equiv i J_A \gamma_{bA}(t)$. Then
the entanglement entropy $S(t)$ is given by Eqs.~(\ref{eq:108}) and (\ref{eq:107}). For the convenience of deriving the fermionic dual below, we rewrite it as
\begin{equation}
    \label{eq:125}
    S(t)=\frac{1}{4}\mathrm{Tr}_A \tilde{h}(C_b(t)),
\end{equation}
where the matrix function
\begin{eqnarray}
    \label{eq:126}
    \tilde{h}(C)&\equiv&\ln | 4C^2-\mathbb{I}_{2L_A}|+2C\ln \left| \frac{2C+\mathbb{I}_{2L_A}}{2C-\mathbb{I}_{2L_A}}\right|\nonumber\\
    &-&(2\ln 2)\,\mathbb{I}_{2L_A},
\end{eqnarray}
with $|X|$ defined for positive- or negative-definite $X$: For positive-definite $X$ we have $|X|=X$ and have $|X|=-X$ for negative-definite $X$.

Now we wish to show that the entanglement dynamics above has a fermionic dual. To this end we first replace the canonical conjugate in Eq.~\eqref{eq:79} by a pair of Majorana fermions: $(\hat{f}_{1r},\hat{f}_{2r})$, so that a $2L$-component operator vector
\begin{equation}
    \label{eq:127}
    \hat{\bf f}=\bigoplus_{r=1}^L\begin{pmatrix}
        \hat{f}_{1r} \\
        \hat{f}_{2r}
    \end{pmatrix}={1\over \sqrt{2}}\bigoplus_{r=1}^{L}\left[\begin{array}{cc}
                                        1 & 1 \\
                                        -i & i
                                      \end{array}
    \right]\left(
                               \begin{array}{c}
                                 \hat{b}_r \\
                                 \hat{b}^\dagger_r \\
                               \end{array}
                             \right)
\end{equation}
results, where $\hat{b}_r,\,\hat{b}^\dagger_r$ are annihilation and creation operators of usual fermions. In the Majorana bases a generic free-fermion Hamiltonian reads
\begin{equation}
    \label{eq:128}
    \hat{H}_f=\frac{i}{2}\,\hat{\bf f}^\mathrm{T}\Omega_f\hat{\bf f},
\end{equation}
where $\Omega_f$ is a real anti-symmetric matrix of size $2L$. There exists a real orthogonal matrix ${\cal G}_{O} \in {\rm O}(2L,\mathbb{R})$ such that
\begin{eqnarray}
    \label{eq:129}
    {\cal G}_{O}^\mathrm{T}\,\Omega_f\, {\cal G}_{O}=D^{\,\prime}\otimes
    \left[\begin{matrix}
        0 & 1 \\
        -1 & 0
    \end{matrix}\right],\qquad\nonumber\\
    D^{\,\prime}=\text{diag}(d_1^{\,\prime},d_2^{\,\prime},\cdots,d_L^{\,\prime}).\qquad
\end{eqnarray}
In the operator bases $\hat{\bf f}^{\,\prime}={\cal G}_O^{-1}\hat{\bf f}$, the Hamiltonian \eqref{eq:128} is written as
\begin{equation}
    \label{eq:130}
    \hat{H}_f=i\sum_{r=1}^L d_r^{\,\prime}\hat{f}^{\,\prime}_{1r}\hat{f}^{\,\prime}_{2r}.
\end{equation}
Since the eigenvlaues of $i\hat{f}^{\,\prime}_{r_1}\hat{f}^{\,\prime}_{r_2}$ are $\pm\frac{1}{2}$, it may be considered as the Majorana dual of the Hamiltonian (\ref{eq:123}), provided the two spectra are identical, i.e.
\begin{equation}\label{eq:131}
  D'=D.
\end{equation}
Under this requirement, we construct ${\cal G}_S\equiv \Omega_b^{-1/2}{\cal G}_O(D^{1/2}\otimes \mathbb{I}_2)$, so that on the bosonic side Eq.~(\ref{eq:122}) is satisfied. Here ${\cal G}_O$ depends on $O$, and must solve the symplectic condition:
\begin{eqnarray}
    \label{eq:132}
    && {\cal G}_S\,J\,{\cal G}_S\,^\mathrm{T}=J \nonumber\\
    &\Rightarrow& \Omega_f={\cal G}_O\,D\otimes \left[\begin{matrix}
        0 & 1 \\
        -1 & 0
    \end{matrix}\right]{\cal G}_O\,^\mathrm{T}=\Omega_b^{1/2}J\Omega_b^{1/2},
\end{eqnarray}
which provides a dual between $\Omega_f$ and $\Omega_b$.

The evolution of a Gaussian state $\psi_f(t)$ is completely determined by the instantaneous covariance matrix $\gamma_f(t) \equiv
\frac{i}{2}\langle{\psi_f(t)}|[\hat{\bf f},\hat{\bf f}]|{\psi_f(t)}\rangle$, which satisfies
\begin{equation}
    \label{eq:133}
    \gamma_f(t)=-S_f(t)JS_f^\mathrm{T}(t),\quad\partial_t S_f(t)=\Omega_fS_f(t).
\end{equation}
Restricting $\gamma_f(t)$ to A so that a matrix $\gamma_{fA}(t)$ results, and defining $C_f(t)\equiv i\gamma_{fA}(t)$, one can follow standard procedures \cite{Peschel_JPA_2003,Vidal_PRL_2003,Peschel_JPA_2009} to find the entanglement entropy
\begin{equation}
    \label{eq:134}
    S(t)=-\frac{1}{4}\mathrm{Tr}_A \tilde{h}(C_f(t)),
\end{equation}
with $\tilde{h}(C)$ given by Eq.~\eqref{eq:126}. (Note that there is an overall minus sign here.)

Equations \eqref{eq:127}, \eqref{eq:128}, \eqref{eq:131}-\eqref{eq:134} constitute a fermion dual of the entanglement dynamics of the boson system Eq.~\eqref{eq:121}. This duality suggests that, formally, the entanglement dynamics of free boson and fermion systems are similar. However, this does not mean that we can find the entanglement dynamics of boson systems from that of fermion systems and {\it vice versa}. Indeed, despite the similarity between Eqs.~\eqref{eq:125} and \eqref{eq:134}, because the
eigenvalue spectrum of $C_b$ and $C_f$ are very different --- the former belongs to $(-\infty,-\frac{1}{2} )\cup(\frac{1}{2},\infty)$ and the latter to $(-\frac{1}{2},\frac{1}{2})$ --- the detailed results differ dramatically.

\begin{figure}
\includegraphics[width=0.95\linewidth]{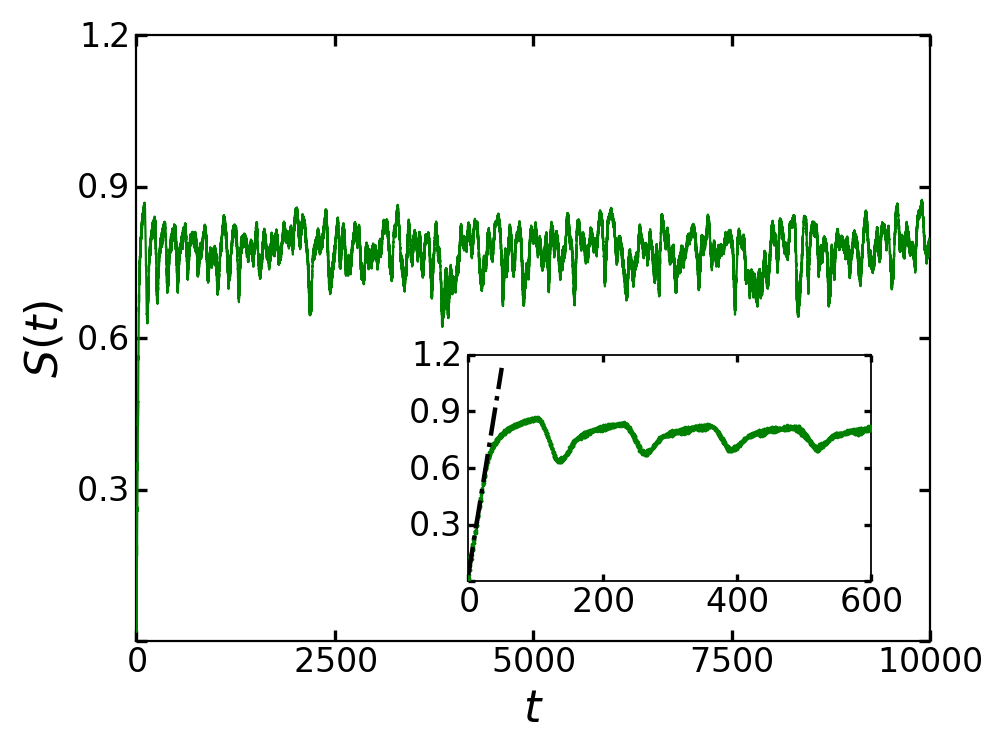}
\caption{A representative simulation result of the time evolution of entanglement entropy. Inset: At short time the evolution displays a linear growth followed by revivals. Main panel: At long time the evolution displays persistent fluctuations. Here $L_A=25$ and $L=124$. The quench protocol is $\omega^2:\,1.5\rightarrow 2.5$ while $K=1$ as in the rest of the work.}
\label{F2}
\end{figure}
\subsection{\label{sec_2_3}Scope of the paper}
The present work is motivated by the above duality, and by numerical simulations of long-time entanglement evolution based on Eqs.~(\ref{EE}) and (\ref{RE}). As shown in the inset of Fig.~\ref{F2}, after the quench, in the initial time window the entanglement displays a linear rise followed by revivals. This behavior has been well described by the Cardy-Calabrese picture of quasiparticles propagating with a finite speed \cite{Calabrese_JSM_2005, Schachenmayer_PRX_2013, Hackl_PRA_2018, Alba_SPP_2018}. However, as clearly shown in the main panel of Fig.~\ref{F2}, after long time evolution (i.e., beyond the revival regime) the entanglement shows a completely different dynamical behavior, namely, displaying persistent fluctuations. These temporal fluctuations resemble noisy behaviors, but they are completely reproducible under the same initial condition and the same quench protocol. This kind of fluctuation phenomena in entanglement dynamics were studied analytically, to the best of our knowledge, in Ref.~\cite{Lih-King_2023} for the first time, but only for free fermion models and spin chains.

The aim of this work is to develop an analytical theory of fluctuation statistics in long-time entanglement dynamics of quantum harmonic oscillator chains, and perform numerical experiments to test analytical predictions. Technically, a key issue to be addressed is the \textit{full} probability distribution of entanglement fluctuations. This is a highly challenging task. Our main strategy is utilizing the Gaussianity of the evolving state reviewed in Sec.~\ref{sec:Gaussianity} and generalizing the temporal entanglement fluctuation theory of free-fermion models \cite{Lih-King_2023} to present free-boson model. Armed with the developed theory, we are able to derive that distribution and study variances of various entanglement probes. We then find that the long-time entanglement dynamics exhibits strictly the same statistical behaviors as that of free-fermion models. This result shows that persistent temporal fluctuations in entanglement dynamics of coupled harmonic oscillator chains not only fall into the same paradigm of mesoscopic fluctuations, as occurring to free-fermion models, but also exhibit an unexpected universality, namely, the boson-fermion universality.

\section{\label{sec_3}Temporal fluctuations and statistical equivalence}
\subsection{A common property of the time dependence of various quantities}
\label{sec:general_properties}
As we discussed in Sec.~\ref{sec_2_2}, various entanglement probes are determined by the time-dependent covariance matrix $\gamma_A(t)$ or its equivalent $C(t)$ defined in Eq.~(\ref{cor_mat}). Given the evolving state $|\psi(t)\rangle$, the latter takes the form of a block-Toeplitz matrix structure, i.e., its element is a $2\times 2$ block defined in the $\hat{x}$-$\hat{p}$ sector, which is introduced by the representation Eq.~(\ref{eq:79}) of $\hat{\r}$, and depends on the entries namely the oscillator indexes $r,\,r'$ only through their displacement $r-r'$. That is, $C_A(t)$ takes the form as
\beq
\left[\begin{array}{cccccc}\tau_0&\tau_1&\tau_2&\ldots&\tau_{L_A-2}&\tau_{L_A-1}\\
\tau_{-1}&\tau_0&\tau_1&\ldots&\tau_{L_A-3}&\tau_{L_A-2}\\
\tau_{-2}&\tau_{-1}&\tau_0&\ldots&\tau_{L_A-4}&\tau_{L_A-3}\\
\vdots& \vdots &\vdots&\ddots &\vdots&\vdots\\
\tau_{-(L_A-2)}& \tau_{-(L_A-3)} &\tau_{-(L_A-4)} &\ldots&\tau_0&\tau_1\\
\tau_{-(L_A-1)}& \tau_{-(L_A-2)} &\tau_{-(L_A-3)} &\ldots&\tau_{-1}&\tau_0\end{array}\right]\!\!,\quad\,\,\,\,
\label{CFM}
\eeq
where the matrix elements $\tau_{l}$, $l=0,\pm 1,\ldots,\pm(L_A-1)$ are block elements. Each element can be separated into two parts, one time-independent and the other time-dependent, i.e.,
\begin{equation}\label{eq:53}
  \tau_l(t)=c_{0,l}+c_{1,l}(t).
\end{equation}
They are given respectively as
\beq
c_{0,l}&=&\left[\begin{array}{cc} 0 &i \bar{p}_{0,l}\\ -i \bar{x}_{0,l}&0 \end{array}\right],\label{eq:64}\\
c_{1,l}(t)&=&\left[\begin{array}{cc}i \bar{m}_l(t)&i \bar{p}_l(t)\\ -i \bar{x}_l(t)&-i \bar{m}_l(t)\end{array}\right],
\label{Ct}
\eeq
with
\beq
\begin{aligned}
&\bar{x}_{0,l} = \frac{1}{L}\sum\limits_k\frac{E_+(k)}{s_k}\cos(kl),\\
&\bar{p}_{0,l} = \frac{1}{L}\sum\limits_k E_+(k)\,s_k\cos(kl)\\
\end{aligned}
\eeq
being independent of time and
\beq
\begin{aligned}
&\bar{x}_l(t) = \frac{1}{L}\sum\limits_k\frac{E_-(k)}{s_k}\cos(kl) \cos(2\omega_k t),\\
&\bar{m}_l(t) = -\frac{1}{L}\sum\limits_kE_-(k)\cos(kl) \sin(2\omega_k t),\\
&\bar{p}_l(t) = -\frac{1}{L}\sum\limits_kE_-(k)\,s_k\cos(kl) \cos(2\omega_k t)
\label{xpm_t}
\end{aligned}
\eeq
depending on time. The derivation of these expressions as well as the explicit form of $E_{\pm}(k)$ are given by Eq.~(\ref{eq:97}) in Appendix~\ref{App_A}. Inheriting from Eq.~(\ref{eq:53}), the separation of $C(t)$ into the time-independent and time-dependent part follows,
\begin{equation}\label{eq:54}
  C(t)=C_0+C_1(t)
\end{equation}
where $C_0$ and $C_1$ are the time-independent and time-dependent part, respectively. Moreover, both $C_0$ and $C_1$ are block-Toeplitz matrix, whose elements are $c_{0,l}$ and $c_{1,l}$, respectively.

Equation (\ref{xpm_t}) shows that the time parameter enters solely as the dynamical phase factor $2\omega_k t$ in the argument of the sine or cosine function in the linear sums. In fact, due to the even parity of the spectrum $\omega_k$ there are only
\begin{equation}\label{eq:68}
  N = L/2 + 1
\end{equation}
distinct frequencies, which we denote as
\begin{equation}\label{eq:76}
  \boldsymbol{\omega}\equiv(\omega_{0}, \omega_{1}, \cdots, \omega_{N-1}).
\end{equation}
Then this time dependence is respected by writing $C(t)$ as a $N$-variable matrix-valued function $\widetilde{C}$, read
\beq
C(t)=\widetilde{C}(\omega_{0} t, \omega_{1} t, \cdots, \omega_{N-1} t)\equiv \widetilde{C}(\boldsymbol{\omega}t).
\label{eq:47}
\eeq
According to Eq.~(\ref{CFM}), we can separate the auxiliary function $\widetilde{C}(\boldsymbol{\varphi})$, $\boldsymbol{\varphi}\equiv (\varphi_0,\varphi_1,\ldots,\varphi_{N-1})$ into $\boldsymbol{\varphi}$-independent and $\boldsymbol{\varphi}$-dependent parts as
\begin{equation}\label{eq:49}
  \widetilde{C}(\boldsymbol{\varphi})=C_0+\widetilde{C}_1(\boldsymbol{\varphi}).
\end{equation}
Here the $\boldsymbol{\varphi}$-dependent part $\widetilde{C}_1(\boldsymbol{\varphi})$ has the same structure as $C_1(t)$, and its matrix elements are
\beq
\begin{aligned}
&\widetilde{x}_l(\boldsymbol{\varphi}) = \frac{2}{L}\sum\limits_{m=0}^{N-1}\bar{\delta}_m\frac{E_-(k_m)}{\omega_{k_m}}\cos{({k_m}l)}\cos(\varphi_m),\\
&\widetilde{m}_l(\boldsymbol{\varphi}) = -\frac{2}{L}\sum\limits_{m=0}^{N-1}\bar{\delta}_m \,E_-(k_m)\cos{({k_m}l)}\sin(\varphi_m),\\
&\widetilde{p}_l(\boldsymbol{\varphi}) = -\frac{2}{L}\sum\limits_{m=0}^{N-1}\bar{\delta}_m\, E_-(k_m)\,\omega_{k_m}\cos{({k_m}l)}\cos(\varphi_m),
\end{aligned}
\label{Cre}
\eeq
corresponding to Eq.~(\ref{xpm_t}), where $\bar{\delta}_m = 1-(\delta_{m,0}+\delta_{m,L/2})/2$ and $k_m=2\pi m/L$.

Because $\widetilde{C}(\boldsymbol{\varphi})$ is $2\pi$-periodic in every argument, Eq.~(\ref{eq:47}) maps the time evolution of $C(t)$ onto a classical motion:
\beq
\boldsymbol{\varphi}=\boldsymbol{\omega}t\in \mathbb{T}^N
\label{eq:45}
\eeq
on a $N$-dimensional torus $\mathbb{T}^{N}$, with constant angular frequencies $\boldsymbol{\omega}$. More generally, an evolving entanglement probe $O(t)$ can be expressed as Eq.~(\ref{eq:108}). From that expression we have
\beq
O(t)=\widetilde{O}(\omega_{0} t, \omega_{1} t, \cdots, \omega_{N-1} t)\equiv \widetilde{O}(\boldsymbol{\omega}t),
\label{eq:48}
\eeq
similar to Eq.~(\ref{eq:47}), where $\widetilde{O}(\boldsymbol{\varphi})$ is the corresponding auxiliary function uniquely determined by $O$, and is determined by $\widetilde{C}(\boldsymbol{\varphi})$ via
\beq\label{new_er}
\widetilde{O}(\boldsymbol{\varphi})=\frac{1}{4}\operatorname{Tr}_A h(\widetilde{C}(\boldsymbol{\varphi})).
\eeq
Here $h$ is the matrix function defined by Eq.~(\ref{eq:107}). From now on we will use the same symbol $C$ for $C(t)$ and $\widetilde{C}(\boldsymbol{\varphi})$.

\subsection{Quasi-periodic oscillations and emergent mesoscopic sample-to-sample fluctuations}

For generic energy spectrum the $N$ frequencies (\ref{eq:76}) carry an important property, namely, the incommensurality:
\begin{equation}\label{eq:69}
  \sum_{k=0}^{N-1} x_k \omega_k=0,\quad x_k\in \mathbb{Z}\, \Rightarrow \,x_1=\ldots=x_{N-1}=0.
\end{equation}
We shall assume this property in the remainder of this work. For incommensurate $\boldsymbol{\omega}$, by a well-known theorem \cite{Samoilenko07} $C(t)$ is quasi-periodic in $t$, and so is $O(t)$. Thus the persistent fluctuations displayed in long-time entanglement fluctuations, exemplified in Fig.~\ref{F2} for $O=S$, are quasi-periodic oscillations. They are reproducible, i.e., given identical initial state and quench protocol the time profiles $O(t)$ are strictly the same.

\begin{figure}
\includegraphics[width=0.8\linewidth]{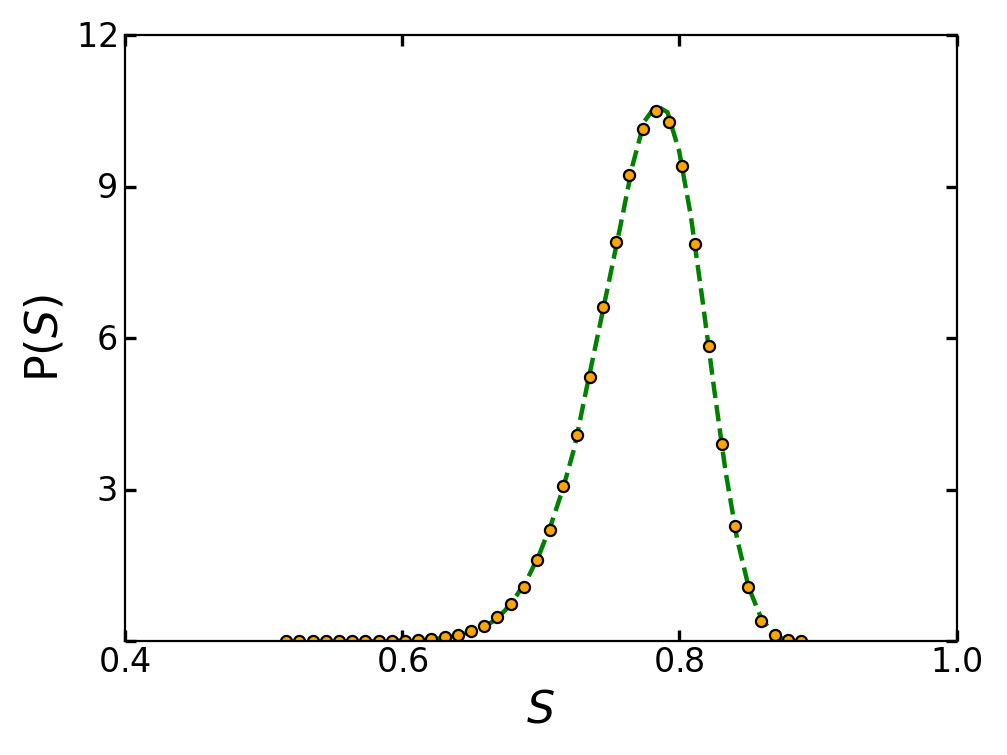}
\caption{Numerical experiments verify the statistical equivalence of two types of entanglement entropy fluctuations. The green dashed line is obtained from the time series $S(t)$, and orange dots are from the random function $\widetilde{S}(\boldsymbol{\varphi})$. $L_A=25$, $L=124$.}
\label{F3}
\end{figure}

Furthermore, because of the incommensurality of $\boldsymbol{\omega}$ the trajectory (\ref{eq:45}) must be ergodic and dense on $\mathbb{T}^{N}$. By the ergodic theorem \cite{Arnold_1989} we have for arbitrary interval $\Delta\subset\mathbb{R}$ and for any entanglement probe $O$,
\begin{equation}\label{eq:50}
  \lim_{T\rightarrow \infty}\int_{O(t)\in \Delta} \frac{dt}{T}
  =\int_{\widetilde{O}(\boldsymbol{\varphi})\in \Delta} {d\boldsymbol{\varphi}\over (2\pi)^N}.
\end{equation}
In words, after long-time evolution the frequency for the time series $O(t)$ to appear in $\Delta$ is identical to the probability  for  $\widetilde{O}(\boldsymbol{\varphi})$ to be in the same interval, i.e., $\mathbf{P}(O(\boldsymbol{\varphi})\in \Delta)$. More generally, for any set ${\cal A}\subset \mathbb{T}^N$, we have
\begin{equation}\label{eq:51}
  \lim_{T\rightarrow \infty}\int_{\boldsymbol{\omega} t\in {\cal A}} \frac{dt}{T}
  =\int_{\boldsymbol{\varphi}\in {\cal A}} {d\boldsymbol{\varphi}\over (2\pi)^N}\equiv \mathbf{P}(\boldsymbol{\varphi}\in {\cal A}).
\end{equation}
Consequently, out-of-equilibrium fluctuations displayed by the quasi-periodic oscillations $O(t)$ must be statistically equivalent to fluctuations of $\widetilde{O}(\boldsymbol{\varphi})$ with $\boldsymbol{\varphi}$, provided $\boldsymbol{\varphi}$ is drawn uniformly from $\mathbb{T}^N$. This statistical equivalence is confirmed numerically, exemplified by $O=S$ in Fig.~\ref{F3}.

Equations (\ref{eq:48}) and (\ref{new_er}) imply that each $\widetilde{C}(\boldsymbol{\varphi})$ describes a virtual mesoscopic --- recall that $N$ is finite --- disordered sample with a disorder realization $\boldsymbol{\varphi}$, and determines various entanglement properties characterized by $\widetilde{O}(\boldsymbol{\varphi})$ by the same token as $C(t)$ determines $O(t)$. Equation (\ref{eq:47}) further maps the quantum time evolution of $C(t)$ onto a classical ergodic trajectory $\boldsymbol{\varphi}=\boldsymbol{\omega}t$, that sweeps out the entire disorder ensemble, i.e., $\mathbb{T}^N$. As such, mesoscopic {\it sample-to-sample} fluctuations $\widetilde{O}(\boldsymbol{\varphi})$, familiar in mesoscopic electronic and optics physics \cite{Sheng_2006, Akkermans_2007}, emerge from entanglement dynamics, and are statistically equivalent to the time series $O(t)$. This was found previously for free fermion systems and interacting spin chains \cite{Lih-King_2023}. Finally, because $\cos(\varphi_0)$, $\cos(\varphi_1)$, ..., $\cos(\varphi_{N-1})$ [or $\sin(\varphi_0)$, $\sin(\varphi_1)$, ..., $\sin(\varphi_{N-1})$] in Eq.~(\ref{Cre}) are $N$ statistically independent random numbers of zero mean \cite{Kac}. As a result, $\tilde{x}_l(\boldsymbol{\varphi})$ in Eq.~(\ref{Cre}) is a sum of $N\sim L$ random numbers, and so are $\tilde{m}_l(\boldsymbol{\varphi})$ and $\tilde{p}_l(\boldsymbol{\varphi})$. So for large $L$ their magnitude scales with the total system size as $\sim 1/\sqrt{L}$. Therefore, remarkably, the disorder strength depends on the system size $L$ and vanishes in the limit $L\rightarrow\infty$: This peculiar property is {\it not} shared by any genuine mesoscopic systems, and is crucially responsible for the arising of new mesoscopic universalities.

\section{\label{sec_3_2}Boson-fermion universality of statistical distribution}

We proceed to study the full statistics of entanglement fluctuations. These fluctuations are characterized by various entanglement probe $O(\boldsymbol{\varphi})$, which are functions of $N$ independent random angles $\varphi_i$ ($i=0,1,\cdots,N-1$). [Hereafter we will use common symbol $O$ for the functions $O(t)$ and $\widetilde{O}(\boldsymbol{\varphi})$, similar to the case of the matrices $C(t)$ and $\widetilde{C}(\boldsymbol{\varphi})$.] These random variables are drawn from the uniform probability measure ${\mathbf P}$ above. We shall demonstrate that for different $O(\boldsymbol{\varphi})$, their statistical distributions hold the same, and most importantly, notwithstanding the present system is bosonic, the distribution is identical to that found in free fermion systems \cite{Lih-King_2023}.

\subsection{Description of the mathematical tool}

\subsubsection{Concentration of the product probability measure}

A fundamental difficulty in addressing the statistics of an entanglement probe $O(\boldsymbol{\varphi})$ is that it depends on the random variables $\boldsymbol{\varphi}$ in a highly nonlinear manner. To overcome this difficulty we observe that ${\mathbf P}$ is a {\it product probability measure} \cite{Ledoux_2001},
\begin{equation}\label{eq:4}
  {\mathbf P}=\mu_0\times\cdots\times\mu_{N-1}
\end{equation}
over the product space
\begin{equation}\label{eq:5}
  \mathbb{T}^N=\mathbb{T}_0\times\cdots\times\mathbb{T}_{N-1},
\end{equation}
where $\mu_i$ is the Lebesgue measure over the one-dimensional torus $\mathbb{T}_i\ni \varphi_i$. This product structure of probability measures lays down a foundation for the occurrence of the phenomenon of concentration of measure \cite{Ledoux_2001,Boucheron_2013}, yielding a \textit{nonasymptotic} probability theory: It can make statistical predictions for any finite $N$, without resorting to a limiting theorem. Specifically, below we use and solve the so-called modified logarithmic Sobolev inequality \cite{Boucheron_2013} to obtain some concentration inequalities describing the tail behaviors of the full distribution of $O(\boldsymbol{\varphi})$.

In general, a concentration inequality takes the following form: For any $\epsilon>0$,
\begin{equation}\label{eq:1}
  P_+(\epsilon)\equiv\mathbf{P}(O-\langle O\rangle\geq\epsilon)\leq \alpha(\epsilon)
\end{equation}
for upward fluctuations $O>\langle O\rangle$, and
\begin{equation}\label{eq:2}
  P_-(\epsilon)\equiv\mathbf{P}(\langle O\rangle-O\geq\epsilon)\leq \alpha(\epsilon)
\end{equation}
for downward fluctuations $O<\langle O\rangle$, where $\langle\cdot\rangle$ denotes the average with respect to $\mathbf{P}$, and $\alpha(\epsilon)$ is the so-called concentration function. It is important to note that $\alpha$ is {\it not} necessarily the same for the upward and downward fluctuations. Were they not be the same (and $\alpha$ is sharp), the distribution is asymmetric,
and $\alpha$ in (\ref{eq:1}) [respectively (\ref{eq:2})] bounds the large deviation behavior of the upper (respectively lower) tail. In particular, if $\alpha(\epsilon)\sim{\rm e}^{-\epsilon^2/2 b}$ with $b>0$, then the concentration is said to be sub-Gaussian, and $b$ is called the {\it variance factor}; if $\alpha(\epsilon)\sim{\rm e}^{-\epsilon/2c}$ then the concentration is said to be sub-exponential, and $c$ is called the {\it scale parameter} \cite{Tao2012}. It is the purpose of this section to find the explicit forms of $\alpha$  and compare them with the results for fermion systems. Below we shall focus mostly on the entanglement entropy, i.e., $O(\boldsymbol{\varphi})=S(\boldsymbol{\varphi})$, and the analysis and the final results hold similarly for other entanglement probes such as $S_n(\boldsymbol{\varphi})$.

\subsubsection{Modified logarithmic Sobolev inequality}
Owing to the product structure of the probability measure Eqs.~(\ref{eq:4}) and (\ref{eq:5}), a generic multivariable random entanglement probe function $O(\boldsymbol{\varphi})$ obeys a very general inequality, the so-called modified logarithmic Sobolev inequality \cite{Boucheron_2013}:
\begin{eqnarray}\label{Sobolev0}
u \left\langle O e^{u O}\right\rangle - \left\langle  e^{u O}\right\rangle \ln \left\langle  e^{u O}\right\rangle\qquad\qquad\nonumber\\
\leq \sum_{m=0}^{N-1}\left\langle e^{u O}    \,\phi(-u(O-O_m))\right\rangle, \quad \forall u\in \mathbb{R},
\end{eqnarray}
where
\begin{equation}\label{eq:6}
  \phi(x)\equiv e^x-x-1
\end{equation}
and $O_m$ is an arbitrary (real) function of $\boldsymbol{\varphi}'_m\equiv (\varphi_0,\varphi_1,\ldots,\varphi_{m-1},\varphi_{m+1},\ldots,\varphi_{N-1}) $:
\begin{eqnarray}\label{eq:84}
  \boldsymbol{\varphi}'_m\equiv (\varphi_0,\ldots,\varphi_{m-1},\varphi_{m+1},\ldots,\varphi_{N-1})\in \mathbb{T}^{N-1} \quad\nonumber\\
  \mapsto O_m (\boldsymbol{\varphi}'_m)
  \in \mathbb{R}. \qquad\qquad\qquad\qquad
\end{eqnarray}
By definition $O_m$ is independent of $\varphi_m$, and is also a random quantity. Then we introduce the logarithm of the moment-generating function
\begin{equation}\label{eq:7}
  G(u)\equiv \ln \langle e^{u (O-\langle O\rangle)}\rangle
\end{equation}
with $O$ shifted by its mean. With this definition, we can rewrite inequality (\ref{Sobolev0}) as
\begin{equation}\label{Sobolev}
\frac{d}{du}\frac{G(u)}{u}\leq H(u),
\end{equation}
where the right-hand side
\begin{equation}\label{eq:11}
  H(u)\equiv\frac{1}{u^2}\frac{\sum_{m=0}^{N-1}\left\langle\left[ e^{u (O-\langle O\rangle)}    \,\phi(-u(O-O_m))\right]\right\rangle}{\left\langle e^{u (O-\langle O\rangle)}\right\rangle}.
\end{equation}
Armed with this inequality, below we follow the method developed in Ref.~\cite{Lih-King_2023} to obtain the form of the concentration functions $\alpha$ in (\ref{eq:1}) and (\ref{eq:2}), that give the sharp bound of the large deviation probability of the upper and the lower tail, respectively.

\subsection{General analysis of distribution tails}

\subsubsection{Upper tail}
We begin with the upper tail analysis. Let $u>0$ and set $O_m$ in the inequality (\ref{Sobolev}) as the infimum of $O(\boldsymbol{\varphi})$ over $\varphi_m$ with the rest of $(N-1)$ variables $\varphi_i$ ($i\neq m$) held fixed \cite{Boucheron_2013}, denoted as $O^+_m$ (cf.~Appendix~\ref{App_C}):
\begin{eqnarray}\label{eq:8}
  O^+_m (\boldsymbol{\varphi}'_m)\equiv \inf_{\varphi_m}O(\boldsymbol{\varphi}).
\end{eqnarray}
This ensures that the variable $u(O-O_m^+)$ in the inequality (\ref{Sobolev}) is nonnegative. 
Next, we Taylor expand $\phi$,
\begin{equation}\label{eq:9}
  \phi\bigl(-u(O-O_m^+)\bigr)={u^2\over 2}(O-O_m^+)^2+\sum_{n=3}^{\infty}{(-u)^n\over n!}(O-O_m^+)^n,
\end{equation}
and make the following separation for $(O-O_m^+)^2$:
\begin{equation}\label{eq:10}
  (O-O_m^+)^2=b_+ +\delta(O-O_m^+)^2,
\end{equation}
with the first term:
\beq\label{var1}
b_+\equiv \sum_{m=0}^{N-1} \left\langle  (O-O_m^+)^2\right\rangle
\eeq
being the average, and the second, $\delta(O-O_m^+)^2\equiv (O-O_m^+)^2-\langle(O-O_m^+)^2\rangle$, being the fluctuations around the average. With the substitution of Eqs.~(\ref{eq:9})-(\ref{var1}), Eq.~(\ref{eq:11}) is rewritten as
\beq\label{eq:12}
H(u)=\frac{b_+}{2}  + \delta F_+(u),
\eeq
where
\begin{eqnarray}\label{eq:13}
  \delta F_+(u)  =  \left\langle  {\rm e}^{u (O-\langle O \rangle)} \right\rangle^{-1} \sum_{m=0}^{N-1} \Big\langle{\rm e}^{u (O-\langle O \rangle)}    \Big(\delta(O-O_m^+)^2\nonumber\\
  +\sum_{n=3}^{\infty}{(-u)^{n-2}\over n!}(O-O_m^+)^{n}\Big)          \Big\rangle.\qquad\qquad\qquad
\end{eqnarray}
So far the treatments of $H(u)$ are rigorous, and the next step is to find a sharp bound of $\delta F_+(u)$.

For this purpose we note that $O=O_m^+$ takes place only at a set of zero (Lebesgue) measure. So
\beq\label{eq:14}
H(u)\stackrel{u\rightarrow+\infty}{\longrightarrow}{1\over u}\frac{\sum_{m=0}^{N-1} \langle\,  {\rm e}^{u (O-\langle O \rangle)}    \,\bigl(O-O_m^+\bigr)          \rangle}{\langle  {\rm e}^{u (O-\langle O \rangle)} \rangle}.
\eeq
As a result, there must exist some constant $const.$ and some sufficiently large $u^*$ (both of which depend on $O$), so that
\begin{equation}\label{eq:15}
H(u)\leq {const.\over u},\quad {\rm for}\,\, u>u^*.
\end{equation}
This implies that $H(u)$ is bounded by the curve $\sim {1\over u}$ in the large $u$ regime. On the other hand, it is easy to show that ${dG(u)\over du}$ has a definite signature for $u>0$, which is positive. It is also easy to show that it grows unboundedly with $u$. Taking these and inequality (\ref{eq:15}) into account, we find that the function $\delta F/{dG\over du}$ must be bounded for $u>0$. Let its supremum be $c_+$. We obtain
\beq \label{sobo3}
\delta F_+(u)\leq c_+ \,  \frac{d G(u)}{d u},\quad {\rm for}\, u>0.
\eeq

Combining Eq.~(\ref{eq:12}) and inequality (\ref{sobo3}), we finally cast the modified Sobolev inequality (\ref{Sobolev}) into the following differential inequality
\beq\label{sobo4}
\frac{d}{du} \frac{G(u)}{u} \leq  \frac{1}{2}b_+ +  c_+ \,  \frac{d G(u)}{du},
\eeq
supplemented by the boundary condition
\begin{equation}\label{eq:16}
  {G(u)\over u}\bigg|_{u=0}=0.
\end{equation}
Rewriting inequality (\ref{sobo4}) as
\beq\label{eq:17}
\frac{d}{du} \left(  \frac{G(u)}{u}-c_+G(u)\right) \leq  \frac{1}{2}b_+.
\eeq
This can be readily integrated to yield
\beq\label{sobo5}
G(u)\leq \frac{b_+}{2}\frac{u^2}{1-c_+ \, u},
\eeq
where the boundary condition (\ref{eq:16}) has been taken into account. It is important that for $c_+>0$ this solution holds only for $u\in [0,\, 1/c_+)$.

Here it is in order to add a remark. In Ref.~\cite{Boucheron_2013}, the concepts of the so-called `weakly' and `strongly' self-bounding property of the multivariable function $O$ are introduce to obtain an inequality similar to inequality (\ref{sobo4}). In essence, the authors of that work demand bound on quantities such as $\sum_m(O-O_m^+)^2$ and $\sum_m(O-O_m^+)$: This condition is much more stringent than that in our treatments.

Now, consider $P_+(\epsilon)$. Thanks to $u>0$,
\begin{equation}\label{eq:18}
  P_+(\epsilon)=\mathbf{P}({\rm e}^{u (O-\langle O \rangle)}\geq {\rm e}^{u \epsilon}).
\end{equation}
By Markov's inequality we obtain
\beq\label{eq:19}
P_+(\epsilon) \leq {\rm e}^{-u \epsilon} \langle {\rm e}^{u (O-\langle O \rangle)}\rangle=
{\rm e}^{-u \epsilon+G(u)}.
\eeq
In combination with inequality (\ref{sobo5}), this gives
\beq\label{eq:20}
P_+(\epsilon) \leq
{\rm e}^{-u \epsilon+\frac{b_+}{2}\frac{u^2}{1-c_+ \, u}}.
\eeq

For $c_+>0$ we minimize the exponent on the right-hand side of inequality (\ref{eq:20}):
\beq\label{eq:21}
{\cal E}(u)\equiv -u \epsilon+\frac{b_+}{2}\frac{u^2}{1-c_+ \, u}
\eeq
in the interval $[0,\,1/c_+)$. Rewrite it as
\beq\label{eq:22}
{\cal E}(u)= -\left({\epsilon\over c_+}+{b_+\over c_+^2}\right)+
\left({\epsilon\over c_+}+{b_+\over 2c_+^2}\right)y+{b_+\over 2c_+^2}{1\over y},\,\,
\eeq
with $y\equiv 1-c_+u\in [0,\,1)$. By the mean value inequality,
\beq\label{eq:23}
{\cal E}(u)\geq -{b_+\over c_+^2}\,h_1 \left({c_+\epsilon\over b_+}\right)
\eeq
with $h_1(x)\equiv 1+x-\sqrt{1+2x}$, and the equality holds only if
\begin{eqnarray}
\label{eq:24}
  && y=\sqrt{b_+\over b_++2c_+\epsilon}\in [0,\,1) \nonumber\\
  &\Leftrightarrow& u={1\over c_+}\left(1-\sqrt{b_+\over b_++2c_+\epsilon}\right)\in [0,\,1/c_+).
\end{eqnarray}
Thus
\beq\label{eq:25}
P_+(\epsilon) \leq
{\rm e}^{-{b_+\over c_+^2}\,h_1 \left({c_+\epsilon\over b_+}\right)}.
\eeq
With the help of the elementary inequality shown in Appendix \ref{sec:inequality}:
\begin{equation}\label{eq:26}
  h_1(x)\geq {x^2\over 2(1+x)},\quad {\rm for}\, x\geq0,
\end{equation}
we simplify inequality (\ref{eq:25}) to
\beq\label{eq:27}
P_+(\epsilon) \leq
{\rm e}^{-\frac{\epsilon^2}{2 (b_+  + c_+ \epsilon)}},
\eeq
giving a sub-Gamma upper tail.

For $c_+\leq 0$ we first simplify inequality (\ref{eq:20}) to
\beq\label{eq:28}
P_+(\epsilon) \leq
{\rm e}^{-u \epsilon+\frac{b_+ u^2}{2}}.
\eeq
The exponent reaches the minimum at $u = \epsilon/b_+$. As a result,
\beq\label{upper}
P_+(\epsilon)\leq
{\rm e}^{-{\epsilon^2\over 2b_+}},
\eeq
giving a sub-Gaussian upper tail. In Appendix \ref{sec:scale_factor} we shall show that, in fact, $P_+(\epsilon)$ vanishes for $\epsilon\geq b_+/2|c_+|$. This is consistent with that an entanglement probe such as $S$ has a maximal value. In Sec.~\ref{sec:BF_universality} we further show that one can actually adjust the value of $b_+$ properly, so that the sub-Gaussian concentration function is sharp, and the maximal value of an entanglement probe does not play roles practically.

\subsubsection{Lower tail}
For the lower tail, we carry out the same analysis with $u<0$. To start we set $O_m=O_m^-$ defined as \cite{Boucheron_2013} (cf.~Appendix~\ref{App_C})
\begin{eqnarray}\label{eq:29}
O^-_m (\boldsymbol{\varphi}'_m)\equiv \sup_{\varphi_m}O(\boldsymbol{\varphi}).
\end{eqnarray}
This ensures that $u(O-O_m^-)$ is nonnegative. With the introduction of
\beq\label{var2}
b_-\equiv \sum_{m=0}^{N-1} \left\langle  (O-O_m^-)^2\right\rangle,
\eeq
we can rewrite Eq.~(\ref{eq:11}) as
\beq\label{eq:30}
H(u)=\frac{b_-}{2}  + \delta F_-(u),
\eeq
similar to Eq.~(\ref{eq:12}), where
\begin{eqnarray}\label{eq:31}
  \delta F_-(u)  =  \left\langle  {\rm e}^{u (O-\langle O \rangle)} \right\rangle^{-1} \sum_{m=0}^{N-1} \Big\langle{\rm e}^{u (O-\langle O \rangle)}    \Big(\delta(O-O_m^-)^2\nonumber\\
  +\sum_{n=3}^{\infty}{(-u)^{n-2}\over n!}(O-O_m^-)^{n}\Big)          \Big\rangle.\qquad\qquad\qquad
\end{eqnarray}
In parallel to inequality (\ref{sobo3}), we further have
\beq \label{sobo3b}
\delta F_-(u)\leq c_- \,  \frac{d G(u)}{d u},\quad {\rm for}\, u<0.
\eeq
From this the inequality:
\beq\label{eq:32}
\frac{d}{du} \frac{G(u)}{u}  \leq  \frac{1}{2}b_- +  c_- \,  \frac{d G(u)}{du}
\eeq
follows.

Similar to the derivations of concentration inequalities (\ref{eq:27}) and (\ref{upper}), from inequality (\ref{eq:32}) we obtain
\beq\label{lower}
P_-(\epsilon)
\leq\left\{
\begin{array}{ll}
{\rm e}^{-\frac{\epsilon^2}{2 (b_-  + |c_-| \epsilon)}},& \textrm{ for\,\, $c_-<0$,} \\
{\rm e}^{-{\epsilon^2\over 2b_-}},&\textrm{ for\,\, $c_-\geq0$.}
\end{array}
\right.
\eeq
Similar to $b_+$ and $c_+$ for $P_+(\epsilon)$, the positive parameters $b_-$ and $|c_-|$ here are the variance factor and the scale parameter, respectively.

\begin{figure}[t]
\includegraphics[width=0.8\linewidth]{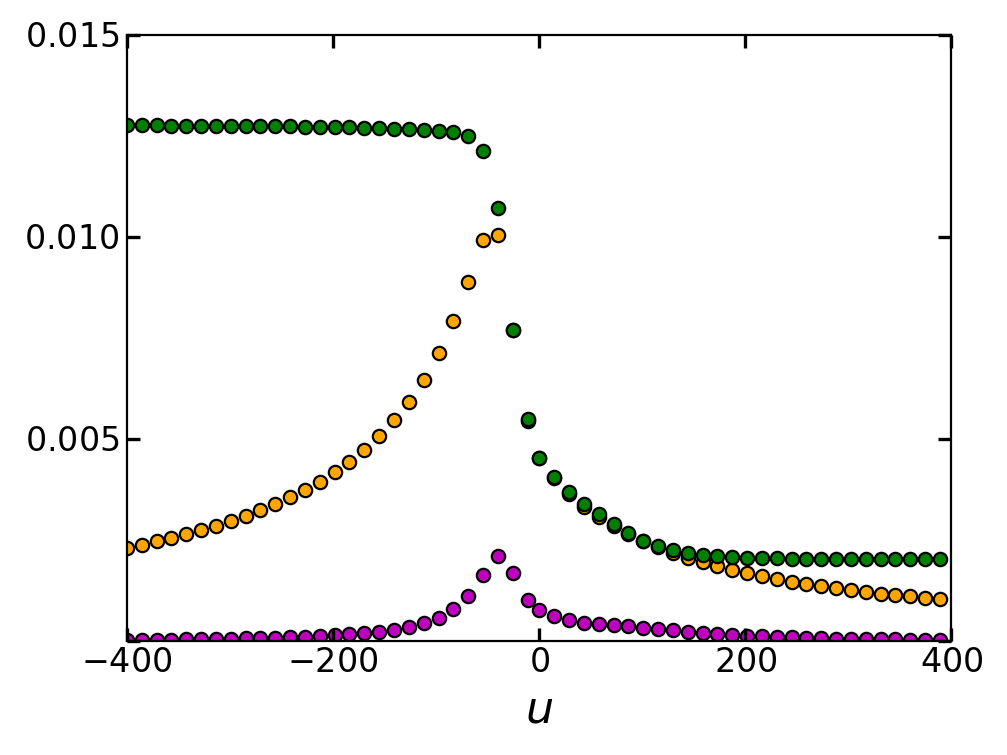}
\caption{The purple and orange curves depict the left-hand side and right-hand side of the modified logarithmic Sobolev inequality~(\ref{Sobolev}) with $O=S$, respectively, scaled by an overall factor $3$ to aid visualization. The green curve depicts the right-hand side of inequality~(\ref{sobo4}) for $u>0$ and that of inequality (\ref{eq:32}) for $u<0$. All curves are generated from \textit{ab initio} data with $r_A=0.2$.  The parameters are determined as: $b_+ = 0.0091$, $b_- = 0.0092$, $c_+ = -0.0246$, and $c_- = -0.0434$; see Appendix~\ref{App_C} for more details.}
\label{F6}
\end{figure}

\subsection{Boson-fermion universality}
\label{sec:BF_universality}
Inequalities (\ref{eq:27}), (\ref{upper}) and (\ref{lower}) show that the tail behaviors are determined by the signature of the parameters $c_\pm$. At this moment we cannot determine this signature analytically. Thus we resort to numerical analysis detailed in Sec.~\ref{Sec:num}. We find that for most the ratio of the subsystem to total system size $r_A\equiv L_A/L$ (except when $r_A$ approaches 0.5) $c_+$ is negative, while $c_-$ is always negative; see Table~\ref{table:1} in Appendix \ref{App_C}. Having determined  the signature of $c_\pm$, we can conclude that the upper tail is sub-Gaussian by inequality (\ref{upper}) and the lower is sub-Gamma by (\ref{lower}). That is,
\beq\label{eq:33}
&&\mathbf{P}(|O-\langle O\rangle|\geq \epsilon)\nonumber\\
&\leq&\left\{
\begin{array}{ll}
{\rm e}^{-{\epsilon^2\over 2b_+}},&\textrm{ for\,\, $O-\langle O\rangle >0$}\\
{\rm e}^{-\frac{\epsilon^2}{2 (b_-  + |c_-| \epsilon)}},& \textrm{ for\,\, $O-\langle O\rangle <0$}
\end{array}
\right..
\eeq
A natural question is: How tight are these concentration functions? Below we address this question.

We have shown that the sub-Gaussian $P_+(\epsilon)$ is essentially determined by inequality (\ref{Sobolev}) in the regime $0\leq u \lesssim 1/2|c_+|$, whereas the sub-Gamma $P_-(\epsilon)$ by inequality (\ref{Sobolev}) in the regime $1/c_-\leq u \leq 0$. In Fig.~\ref{F6} we compare the right-hand side of inequality (\ref{Sobolev}), namely, $H(u)$ given in Eq.~(\ref{eq:11}) with that of (\ref{sobo4}) for $u>0$, and with that of (\ref{eq:32}) for $u<0$, and the moment generating function $\langle {\rm e}^{u (O-\langle O\rangle)}\rangle={\rm e}^{G(u)}$ is constructed from the numerically generated \textit{ab initio} data. We find that within the range of interests  $H(u)$ is pretty close to the right-hand side of inequality of (\ref{sobo4}) for $u>0$, and to that of (\ref{eq:32}) for $u<0$. Moreover, as shown in that figure also, when the left-hand side of inequality (\ref{Sobolev}) is multiplied by some numerical constant, it becomes very close to $H(u)$. This implies that one can upgrade the modified logarithmic Sobolev inequality (\ref{Sobolev}) to an approximate equality, read
\beq\label{eq:35}
{d\over du}{G(u)\over u}\approx \left\{
\begin{array}{ll}
\frac{1}{2}\mathfrak{b}_+ +  \mathfrak{c}_+ \,  \frac{d G(u)}{du},&\textrm{ for\,\, $u>0$}\\
\frac{1}{2}\mathfrak{b}_- +  \mathfrak{c}_- \,  \frac{d G(u)}{du},& \textrm{ for\,\, $u<0$}
\end{array}
\right.,
\eeq
with the numerical constants $\mathfrak{b}_\pm,\,\mathfrak{c}_\pm$ satisfying
\begin{equation}\label{eq:36}
  {\mathfrak{b}_+\over b_+}={\mathfrak{c}_+\over c_+}={\mathfrak{b}_-\over b_-}={\mathfrak{c}_-\over c_-}.
\end{equation}
Finally, as shown in Fig.~\ref{F6} $H(u)$ changes only slightly within the regime $0\leq u \leq 1/|c_+|$, and thus we may approximate it as $b_+/2$, i.e., set $c_+=0$. Taking this and Eqs.~(\ref{eq:35}) and (\ref{eq:36}) into account, we arrive at
\beq\label{eq:34}
{d\over du}{G(u)\over u}\approx \left\{
\begin{array}{ll}
\frac{1}{2}\mathfrak{b}_+,&\textrm{ for\,\, $u>0$}\\
\frac{1}{2}\mathfrak{b}_- +  \mathfrak{c} \,  \frac{d G(u)}{du},& \textrm{ for\,\, $u<0$}
\end{array}
\right.
\eeq
with $\mathfrak{c}\equiv \mathfrak{c}_-.$

Equations (\ref{eq:34}) can be solved readily, giving for large $\epsilon$:
\beq\label{confunc}
&&\mathbf{P}(|O-\langle O\rangle|\geq \epsilon)\nonumber\\
&=&\left\{
\begin{array}{ll}
{\rm e}^{-{\epsilon^2\over 2\mathfrak{b}_+}},&\textrm{ for\,\, $O-\langle O\rangle >0$}\\
{\rm e}^{-\frac{\epsilon^2}{2 (\mathfrak{b}_-  + \mathfrak{c}\epsilon)}},& \textrm{ for\,\, $O-\langle O\rangle <0$}
\end{array}
\right.,
\eeq
which is asymmetric: The upper tail is sub-Gaussian whereas the lower is sub-Gamma. This distribution has been found previously for mesoscopic fluctuations in entanglement dynamics of free fermions \cite{Lih-King_2023}. Thus we have established the boson-fermion universality of the statistical distribution (\ref{confunc}) in free systems. In fact, that distribution holds in a much broader context. For example, it has been shown and numerically confirmed \cite{Lih-King_2023} that the distribution (\ref{confunc}) holds for mesoscopic fluctuations in entanglement dynamics of interacting spin chains also. In addition, in an earlier numerical experiment on entanglement dynamics in lattice fermions with interactions, similar temporal fluctuations were observed \cite{Faiez_2020}. From the technical viewpoint this universality, displayed by statistical distribution of various $O$, arise from two mathematical structures, which are shared by both free bosonic and free fermionic systems: (i) As described by Eqs.~(\ref{eq:4}) and (\ref{eq:5}), the probability measure carries a product structure; and (ii) an entanglement probe $O(\boldsymbol{\varphi})$ is a highly nonlinear function on the product space $\mathbb{T}^N$. These two structures render the observed universality falling into the class of concentration-of-measure phenomena in modern probability theory.

\section{\label{sec_3_3} Boson-fermion universality of the scaling law of variances}
The studies in last section are mute on the properties of the variance of various entanglement probes. In this section we shall show that boson-fermion universality holds for the variance also.
\subsection{A general formula of $\boldsymbol{{\rm Var}(S)}$}
Equation (\ref{confunc}) show that $\mathfrak{b}_\pm$ or $b_\pm$ give the variance of probe $O$. In Ref.~\cite{Lih-King_2023} we further showed for free fermion systems that $b_\pm \propto \left\langle |\partial_{\boldsymbol{\varphi}}O|^2\right\rangle $, where the latter $|\partial_{\boldsymbol{\varphi}}O|^2=\sum_{m}(\partial_{\varphi_m}O)^2$ (the Euclidean norm squared) takes the form of Lipschitz-like constant, in the sense that it is the average of the gradient squared rather than the supremum of the gradient squared used in the concentration-of-measure theory \cite{Milman_1982, Ledoux_2001, Fang_2018}.  We then have
\begin{equation}\label{lip1}
{\rm Var}(O)={\cal C}\left\langle |\partial_{\boldsymbol{\varphi}}O|^2\right\rangle,
\end{equation}
where ${\cal C}$ is an overall numerical coefficient found to be $1/2$ numerically (see Sec.~\ref{Sec:num}).
The derivation of Eq.~(\ref{lip1}) is carried out in Ref.~\cite{Lih-King_2023}, and holds for the present system. For completeness we repeat it in Appendix~\ref{App_B}.

To calculate Eq.~(\ref{lip1}) we use Eq.~(\ref{new_er}). Recall that we drop out `$\sim$' in symbols $\widetilde{O}(\boldsymbol{\varphi})$ and $\widetilde{C}(\boldsymbol{\varphi})$. In the following we use the entanglement entropy $O=S$ as an example, and the results hold for other entanglement probes also. Consider a component of $\partial_{\boldsymbol{\varphi}}S$, say $\partial_{{\varphi}_m}S$ without loss of generality. Using the identity shown in Appendix~\ref{sec:derivative_matrix_function}, we find from Eq.~(\ref{new_er}) that
\beq\label{comp}
\partial_{\varphi_m} S(\boldsymbol{\varphi}) = \frac{1}{2} \operatorname{Tr}_A \left[ \ln{\left( \frac{C(\boldsymbol{\varphi})+{1\over 2}\,\mathbb{I}_{2L_A}}{C(\boldsymbol{\varphi}) - {1\over 2}\,\mathbb{I}_{2L_A}} \right)}\partial_{\varphi_m} C(\boldsymbol{\varphi}) \right].\nonumber\\
\eeq

Let us recall discussions in the end of Sec.~\ref{sec_3}, according to which $C_1(\boldsymbol{\varphi})$ is small. So we can expand Eq.~(\ref{comp}) up to the second order to obtain
\beq
\partial_{\varphi_m}S=(\partial_{\varphi_m}S)_1+(\partial_{\varphi_m}S)_2,
\label{psm}
\eeq
where
\beq
(\partial_{\varphi_m}S)_1 =\frac{1}{2} \operatorname{Tr}_A\left[  \ln{\left( \frac{C_0+{1\over 2}\,\mathbb{I}_{2L_A}}{C_0 - {1\over 2}\,\mathbb{I}_{2L_A}} \right)}\, \partial_{\varphi_m} C_1 \right],
\label{ps2a}
\eeq
\beq
(\partial_{\varphi_m}S)_2 &=&-\frac{1}{2}\operatorname{Tr}_A\left[\left(C_0^2-{1\over 4}\,\mathbb{I}_{2L_A}\right)^{-1}\, C_1\, \partial_{\varphi_m}C_1\right]\nonumber\\
&+& \ldots.
\label{ps2}
\eeq
The extra terms $\ldots$ in Eq. (\ref{ps2}) denote the corrections from commutators $[C_0, C_1]$, which do not give different overall scaling behaviors. Hence we drop them from the discussion onwards.
With the substitution of Eq.~(\ref{psm}), Eq.~(\ref{lip1}) gives three terms. One can show that the crossing term vanishes upon averaging over $\boldsymbol{\varphi}$. As a result, $\textrm{Var}(S)$ can be separated into two contributions,
\beq
\operatorname{Var}(S)=\operatorname{Var}_1(S)+\operatorname{Var}_2(S),
\label{two_part}
\eeq
with
\beq
\operatorname{Var}_1(S) &=& {\cal C}\sum\limits_{m=0}^{N-1} \bigl\langle\bigl((\partial_{\varphi_m}S)_1\bigr)^2\, \bigr\rangle, \label{tpva}\\
\operatorname{Var}_2(S) &=&{\cal C} 
\sum\limits_{m=0}^{N-1} \bigl\langle\bigl((\partial_{\varphi_m}S)_2\bigr)^2\, \bigr\rangle.
\label{tpv}
\eeq
In the following we calculate Eqs.~(\ref{tpva}) and (\ref{tpv}) separately.

\subsection{First term $\boldsymbol{\textrm{Var}_1(S)}$}
Recall that the matrices $C,\,C_0,\,C_1$ are defined on two sectors, labelled by the oscillator index $r=1,\ldots, L_A$ and the $\hat{x}$-$\hat{p}$ sector index $\sigma=\bar{A},\bar{B}$. Thus the trace operation in Eq.~(\ref{ps2a}) can be cast to ${\rm Tr}_A(\cdot)=\sum_{r=1}^{L_A}\sum_{\sigma}(\cdot)\equiv \sum_{r=1}^{L_A}\operatorname{tr}(\cdot)$. Because the matrix elements $(C_0)_{r\sigma,r'\sigma'}$ decay exponentially with the distance $|r-r'|$, as confirmed by numerical simulations, the summation over $r$ can be divided into the summation over $r$ belonged respectively to the bulk and the edge of the subsystem. The bulk has a size $\approx L_A$, while the size of the edge is determined by the decay length $L_e$ of $C_0$, where $L_e$ depends on the parameters of the Hamiltonian. We use $A_b$ and $A_e$ to denote the bulk and the edge, respectively, so that $A_b=\{r| L_e \leq r \leq L_A-L_e\}$ and $A_e=\{r| r \leq L_e \, \operatorname{or} \, r \geq L_A-L_e\}$. Taking these into account, we write the trace in Eq.~(\ref{ps2a}) as
\begin{equation}\label{eq:55}
  \operatorname{Tr}_A\left[\ln{\left( \frac{C_0+{1\over 2}\mathbb{I}_{2L_A}}{C_0 - {1\over 2}\mathbb{I}_{2L_A}} \right)} \partial_{\varphi_m}C_1 \right]=T_{b,m}+T_{e,m},
\end{equation}
where
\beq
T_{X,m}=\sum_{r\in A_{X}}\sum_{\sigma}\left(\ln{\left( \frac{C_0+{1\over 2}\mathbb{I}_{2L_A}}{C_0 - {1\over 2}\mathbb{I}_{2L_A}} \right)} \partial_{\varphi_m}C_1 \right)_{r\sigma,r\sigma}\,\,
\label{b_and_e}
\eeq
with $ X=b,\,e$.

For the bulk term $T_{b,m}$, one may extend the oscillator indices involved in the matrix product from the set: $\{1,2,...,L_A\}$ to $\{1,2,...,L\}$ , since $(C_0)_{r\sigma,r'\sigma'}$ decays exponentially with $|r-r'|$. Thus Eq.~(\ref{b_and_e}) is reduced to
\beq
T_{b,m}\approx L_A \operatorname{tr}\left[\ln{\left( \frac{\tilde{c}_{0}(k_m)+\mathbb{I}_2/2}{\tilde{c}_{0}(k_m) - \mathbb{I}_2/2} \right)}\tilde{c}'_{1}(k_m)\right].
\label{bulk}
\eeq
Here $\tilde{c}_{0}(k_m)$ and $\tilde{c}_{1}(k_m)$ are, respectively, the Fourier transformation of $c_{0,l}$ and $c_{1,l}$ with respect to the subscript $l$ [see Eq.~(\ref{Ct})], and $\tilde{c}'_{1}(k_m)$ denotes the first derivative of $\tilde{c}_{1}(k_m)$ with respect to $\varphi_m$. They are given by
\begin{equation}
\tilde{c}_{0}(k_m)= i E_+(k_m)\left[\begin{array}{cc}0&s_{k_m}\\ -1/s_{k_m}&0\end{array}\right]
\end{equation}
and
\begin{equation}\label{eq:110}
  \tilde{c}'_{1}(k_m) =\frac{2i\bar{\delta}_m}{L} E_-(k_m) \left[\begin{array}{cc}-\cos \varphi_m & s_{k_m}\sin \varphi_m \\ \frac{\sin \varphi_m}{s_{k_m}} & \cos \varphi_m\end{array}\right],\qquad
\end{equation}
respectively. To arrive at Eq.~(\ref{bulk}), we have made use of the Toeplitz structure of $C$.
Moreover, we have made use of the identity: ${1\over L}\sum_{r=1}^L e^{i (k_m-k_n)r}=\delta_{k_mk_n}$. Using
\beq
\ln{\left( \frac{1+x/2}{1 - x/2} \right)}=\sum_{n=0}^{\infty}\, \frac{1}{4^{n}\, (2n+1)}\, x^{2n+1}
\label{eq:65}
\eeq
and $\tilde{c}_0^{-1}=\tilde{c}_0/E_+^2$, we reduce Eq.~(\ref{bulk}) to
\beq
T_{b,m}= {L_A\over E_+}\ln{\left( \frac{2E_++1}{2E_+ - 1}\right)}\operatorname{tr}\left[\tilde{c}_0(k_m)\tilde{c}'_{1}(k_m)\right]=0.\quad
\label{eq:60}
\eeq
Note that by inequality (\ref{eq:59}) $2E_+ - 1>0$.

For the edge term $T_{e,m}$, substituting Eqs.~(\ref{eq:49}) and (\ref{Cre}) into Eq.~(\ref{b_and_e}) and taking into account that $(C_0)_{r\sigma,r'\sigma'}$ decays exponentially with $|r-r'|$, we obtain
\beq
T_{e,m}&\approx& i{2L_e\over L}\bar{\delta}_m E_-(k_m)\nonumber\\
&\times&\operatorname{tr}\left(\varsigma
\left[
  \begin{array}{cc}
    -\cos(\varphi_m) & s_{k_m}\sin(\varphi_m) \\
    {\sin(\varphi_m)\over s_{k_m}} & \cos(\varphi_m) \\
  \end{array}
\right]\right),
\label{eq:61}
\eeq
where
\begin{eqnarray}
\label{eq:62}
\varsigma=\ln{\left(\frac{C_0+{1\over 2}\mathbb{I}_{2L_A}}{C_0 - {1\over 2}\mathbb{I}_{2L_A}}\right)}_{r r}
\end{eqnarray}
is the block diagonal  element and is, in fact, independent of $r$. By using Eqs.~(\ref{eq:64}) and (\ref{eq:65}) it is easy to show that $\varsigma$ is purely imaginary. Let $\varsigma=i\zeta$ with $\zeta$ being real. We have
\begin{widetext}
\beq
\left\langle T^2_{e,m}\right\rangle={2L^2_e\over L^2}(\bar{\delta}_m E_-(k_m))^2
\left(\left(\operatorname{tr}\left[\zeta
\left[
  \begin{array}{cc}
    -1 & 0 \\
    0 & 1 \\
  \end{array}
\right]\right]\right)^2+\left(\operatorname{tr}\left[\zeta
\left[
  \begin{array}{cc}
    0 & s_{k_m} \\
    {1\over s_{k_m}} & 0 \\
  \end{array}
\right]\right]\right)^2\right),
\label{eq:63}
\eeq
Substituting it into Eq.~(\ref{tpva}) and passing to the continuum limit: ${1\over L}\sum_k\rightarrow \int {dk\over 2\pi}$ for large $L$, we obtain
\beq
\textrm{Var}_1(S) = \frac{a}{L},
\label{eq:71}
\eeq
where the proportionality coefficient
\beq\label{eq:66}
  a={{\cal C}L^2_e\over 8}\int_{-\pi}^{\pi}{dk\over 2\pi}(E_-(k))^2 \left(\left(\operatorname{tr}\left[\zeta
\left[
  \begin{array}{cc}
    -1 & 0 \\
    0 & 1 \\
  \end{array}
\right]\right]\right)^2+\left(\operatorname{tr}\left[\zeta
\left[
  \begin{array}{cc}
    0 & s_{k} \\
    {1\over s_{k}} & 0 \\
  \end{array}
\right]\right]\right)^2\right).
\eeq
Recall that $\varsigma$ is a constant $2\times 2$ matrix determined completely by system's microscopic parameters. Thus $a$ depends neither on $L$ (for $C_0$ has a well-defined large-$L$ limit) nor on $L_A$. Consequently, ${\rm Var}_1(S)$ scales with $L$ as $\sim 1/L$ and is independent of $L_A$.

\subsection{Second term $\boldsymbol{\textrm{Var}_2(S)}$\label{secvar}}
Because $(C_{0})_{r\sigma,r'\sigma'}$ decays rapidly with $|r-r'|$,
Eq.~(\ref{ps2}) can be approximated by
\beq
(\partial_{\varphi_m}S)_2 \approx \sum\limits_{r,r'=1}^{L_A}\, \mathrm{tr}\left(\kappa c_{1,r-r'}\partial_{\varphi_m} c_{1,r'-r}\right).
\label{eq:67}
\eeq
Here $\kappa$ is a $2 \times 2$ matrix in the $\hat{x}$-$\hat{p}$ sector, whose elements $\kappa_{\sigma\sigma'}=(C_0^2-{1\over 4}\mathbb{I}_{2L_A})^{-1}_{r\sigma,r\sigma'}$ are independent of $r$. With the substitution of Eq.~(\ref{Cre}) it can be rewritten as

\beq
(\partial_{\varphi_m}S)_2&=&\frac{4}{L^2} \sum_{r,r'=1}^{L_A}\sum_{n=0}^{N-1} \bar{\delta}_m\bar{\delta}_n \,\mathfrak{O}_{m,r-r'}\mathfrak{O}_{n,r-r'}E_-(k_n)E_-(k_m)\nonumber\\
&&\times \bigr(\operatorname{tr}\left(\kappa{\check{\mathcal{I}}_{cs}}(k_n,k_m)\right)\cos\varphi_n\sin\varphi_m
+\operatorname{tr}\left(\kappa{\check{\mathcal{I}}_{ss}}(k_n,k_m)\right)\sin\varphi_n \sin \varphi_m\nonumber\\
&&\,\,+\operatorname{tr}\left(\kappa{\check{\mathcal{I}}_{cc}}(k_n,k_m)\right)\cos\varphi_n\cos\varphi_m
+\textrm{tr}\left(\kappa{\check{\mathcal{I}}_{sc}}(k_n,k_m)\right)\sin\varphi_n\cos\varphi_m \bigr),
\label{appopS2}
\eeq
where $\mathfrak{O}_{n,l}\equiv\cos(k_nl)$ and
\beq
\begin{aligned}
{\check{\mathcal{I}}}_{cs} (k_n,k_m)
= \left[\begin{array}{cc}\frac{s_{k_n}}{s_{k_m}}&0\\ 0&\frac{s_{k_m}}{s_{k_n}}\end{array}\right],\,
\check{I}_{ss} (k_n,k_m) = \left[\begin{array}{cc}0&s_{k_m}\\ -\frac{1}{s_{k_m}}&0\end{array}\right],\,
\check{I}_{cc} (k_n,k_m) =  \left[\begin{array}{cc}0&s_{k_n}\\ -\frac{1}{s_{k_n}}&0\end{array}\right],\,
\check{I}_{sc} (k_n,k_m) = -\mathbb{I}_2.
\end{aligned}
\quad
\label{eq:70}
\eeq
Substituting Eq.~(\ref{appopS2}) into Eq.~(\ref{tpv}) and carrying the averaging over $\boldsymbol{\varphi}$, we obtain
\beq
{\rm Var}_2(S)&=&\frac{\cal{C}}{L^4}\sum_{r,r'=1}^{L_A}\sum_{s,s'=1}^{L_A}\sum_{m\neq n}
\bar{\delta}_m^2\bar{\delta}_n^2 E^2_-(k_n) E^2_-(k_m)
\mathfrak{O}_{n,r-r'}\mathfrak{O}_{m,r-r'}\mathfrak{O}_{n,s-s'}\mathfrak{O}_{m,s-s'}\nonumber\\
&\times& \left[\left(\operatorname{tr}\left(\kappa{\check{\mathcal{I}}_{cs}}(k_n,k_m)\right)\right)^2
+\left(\operatorname{tr}\left(\kappa{\check{\mathcal{I}}_{ss}}(k_n,k_m)\right)\right)^2
+\left(\operatorname{tr}\left(\kappa{\check{\mathcal{I}}_{cc}}(k_n,k_m)\right)\right)^2
+\left(\textrm{tr}\left(\kappa{\check{\mathcal{I}}_{sc}}(k_n,k_m)\right)\right)^2 \right].
\eeq
In fact there is a contribution from the $m=n$ terms, but it is smaller by an order of $1/L$ and thus is ignored. The summation over $r,r', s,s'$ is dominated by the configurations: $r-r'=\pm(s-s')$. As a result,
\beq
{\rm Var}_2(S)&=&\frac{{\cal C}L_A^3}{2L^4}\sum_{m\neq n}
\bar{\delta}_m^2\bar{\delta}_n^2 E^2_-(k_n) E^2_-(k_m)\nonumber\\
&\times&\left[\left(\operatorname{tr}\left(\kappa{\check{\mathcal{I}}_{cs}}(k_n,k_m)\right)\right)^2
+\left(\operatorname{tr}\left(\kappa{\check{\mathcal{I}}_{ss}}(k_n,k_m)\right)\right)^2
+\left(\operatorname{tr}\left(\kappa{\check{\mathcal{I}}_{cc}}(k_n,k_m)\right)\right)^2
+\left(\textrm{tr}\left(\kappa{\check{\mathcal{I}}_{sc}}(k_n,k_m)\right)\right)^2 \right].
\eeq
Passing to the continuum limit this gives
\beq
\textrm{Var}(S)_2=b\frac{L_A^3}{L^2}
\label{eq:72}
\eeq
where the proportionality coefficient
\beq
b=\frac{{\cal C}}{8}\int\!\!\!\!\int_{-\pi}^{\pi}{dkdk'\over (2\pi)^2}\ E^2_-(k) E^2_-(k')
\left[\left(\operatorname{tr}\left(\kappa{\check{\mathcal{I}}_{cs}}(k,k')\right)\right)^2
+\left(\operatorname{tr}\left(\kappa{\check{\mathcal{I}}_{ss}}(k,k')\right)\right)^2
+\left(\operatorname{tr}\left(\kappa{\check{\mathcal{I}}_{cc}}(k,k')\right)\right)^2
+\left(\textrm{tr}\left(\kappa{\check{\mathcal{I}}_{sc}}(k,k')\right)\right)^2 \right]\nonumber\\
\eeq
\end{widetext}
depends neither on $L$ nor $L_A$. Consequently, ${\rm Var}_2(S)$ scales with $L$ as $\sim 1/L^2$ and with $L_A$ as $\sim L_A^3$. The latter scaling is faster than the linear scaling $\sim L_A$, expected for the variance of an extensive quantity by the central limit theorem.

\subsection{Boson-fermion universality of scaling behaviors}
Combining Eqs.~(\ref{eq:71}) and (\ref{eq:72}) we arrive at the scaling law for the variance:
\beq
{\rm Var} (S)= \frac{a}{L}+\frac{bL_A^3}{L^2}
\label{var_1}
\eeq
for $L\gg L_A\gg 1$. We emphasize that the coefficients $a,b$ depend on system's microscopic parameters, not on $L,L_A$. As we shall show in the next section, this scaling law also holds for more entanglement probes such as $S_n$. Of course, $a,b$ depend on the entanglement probe $O$. Let us rescale $O$ by $\sqrt{ab}$ and $L, L_A$ by $\sqrt{a/b}$. Then we obtain
\beq
{\rm Var} (O)= \frac{1}{L}+\frac{L_A^3}{L^2},
\label{eq:73}
\eeq
which are independent of system's microscopic parameters and $O$. Surprisingly, the scaling law (\ref{eq:73}) and its universality were previously found for free fermion models \cite{Lih-King_2023}. Since this scaling law is now shown to be independent of the particle statistics, its universality is broadened further. According to Eq.~(\ref{eq:73}), when $L_A$ increases while $L$ is fixed, ${\rm Var} (O)$ displays a crossover from the scaling $\sim 1/L$ to $\sim L_A^3/L^2$ at $L_A\sim L^{1/3}$. Furthermore, we have seen from the derivations above that the first term in Eq.~(\ref{eq:73}) arises from the edge of the subsystem, while the second from subsystem's bulk.

\begin{figure}[t]
\includegraphics[width=0.8\linewidth]{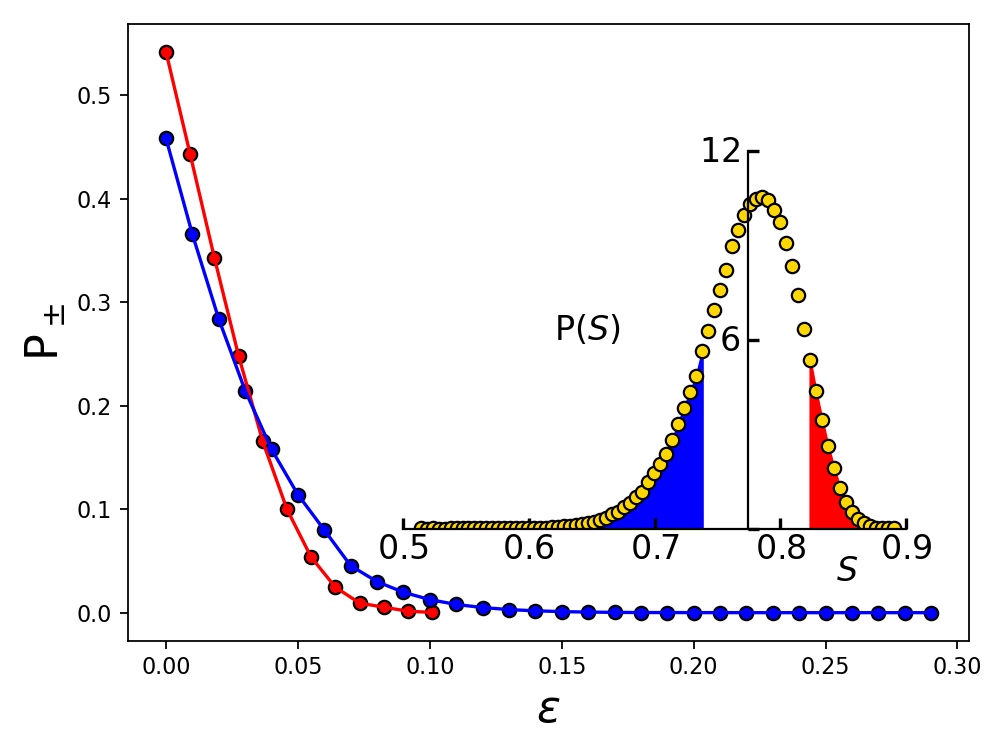}
\includegraphics[width=0.8\linewidth]{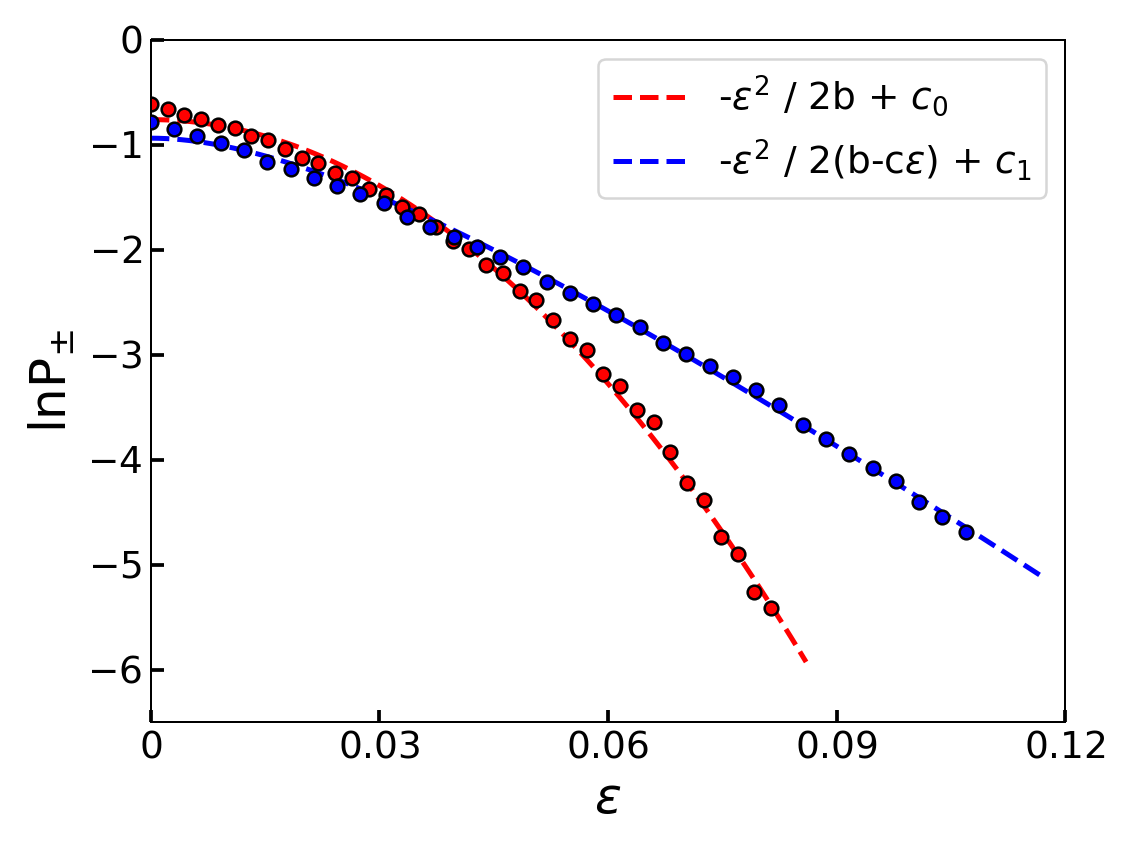}
\caption{Numerical verifications of Eq.~(\ref{confunc}) for $O=S$. (a) Numerically, the deviation probability $P_+(\epsilon)$ [respectively $P_-(\epsilon)$] is given by the area of the red (respectively blue) region under the distribution of $S$, and is represented by the red (respectively blue) dots. (b) Same plot on a semi-log scale. The dashed (blue and red) lines are fitting functions given in the inset.}
\label{F4}
\end{figure}

\begin{figure}
\includegraphics[width=0.9\linewidth]{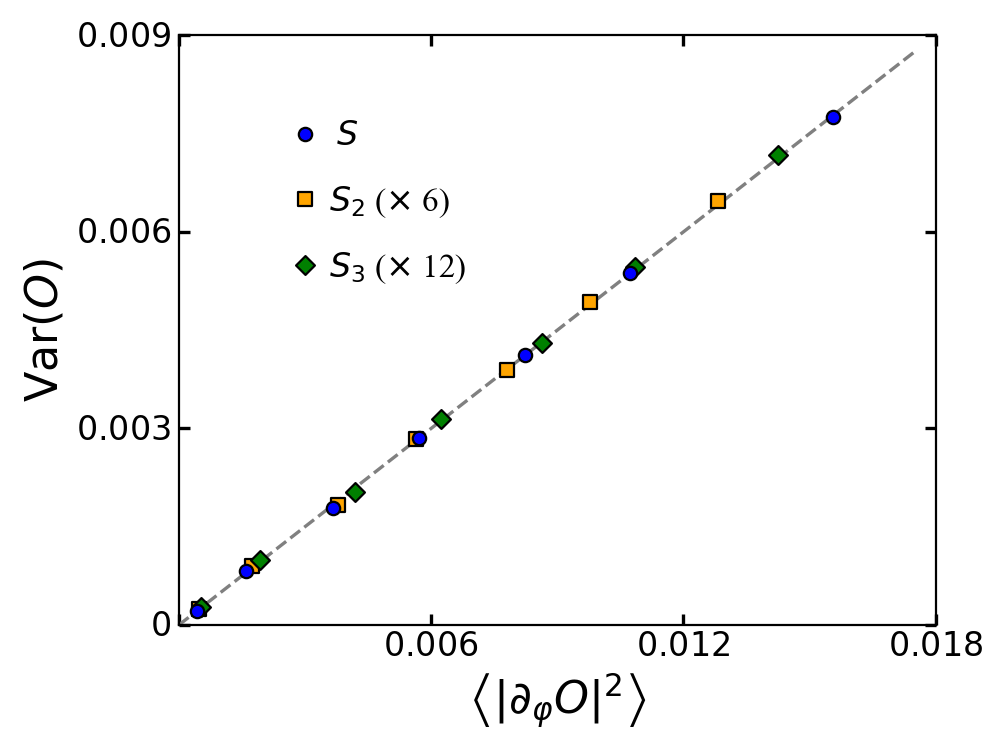}
\caption{We perform numerical simulations for different quenches, different system lengths and different entanglement probes $O$, and obtain the variance of the time series $O(t)$ and Lipschitz-like constant $\langle\,|\partial_{\boldsymbol{\varphi}}O|^2\rangle$. The simulation results verify the theoretical prediction Eq.~(\ref{lip1}).}
\label{F8}
\end{figure}

\section{Numerical verifications\label{Sec:num}}
First, we determine the deviation probabilities $P_{\pm}(\epsilon)$ from the simulated full distribution. As shown in Fig.~\ref{F4}(a), they correspond to the area under the curve representing the statistical distribution, which are shaded in red and blue, respectively. Their explicit forms confirm the analytical prediction Eq.~(\ref{confunc}). More precisely, $P_+(\epsilon)$ can be fitted with a sub-Gaussian form and $P_-(\epsilon)$ by a sub-Gamma form, as shown in Fig.~\ref{F4}(b).

Next, we verify that the variance of an entanglement probe $O$ is proportional to $\langle\,|\partial_{\boldsymbol{\varphi}}O|^2\rangle$, a key result which allows studying fluctuations analytically. In Fig.~\ref{F8}, we simulate values for both the left-hand and right-hand sides of Eq.~(\ref{lip1}) for different quench parameters. The results confirm the proportionality behavior, and the proportionality coefficient is determined as
\begin{equation}\label{eq:74}
  {\cal C}\approx 1/2,
\end{equation}
instead of $\approx 1/8$ for free fermions \cite{Lih-King_2023}. Also, the result is independent of entanglement probes, as shown for $O=S, S_2, S_3$.

\begin{figure*}[ht]
\includegraphics[width=0.31\linewidth]{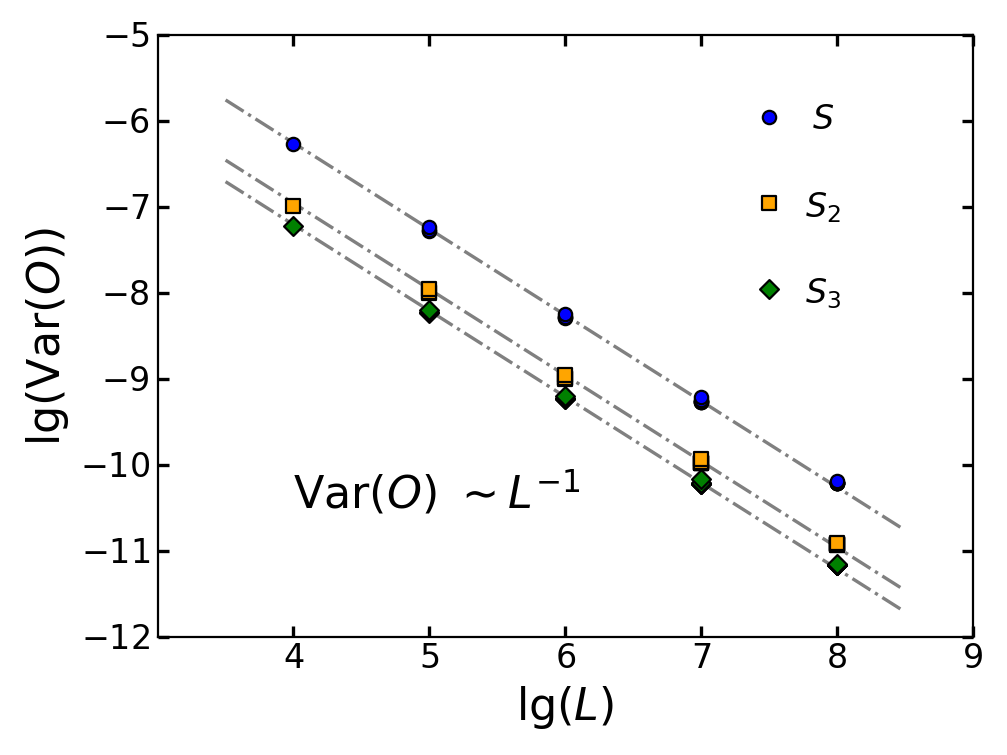}
\includegraphics[width=0.31\linewidth]{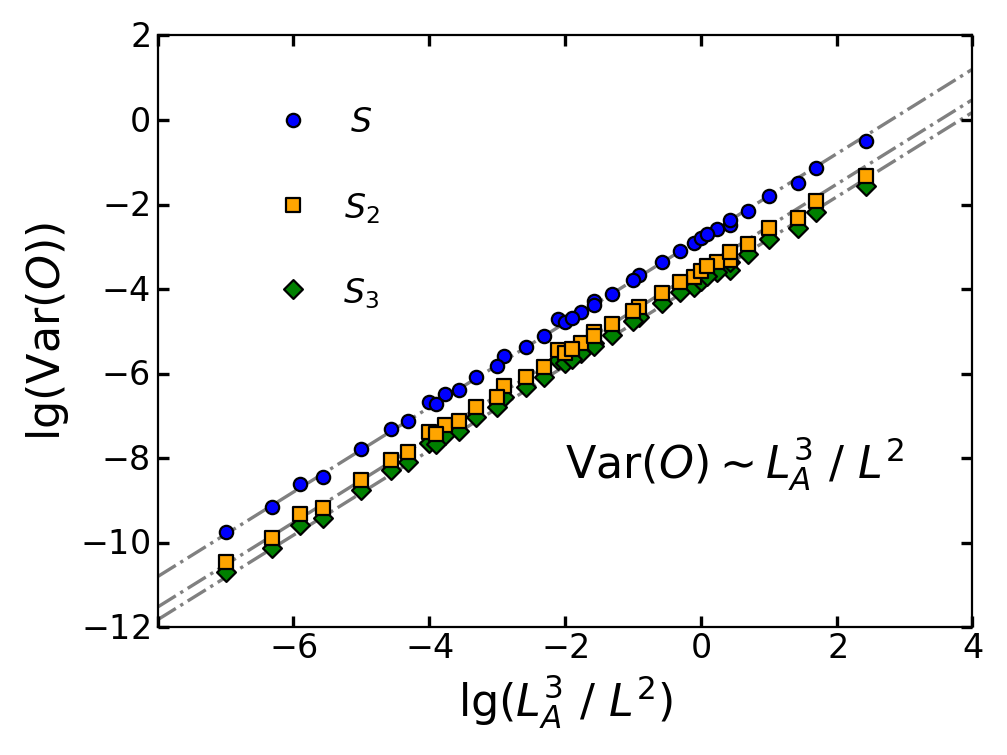}
\includegraphics[width=0.31\linewidth]{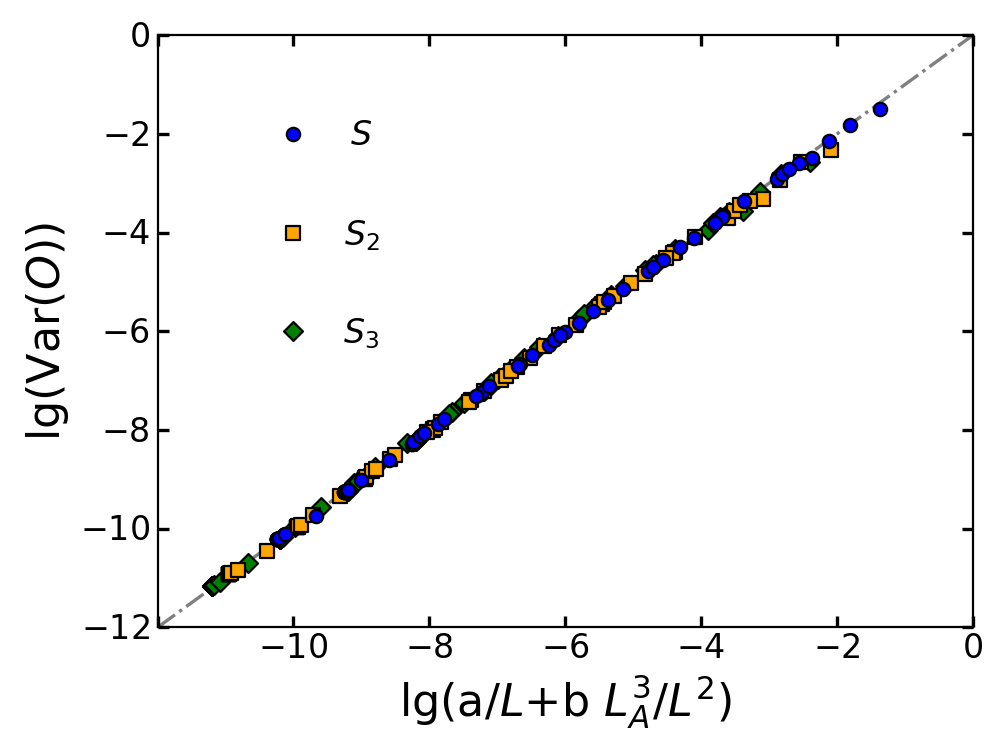}
\caption{Numerical verifications of the scaling law of the variance. (a) and (b) The two scaling regimes correspond to the first and second terms in Eq.~(\ref{eq:73}), respectively. (c) All data collapse into a single curve described by Eq.~(\ref{eq:73}) after rescaling.}
\label{F9}
\end{figure*}

Finally, we perform an extensive simulations of the variance for system length from $L_A=10^1$ to $10^3$ and from $L=10^4$ to $10^8$, for different entanglement probes $O=S, S_2, S_3$. In Fig.~\ref{F9}, we verify the two scaling behaviors displayed by Eq.~(\ref{eq:73}): In the regime $L_A \lesssim L^{1/3}$ we have ${\rm Var}(O)\sim 1/L$ (a) whereas in the regime $L_A \gtrsim L^{1/3}$ we have ${\rm Var}(O)\sim L_A^3/L^2$ (b). Upon proper rescaling, we find that the data all collapse into a single curve (c). We thus have numerically verified the scaling law of the variance, and found that the agreement is excellent.

\section{\label{sec_4}Conclusions and outlook}

In this work, we studied the long-time entanglement dynamics of the coupled harmonic oscillator chain, which is quenched from the ground state and undergoes unitary evolution. We generalized the previous theory for temporal fluctuations in long-time entanglement dynamics of free-fermion models \cite{Lih-King_2023} to the present bosonic model. Despite that the particle statistics is completely different, we found that the statistics of temporal entanglement fluctuations is strictly the same as that in free-fermion models: First, irrespective of entanglement probes $O$ and microscopic parameters of the chain, the statistical distribution of entanglement fluctuations is universal. That distribution is asymmetric: Its upper tail gives the sub-Gaussian probability: ${\rm e}^{-{\epsilon^2\over 2 \mathfrak{b}_+}}$ for upward fluctuations $\epsilon=O-\langle O\rangle>0$, while its lower tail gives the sub-Gamma probability: ${\rm e}^{-\frac{\epsilon^2}{2 (\mathfrak{b}_-  + \mathfrak{c}\epsilon)}}$ for downward fluctuations $\epsilon=\langle O\rangle-O>0$. Second, for distinct probes $O$ their variances all obey the scaling law (\ref{eq:73}), which depends only on the (rescaled) subsystem size $L_A$ and the total system size $L$. In particular, for small $L_A\ll L^{1/3}$ the scaling law reduces to $\sim 1/L$, which is an edge (between the subsystem $A$ and its complement $B$) effect; while for large $(L/2\geq)\, L_A\gg L^{1/3}$ it reduces to $\sim L_A^3/L^2$, which is a bulk (of $A$) effect. These findings show that temporal entanglement fluctuations differ completely from usual thermodynamic fluctuations.

The found boson-fermion universality has a deep origin, i.e., the fluctuation statistics for entanglement dynamics of both free-boson and free-fermion models arises from common probabilistic structures: (i) As we have shown, despite the out-of-equilibrium  nature of entanglement fluctuations, at each instant the many-body wavefunction describes a virtual mesoscopic `disordered' sample, with its `disorder realization' essentially given by $N\gg 1$ dynamical phases $\boldsymbol{\varphi}$. Consequently, the temporal fluctuations of entanglement accompanying the wavefunction evolution are statistically equivalent to mesoscopic sample-to-sample fluctuations. (ii) The probability measure of the ensemble of these `virtual' mesoscopic disordered samples has a product structure, which renders entanglement fluctuations displaying the phenomenon of concentration of measure. In addition, on physical grounds the temporal entanglement fluctuations might be attributed to certain fluctuation effects of the motion of the Calabrese-Cardy entangled pairs in a finite system. Since the considerations above are very general, we conjecture that the boson-fermion universality might be further extended to other particle statistics, such as in anyon and supersymmetric systems.

Our theory requires that the pre-quench state to be Gaussian, since the Gaussianity allows us to express an evolving entanglement probe in terms of a functional of the evolving matrix $C(t)$ Eq.~(\ref{cor_mat}). In other words, the probe is completely determined by the instant correlation of $\hat{\r}$. In this work the theory is formulated for the pre-quench state chosen to be a vacuum state. There are no principal difficulties to generalize it to other pre-quench pure states, provided they are Gaussian. However, it remains challenging to extend the present theory to non-Gaussian pre-quench states, which we leave for future studies.

Quantum entanglement is central to physics of Hawking radiation. Therefore, a rich avenue to explore in the future is the relations of out-of-equilibrium entanglement fluctuations in bosonic models to dynamics of Hawking radiation in analogue black hole systems \cite{Isoard_2021}, as realized in cold-atomic Bose-Einstein condensates \cite{Steinhauer_2016} and superconducting qubits \cite{Shi_2023}.

\begin{acknowledgments}
C.T. is supported by NSFC (Grants No. 11925507, No. 12475043 and No. 12047503). T.S. is supported by NSFC (Grants No. 12135018 and No.12047503), by National Key Research and Development Program of China (Grant No. 2021YFA0718304), and by CAS Project for Young Scientists in Basic Research (Grant No. YSBR-057). L.-K.L. is supported by NSFC (Grant No. 11974308). 
\end{acknowledgments}

\appendix
\section{\label{App_A}Explicit form of matrices $\boldsymbol{\gamma(t)}$ and $\boldsymbol{C_A(t)}$}

For the convenience of the readers here we give the explicit expression of the time-dependent covariance matrix $\gamma(t)$, i.e., the real part of the matrix $G(t)$ defined by Eq.~(\ref{eq:77}). The derivations are standard and the details can be found in, e.g., Ref.~\cite{Calabrese07}. In the block-matrix form, it takes the form
\beq
\gamma(t)
=\left[\begin{array}{cc} X(t) & M(t) \\ M^{\rm T}(t) & P(t) \end{array}\right].
\label{eq:82}
\eeq
Note that this $2\times 2$ matrix structure corresponds to the representation Eq.~(\ref{eq:79}) of the vector $\hat{\r}$. Recall that $X=X^{\rm T}$ and $P=P^{\rm T}$.

To find the elements of matrix blocks $X(t)$, $P(t)$ and $M(t)$, firstly, their time-dependence can be cast using operators in the Heinsenberg picture giving 
\beq
\hat{x}_r(t)&=&\sum_k \frac{1}{\sqrt{2 L s_k}} \bigl( \hat{\alpha}_k^\dag\,e^{-i(k x_r - \omega_k t)}+\textrm{h.c.} \bigr),\\
\hat{p}_r(t)&=&i \sum_k \sqrt{\frac{s_k}{2 L}} \bigl( \hat{\alpha}_k^\dag\,e^{-i(k x_r - \omega_k t)} -\textrm{h.c.}\bigr).
\eeq
While the initial vacuum state is annihilated by pre-quench normal modes of all momenta (designated by $\hat{\alpha}_k'$ in this section), these modes are related to the post-quench normal modes via
\beq
\hat{\alpha}_k=     \sqrt{\frac{s_k^2-\bar{s}_k^2}{4 s_k \bar{s}_k}}\,   \hat{\alpha}_{-k}'^{\,\dag}  + \sqrt{\frac{s_k^2+\bar{s}_k^2}{4 s_k \bar{s}_k}}\,\hat{\alpha}'_k,
\eeq 
with $\bar{s}_{k} = \sqrt{1+{4K_0\over \omega_0^2}\sin^2{k\over 2}}$ corresponding to the pre-quench Hamiltonian parameters $(\omega_0,\,K_0)$ (correspondingly, $s_k=\sqrt{1+{4K\over \omega^2}\sin^2{k\over 2}}$ for the post-quench Hamiltonian parameters $(\omega,\,K)$, and with $\omega_k= \omega s_k$). Then, we obtain

\begin{eqnarray}
X_{rs}(t) &=& \left\langle \psi(t)| \hat{x}_r \hat{x}_s|\psi(t)\right\rangle = \left\langle \psi_0| \hat{x}_r(t) \hat{x}_s(t)|\psi_0\right\rangle \nonumber\\
&=& \frac{1}{L}\sum\limits_{k} \frac{{\rm e}^{-i k(s-r)} }{s_k}\left(E_+(k) + E_-(k)\cos{(2\omega_k t)}\right)\nonumber\\
&\equiv&\bar{x}_{r-s}(t),
\label{mat_ele1}
\end{eqnarray}
\begin{eqnarray}
P_{rs}(t) &=& \left\langle \psi(t)| \hat{p}_r \hat{p}_s|\psi(t)\right\rangle= \left\langle \psi_0| \hat{p}_r(t) \hat{p}_s(t)|\psi_0\right\rangle\nonumber\\
&=&\frac{1}{ L}\sum\limits_{k} {\rm e}^{-i k(s-r)}s_k(E_+(k) - E_-(k) \cos{(2\omega_k t)})\nonumber\\
&\equiv& \bar{p}_{r-s}(t),
\label{mat_ele2}
\end{eqnarray}
and
\begin{eqnarray}
M_{rs}(t) &=& \left\langle \psi_0| \hat{x}_r(t) \hat{p}_s(t)|\psi_0\right\rangle - i\delta_{r,s}/2\nonumber\\
&=&- \frac{1}{L}\sum\limits_{k}{\rm e}^{-i k(s-r)} E_-(k) \sin{(2\omega_k t)}\nonumber\\
&\equiv&\bar{m}_{r-s}(t).
\label{mat_ele3}
\end{eqnarray}
Here
\begin{equation}\label{eq:97}
 E_\pm(k) = {1\over 4}({\omega s_k\over\omega_0 \bar{s}_{k}} \pm {\omega_0\bar{s}_{k}\over \omega s_k})
\end{equation}
are even functions of the Bloch momenta $k$.
Note that
\begin{equation}\label{eq:59}
  E_+(k)\geq 1/2,
\end{equation}
and the equality holds only if $\omega s_k=\omega_0\bar{s}_{k}$. It is easy to see that $\bar{x}_{l},\,\bar{p}_{l}$ and $\bar{m}_{l}$ are all real.

Then, combining Eqs.~(\ref{mat_ele1})-(\ref{mat_ele3}) and Eq.~(\ref{cor_mat}), we find that the matrix $C_A(t)$ takes the block-Toeplitz form Eq.~(\ref{CFM}), with the $2\times2$ block-matrix element

\beq
\tau_{l}=\left[\begin{array}{cc}i \bar{m}_l(t)&i \bar{p}_l(t)\\-i \bar{x}_l(t)&-i \bar{m}_l(t)\end{array}\right].
\label{TC}
\eeq
Note that this $2\times 2$ matrix structure corresponds to the representation Eq.~(\ref{eq:79}), similar to Eq.~(\ref{eq:82}).

\begin{figure}[b]
\includegraphics[width=0.8\linewidth]{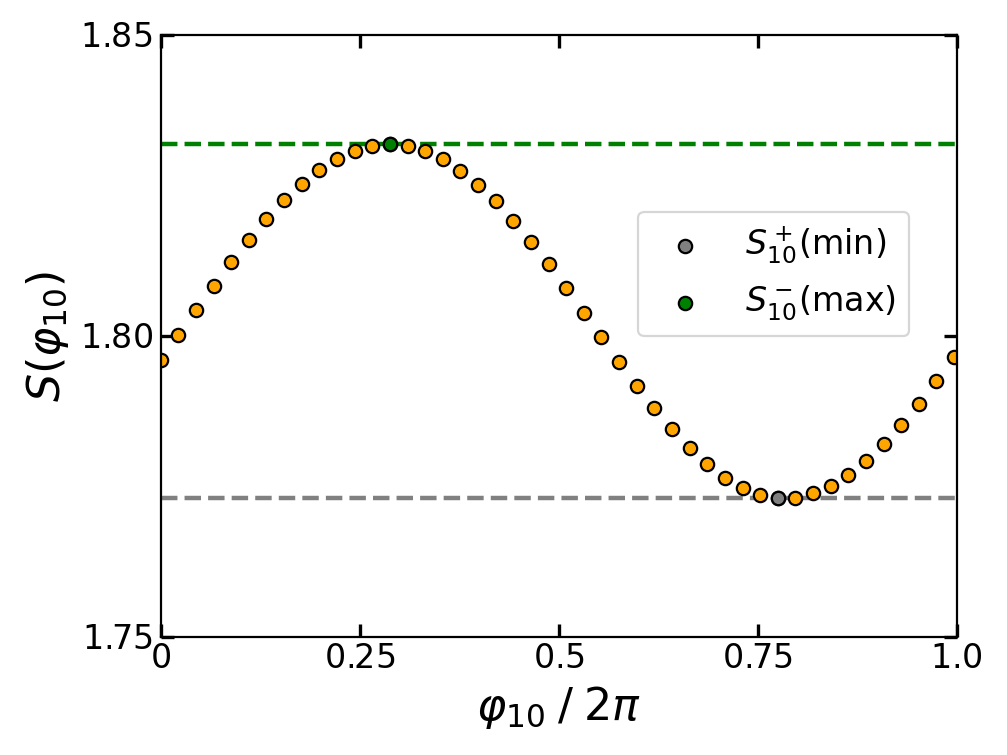}
\includegraphics[width=0.8\linewidth]{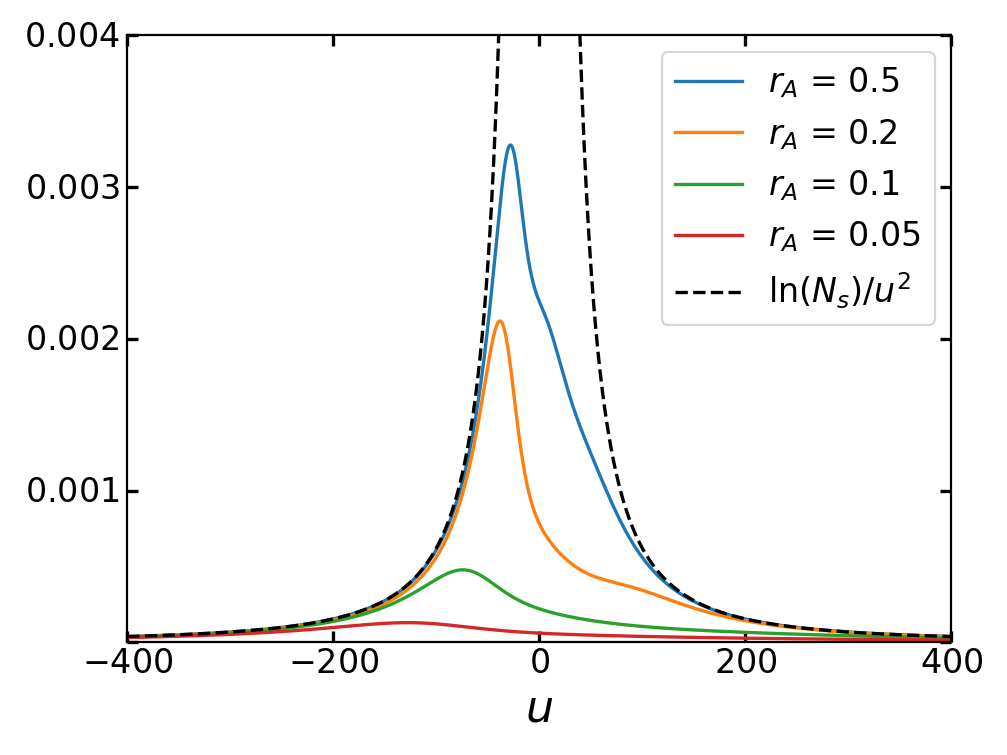}
\caption{(a) Variation of $S(\boldsymbol{\varphi})$ with the argument $\varphi_{10}$. The other arguments, i.e., the random realization $\boldsymbol{\varphi}'_{10}$ is fixed. The minimum (gray dot) gives $S^+_{10}$ and the maximum (green dot) gives $S^-_{10}$.  $L=124$. (b) The left-hand side of the modified logarithmic Sobolev inequality~(\ref{Sobolev}) for different ratios $r_A$ with fixed $L_A=25$ generated with \textit{ab initio} data. All lines approach the asymptote (dash line) $\ln{N_s}/u^2$ at large $u$, where $N_s = 500$ is the sampling size used.}
\label{F5}
\end{figure}

\section{\label{App_C} More numerical results}
In Fig.~\ref{F5}(a) we show an example of the numerical determination of $O_m^{\pm}$, the average of the extreme values of $O(\boldsymbol{\varphi})$ with respect to the argument $\varphi_m$. Note that the function $O(\boldsymbol{\varphi})$ is $2\pi$-periodic in each argument.

In Fig. \ref{F5}(b) we simulate the left-hand side of the modified logarithmic Sobolev inequality~(\ref{Sobolev}), ${d\over du}{G\over u}$, for different ratios $r_A$ with \textit{ab initio} data. The peak observed in $u<0$ region generally reflects the asymmetry between the upper and lower tails of the distribution. The position of the peak is comparable to the parameter $c_-$ in the sub-Gamma lower tail, wherein a true divergence at $u=-1/|c_-|$ is rounded due to data truncation. On the other hand, the value of ${d\over du}{G\over u}$ at origin is related to the variance factor. Note that given a Gaussian distribution, with truncated data we would have a symmetric flat top curve around the origin, with half height giving the variance. We also note that, independent of the different data sets, all curves follow the same large-$u$ asymptote given by a $1/u^2$-curve. This is an artifact of the exponential dependence of the characteristic function for large $u$, arising from that the average is dominated by rare events of the given data set. Then the quantity ${d\over du}{G\over u}$ is independent of the details with the sample size as the only relevant parameter.

\begin{figure}[b]
\includegraphics[width=0.8\linewidth]{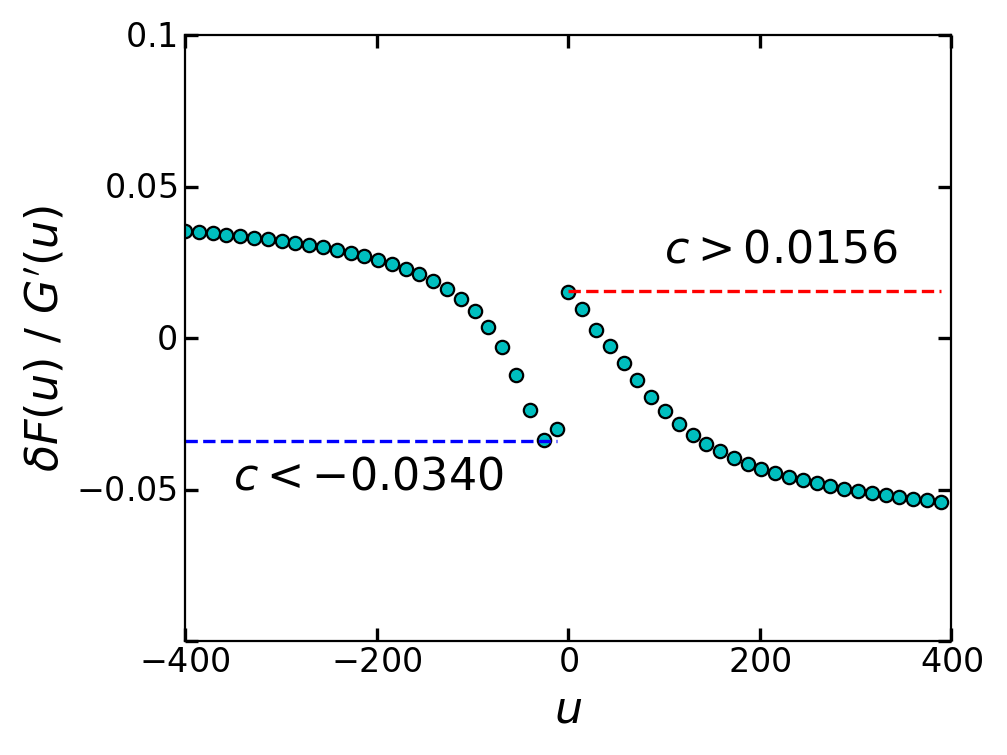}
\caption{Numerical determination of $c_\pm$ from $\delta F/\frac{dG}{du}$ versus $u$ line, according to inequalities (\ref{sobo3}) and (\ref{sobo3b}).}
\label{Fc}
\end{figure}

Finally, in Fig.~\ref{Fc} we show an example of the numerical determination of the scale parameter $c_{\pm}$ defined in Eqs.~(\ref{sobo3}) and (\ref{sobo3b}). Table~\ref{table:1} summarizes the numerical values of $(b_{\pm},c_{\pm})$ for different values of the ratio $r_A$.

\begin{table}
\centering
  \caption{The value of $b_\pm$ and $c_\pm$ for various ratios $r_A$, determined for probes $S$ and $S_2$, respectively.}\label{table:1}
\begin{tabular}{c|cccc}
     \hline
     \hline
         & & $S$ & $L_A=25$  &
\\
     $\quad$$r_A\quad$ &   $\quad b_+ \quad $  & $\quad c_+ \quad $ & $\quad b_-\quad$ & $\quad c_-\quad$ \\
     \hline
     $0.05$&$\quad 0.0007$ &$\quad -0.0098 \quad$&$0.0007$&-0.0130\\
     $0.1$ & $\quad 0.0026$ &$-0.0184$&$0.0025$& -0.0222\\
     $0.2$ &$\quad 0.0091$ &$-0.0246$&$0.0092$&-0.0434\\
     $0.3$ & $\quad 0.0169$&$-0.0170$&$0.0171$&-0.0410\\
     $0.4$ & $\quad 0.0241$&$-0.0030$&$0.0230$&-0.0328\\
     $0.5$ &  $\quad 0.0256$&$0.0156$&$0.0234$&-0.0340\\
     \hline
     \hline
    \end{tabular}

\begin{tabular}{c}

\\

\end{tabular}

\begin{tabular}{c|cccc}
     \hline
     \hline
         & &$S_2$  & $L_A=25$  &
\\
      $\quad$$r_A\quad$ &   $\quad b_+ \quad $  & $\quad c_+ \quad $ & $\quad b_-\quad$ & $\quad c_-\quad$ \\
     \hline
     $0.05$&$\quad 0.0001$ &$\quad -0.0044 \quad$&$0.0001$&-0.0058\\
     $0.1$ & $\quad 0.0004$ &$-0.0076$&$0.0004$&-0.0094\\
     $0.2$ &$\quad 0.0014$ &$-0.0088$&$0.0015$&-0.0172\\
     $0.3$ & $\quad 0.0024$&$-0.0046$&$0.0026$&-0.0152\\
     $0.4$ & $\quad 0.0032$&$0.0014$&$0.0033$&-0.0120\\
     $0.5$ &  $\quad 0.0033$&$0.0072$&$0.0033$&-0.0122\\
     \hline
     \hline
    \end{tabular}
\end{table}

\section{Proof of inequality (\ref{eq:26})}
\label{sec:inequality}

Using the elementary inequality: $1-{z\over 2}\geq \sqrt{1-z},\, z\geq 0$ we find that
\begin{eqnarray}
\label{eq:37}
1-{1\over 2}(1-y)^2&\geq& \sqrt{1-(1-y)^2},\nonumber\\
&=& \sqrt{2y-y^2},\quad {\rm for}\, 0\leq y\leq 1.
\end{eqnarray}
Let us write $y$ as $y=1/(1+x),\, x\geq 0$. With its substitution the inequality above is rewritten as
\begin{eqnarray}
\label{eq:38}
&&1-{1\over 2}\left(x\over 1+x\right)^2\geq {\sqrt{1+2x}\over 1+x}\nonumber\\
&\Leftrightarrow& 1+x-{x^2\over 2(1+x)}\geq \sqrt{1+2x},
\end{eqnarray}
which is inequality (\ref{eq:26}). This inequality was given in Ref.~\cite{Boucheron_2013} without proof.

\section{Discussions on negative $\boldsymbol{c_\pm}$}
\label{sec:scale_factor}

In this Appendix we focus on negative $c_+$. Discussions on negative $c_-$ are similar. For $c_+<0$ we have
\beq\label{eq:39}
P_+(\epsilon) \leq
{\rm e}^{\widetilde{{\cal E}}(u)},
\eeq
where the exponent is given by
\begin{equation}\label{eq:40}
  \widetilde{{\cal E}}(u)\equiv -u \epsilon+\frac{b_+}{2}\frac{u^2}{1+|c_+| \, u}.
\end{equation}
We wish to minimize $\widetilde{{\cal E}}(u)$ for $u>0$. Let us rewrite it as
\beq\label{eq:41}
\widetilde{{\cal E}}(u)= \left({\epsilon\over |c_+|}-{b_+\over c_+^2}\right)+
\left({b_+\over 2c_+^2}-{\epsilon\over |c_+|}\right)y+{b_+\over 2c_+^2}{1\over y}\,\,\,\,\,\,\,
\eeq
with $y\equiv 1+|c_+|u\geq 1$. Note that
\begin{equation}\label{eq:42}
  {\rm for}\,\epsilon>{{b_+\over 2|c_+|}}:\, {b_+\over 2c_+^2}-{\epsilon\over |c_+|}<0.
\end{equation}
In this case, we have
\begin{equation}\label{eq:43}
  \widetilde{{\cal E}}(u) \stackrel{u\rightarrow\infty}{\longrightarrow} -\infty,
\end{equation}
from which
\beq\label{eq:44}
P_+(\epsilon>{{b_+\over 2|c_+|}}) =0
\eeq
follows.

\section{\label{App_B}The derivation of Eq.~(\ref{lip1})}
In this Appendix we derive Eq.~(\ref{lip1}). The derivations follow closely those in our previous work \cite{Lih-King_2023}. According to Sec.~\ref{sec_3_3}, the variance factor $b_\pm$ is proportional to ${\rm Var}(O)$. So, we simply have to show that $b_\pm\propto\left\langle |\partial_{\boldsymbol{\varphi}}O|^2\right\rangle$.

For the convenience of the readers let us rewrite here Eqs.~(\ref{var1}) and (\ref{var2}), that defined $b_\pm$,
\beq
b_\pm\equiv \sum_{m=0}^{N-1} \left\langle  (O-O_m^\pm)^2\right\rangle.
\eeq
Note that $(O-O_m^\pm)^2$ is a function of $\boldsymbol{\varphi}$. Let $\varphi_m^\pm$ be the value of $\varphi_m$ at which $O(\boldsymbol{\varphi})=O^\pm_m (\boldsymbol{\varphi}'_m)$.  By the mean value theorem there exists ${\bar \varphi}_m^\pm$ between $\varphi_m$ and $\varphi_m^\pm$ which depends on $\boldsymbol{\varphi}$, so that
\begin{equation}\label{mean}
  (O-O_m^\pm)^2(\boldsymbol{\varphi})
  =(\varphi_m-\varphi_m^\pm)^2\left(\partial_{{\varphi}_m}O|_{{\varphi}_m=\bar{\varphi}_m^\pm}\right)^2.
\end{equation}
Here we use the suppressed notion $\bar{\varphi}_m^\pm$ to stand for $(\varphi_0,\ldots,\bar{\varphi}_m^\pm,\ldots,\varphi_{N-1})$. We thus obtain
\begin{equation}\label{bpm_1}
  b_\pm=\sum_{m=0}^{N-1}\left\langle(\varphi_m-\varphi_m^\pm)^2\left(\partial_{{\varphi}_m}O|_{{\varphi}_m=\bar{\varphi}_m^\pm}\right)^2\right\rangle.
\end{equation}
Then we do the Fourier expansion of $\partial_{{\varphi}_m}O$ with respect to $\varphi_m$. As exemplified by $C(\boldsymbol{\varphi})$ in Sec.~\ref{sec_3}, the random part of the general expression of $O$ decays with the total system size $L$ according to certain power law. Thus we can keep the Fourier expansion up to the second order to get
\begin{equation}\label{fourier}
\partial_{{\varphi}_m}O=A_0+A_1\sin(\varphi_m+\phi_{m1})+A_2\sin(2\varphi_m+\phi_{m2}),
\end{equation}
where $A_i$ and $\phi_{mi}$ are constants independent of $\varphi_m$.
With the use of the following relations,
\begin{equation}\label{ieq}
  (\partial_{{\varphi}_m}O)^2\leq (|A_0|+|A_1|+|A_2|)^2\leq 3(A_0^2+A_1^2+A_2^2)
\end{equation}
and
\begin{equation}\label{inte}
  \int\frac{d\varphi_m}{2\pi}(\partial_{{\varphi}_m}O)^2=A_0^2+\frac{1}{2}(A_1^2+A_2^2),
\end{equation}
we obtain
\begin{equation}\label{bound}
(\partial_{{\varphi}_m}O)^2\leq 6\int\frac{d\varphi_m}{2\pi}(\partial_{{\varphi}_m}O)^2.
\end{equation}
This bound suggests that $(\partial_{{{\varphi}}_m}S)^2$ at given ${\varphi}_m$ is the same order as its average, i.e.,
\begin{equation}
(\partial_{{\varphi}_m}O)^2=r_m(\boldsymbol{\varphi})\int\frac{d\varphi_m}{2\pi}(\partial_{{\varphi}_m}O)^2, \quad r_m(\boldsymbol{\varphi})={\cal O}(1).
\end{equation}
Substituting it into Eq.~(\ref{bpm_1}) we obtain
\begin{equation}\label{eq:58}
b_\pm=\sum_{m=0}^{N-1}\left\langle\left((\varphi_m-\varphi_m^\pm)^2
r_m(\boldsymbol{\varphi}_m^{\pm*})\right)\int\frac{d\varphi_m}{2\pi}(\partial_{{\varphi}_m}O)^2\right\rangle,
\end{equation}
where $\boldsymbol{\varphi}_m^{\pm*}\equiv (\varphi_0,\ldots,\bar{\varphi}^\pm_m,\ldots,\varphi_{N-1})$. Because the first factor characterizes the distance between $\varphi_m$ and $\varphi_m^\pm$ and the second describes the $\varphi_m$ average of $(\partial_{{\varphi}_m}O)^2$, we assume that they are statistically uncorrelated and obtain
\begin{equation}\label{bpm_2}
b_\pm=\sum_{m=0}^{N-1}\left\langle (\varphi_m-\varphi_m^\pm)^2  r_m(\boldsymbol{\varphi}_m^{\pm*})\right\rangle\left\langle\int\frac{d\varphi_m}{2\pi}(\partial_{{\varphi}_m}O)^2\right\rangle.
\end{equation}
Furthermore, by symmetry $\left\langle (\varphi_m-\varphi_m^\pm)^2
r_m(\boldsymbol{\varphi}_m^{\pm*})\right\rangle$ must be the same for distinct $m$, at least approximately. As a result, we reduce Eq.~(\ref{bpm_2}) to
\begin{equation}\label{bpm_3}
b_\pm=R^\pm\left\langle |\partial_{\boldsymbol{\varphi}}O|^2\right\rangle,
\end{equation}
where
\begin{equation}\label{rpm}
R^\pm=\frac{1}{N}\sum_{m=0}^{N-1}\left\langle (\varphi_m-\varphi_m^\pm)^2
r_m(\boldsymbol{\varphi}_m^{\pm*})\right\rangle
\end{equation}
is an overall numerical coefficient. Combining with the fact that $b_\pm\propto{\rm Var}(O)$, we finally arrive at Eq.~(\ref{lip1}).

\section{An identity for the parameter derivative of matrix functions}
\label{sec:derivative_matrix_function}

Let $M(s)$ be a matrix depending on the parameter $s \in \mathbb{R}$. We assume that it is
diagonalizable, i.e. there exists a matrix $U$ such that
\begin{equation}
    \label{eq:112}
     U^{-1}(s) M(s) U(s)=
    \left[
    \begin{array}{ccc}
    d_1(s) & \phantom{} & \phantom{}  \\
    \phantom{} & d_2(s) & \phantom{}  \\
    \phantom{} & \phantom{} & \ddots  \\
    \end{array}\right]\equiv D(s)
     .
\end{equation}
Note that both $U$ and the eigenvalues $d_i$ depend on $s$ in general. From now on we suppress $s$ in $M,\,U,\,D$ and the eigenvalue spectrum $\{d_i\}$ in order to make formulas compact.

\par Consider a generic matrix function $f(M)$, with $f(x)$ being differentiable for $x \in \mathbb{R}$. By
Eq.~\eqref{eq:112}, we have
\begin{eqnarray}
    \label{eq:113}
    f(M)=U
    \left[
    \begin{array}{ccc}
        f(d_1) & \phantom{} & \phantom{}  \\
        \phantom{} & f(d_2) & \phantom{}  \\
        \phantom{} & \phantom{} & \ddots  \\
        \end{array}\right]
    U^{-1}=Uf(D)U^{-1}.\quad\,\,
\end{eqnarray}
Taking the derivative with respect to $s$ on both sides, we obtain
\begin{eqnarray}
    \label{eq:114}
    \partial_s f(M)&=&(\partial_s U) f(D) U^{-1}
    +Uf(D) (\partial_s U^{-1})\nonumber\\
    &+&U (f'(D) \partial_s D) U^{-1}.
\end{eqnarray}
When we take its trace, thanks to the cyclic relation the first two terms cancel out. As a result,
\begin{equation}
    \label{eq:115}
    \text{tr}(\partial_s f(M))=\text{tr} (f'(D) \partial_s D).
\end{equation}
On the other hand,
\begin{eqnarray}
    \label{eq:116}
    \text{tr}(f^{\prime}(M)\partial_sM)=\text{tr}\left(f^{\prime}(UDU^{-1})\partial_s(UDU^{-1})\right).
\end{eqnarray}
With the substitution of $f^{\prime}(UDU^{-1})=Uf^{\prime}(D)U^{-1}$ and
\begin{equation}
    \label{eq:117}
    \partial_s(UDU^{-1})=(\partial_s U)DU^{-1}+UD(\partial_s U^{-1})+U(\partial_s D)U^{-1},
\end{equation}
Eq.~\eqref{eq:116} is simplified as
\begin{eqnarray}
    \label{eq:118}
    \text{tr}(f^{\prime}(M)\partial_s M)
    &=&\text{tr} \Big(
    f^{\prime}(D)U^{-1}(\partial_s U)D \nonumber\\
    &+&Uf^{\prime}(D)D(\partial_s U^{-1})+f^{\prime}(D)\partial_s D\Big).\quad
\end{eqnarray}
By the cyclic relation we find that the first two terms cancel out. Thus Eq.~\eqref{eq:117} reduces to
\begin{equation}
    \label{eq:119}
    \text{tr}(f^{\prime}(M)\partial_s M)=\text{tr}(f^{\prime}(D)\partial_s D).
\end{equation}
Comparing Eq.~\eqref{eq:115} and \eqref{eq:119}, we arrive at the identity:
\begin{equation}
    \label{eq:120}
    \text{tr}(\partial_s f(M))=\text{tr}(f^{\prime}(M)\partial_s M).
\end{equation}
Note that for this identity the trace operation is essential.


%


\nocite{*}


\begin{thebibliography}{62}%
\makeatletter
\providecommand \@ifxundefined [1]{%
 \@ifx{#1\undefined}
}%
\providecommand \@ifnum [1]{%
 \ifnum #1\expandafter \@firstoftwo
 \else \expandafter \@secondoftwo
 \fi
}%
\providecommand \@ifx [1]{%
 \ifx #1\expandafter \@firstoftwo
 \else \expandafter \@secondoftwo
 \fi
}%
\providecommand \natexlab [1]{#1}%
\providecommand \enquote  [1]{``#1''}%
\providecommand \bibnamefont  [1]{#1}%
\providecommand \bibfnamefont [1]{#1}%
\providecommand \citenamefont [1]{#1}%
\providecommand \href@noop [0]{\@secondoftwo}%
\providecommand \href [0]{\begingroup \@sanitize@url \@href}%
\providecommand \@href[1]{\@@startlink{#1}\@@href}%
\providecommand \@@href[1]{\endgroup#1\@@endlink}%
\providecommand \@sanitize@url [0]{\catcode `\\12\catcode `\$12\catcode
  `\&12\catcode `\#12\catcode `\^12\catcode `\_12\catcode `\%12\relax}%
\providecommand \@@startlink[1]{}%
\providecommand \@@endlink[0]{}%
\providecommand \url  [0]{\begingroup\@sanitize@url \@url }%
\providecommand \@url [1]{\endgroup\@href {#1}{\urlprefix }}%
\providecommand \urlprefix  [0]{URL }%
\providecommand \Eprint [0]{\href }%
\providecommand \doibase [0]{https://doi.org/}%
\providecommand \selectlanguage [0]{\@gobble}%
\providecommand \bibinfo  [0]{\@secondoftwo}%
\providecommand \bibfield  [0]{\@secondoftwo}%
\providecommand \translation [1]{[#1]}%
\providecommand \BibitemOpen [0]{}%
\providecommand \bibitemStop [0]{}%
\providecommand \bibitemNoStop [0]{.\EOS\space}%
\providecommand \EOS [0]{\spacefactor3000\relax}%
\providecommand \BibitemShut  [1]{\csname bibitem#1\endcsname}%
\let\auto@bib@innerbib\@empty
\bibitem [{\citenamefont {Anderson}(2008)}]{Anderson_1973}%
  \BibitemOpen
  \bibfield  {author} {\bibinfo {author} {\bibfnamefont {P.~W.}\ \bibnamefont
  {Anderson}},\ }\bibfield  {title} {\bibinfo {title} {Resonating valence
  bonds: a new kind of insulator?},\ }\href
  {https://doi.org/10.1016/0025-5408(73)90167-0} {\bibfield  {journal}
  {\bibinfo  {journal} {Mat. Res. Bull.}\ }\textbf {\bibinfo {volume} {8}},\
  \bibinfo {pages} {153} (\bibinfo {year} {2008})}\BibitemShut {NoStop}%
\bibitem [{\citenamefont {Li}\ and\ \citenamefont
  {Haldane}(2008)}]{Li_PRL_2008}%
  \BibitemOpen
  \bibfield  {author} {\bibinfo {author} {\bibfnamefont {H.}~\bibnamefont
  {Li}}\ and\ \bibinfo {author} {\bibfnamefont {F.~D.~M.}\ \bibnamefont
  {Haldane}},\ }\bibfield  {title} {\bibinfo {title} {Entanglement spectrum as
  a generalization of entanglement entropy: identification of topological order
  in non-abelian fractional quantum hall effect states},\ }\href
  {https://doi.org/10.1103/PhysRevLett.101.010504} {\bibfield  {journal}
  {\bibinfo  {journal} {Phys. Rev. Lett.}\ }\textbf {\bibinfo {volume} {101}},\
  \bibinfo {pages} {010504} (\bibinfo {year} {2008})}\BibitemShut {NoStop}%
\bibitem [{\citenamefont {Coleman}(2015)}]{Coleman2015}%
  \BibitemOpen
  \bibfield  {author} {\bibinfo {author} {\bibfnamefont {P.}~\bibnamefont
  {Coleman}},\ }\href {https://doi.org/10.1017/CBO9781139020916} {\emph
  {\bibinfo {title} {Introduction to Many-Body Physics}}}\ (\bibinfo
  {publisher} {Cambridge University Press},\ \bibinfo {year}
  {2015})\BibitemShut {NoStop}%
\bibitem [{\citenamefont {Kitaev}(2003)}]{Kitaev_AP_2003}%
  \BibitemOpen
  \bibfield  {author} {\bibinfo {author} {\bibfnamefont {A.}~\bibnamefont
  {Kitaev}},\ }\bibfield  {title} {\bibinfo {title} {Fault-tolerant quantum
  computation by anyons},\ }\href
  {https://doi.org/10.1016/S0003-4916(02)00018-0} {\bibfield  {journal}
  {\bibinfo  {journal} {Ann. Phys. (N.Y.)}\ }\textbf {\bibinfo {volume}
  {303}},\ \bibinfo {pages} {2} (\bibinfo {year} {2003})}\BibitemShut {NoStop}%
\bibitem [{\citenamefont {Chen}\ \emph {et~al.}(2010)\citenamefont {Chen},
  \citenamefont {Gu},\ and\ \citenamefont {Wen}}]{Chen_2010}%
  \BibitemOpen
  \bibfield  {author} {\bibinfo {author} {\bibfnamefont {X.}~\bibnamefont
  {Chen}}, \bibinfo {author} {\bibfnamefont {Z.-C.}\ \bibnamefont {Gu}},\ and\
  \bibinfo {author} {\bibfnamefont {X.-G.}\ \bibnamefont {Wen}},\ }\bibfield
  {title} {\bibinfo {title} {Local unitary transformation, long-range quantum
  entanglement, wave function renormalization, and topological order},\ }\href
  {https://doi.org/10.1103/PhysRevB.82.155138} {\bibfield  {journal} {\bibinfo
  {journal} {Phys. Rev. B}\ }\textbf {\bibinfo {volume} {82}},\ \bibinfo
  {pages} {155138} (\bibinfo {year} {2010})}\BibitemShut {NoStop}%
\bibitem [{\citenamefont {Islam}\ \emph {et~al.}(2015)\citenamefont {Islam}
  \emph {et~al.}}]{Islam_2015}%
  \BibitemOpen
  \bibfield  {author} {\bibinfo {author} {\bibfnamefont {R.}~\bibnamefont
  {Islam}} \emph {et~al.},\ }\bibfield  {title} {\bibinfo {title} {Measuring
  entanglement entropy in a quantum many-body system},\ }\href
  {https://doi.org/10.1038/nature15750} {\bibfield  {journal} {\bibinfo
  {journal} {Nature}\ }\textbf {\bibinfo {volume} {528}},\ \bibinfo {pages}
  {77} (\bibinfo {year} {2015})}\BibitemShut {NoStop}%
\bibitem [{\citenamefont {Satzinger}\ \emph {et~al.}(2021)\citenamefont
  {Satzinger} \emph {et~al.}}]{Satzinger_2021}%
  \BibitemOpen
  \bibfield  {author} {\bibinfo {author} {\bibfnamefont {K.~J.}\ \bibnamefont
  {Satzinger}} \emph {et~al.},\ }\bibfield  {title} {\bibinfo {title}
  {Realizing topologically ordered states on a quantum processor},\ }\href
  {https://doi.org/10.1126/science.abi8378} {\bibfield  {journal} {\bibinfo
  {journal} {Science}\ }\textbf {\bibinfo {volume} {374}},\ \bibinfo {pages}
  {1237} (\bibinfo {year} {2021})}\BibitemShut {NoStop}%
\bibitem [{\citenamefont {Semeghini}\ \emph {et~al.}(2021)\citenamefont
  {Semeghini} \emph {et~al.}}]{Semeghini_2021}%
  \BibitemOpen
  \bibfield  {author} {\bibinfo {author} {\bibfnamefont {G.}~\bibnamefont
  {Semeghini}} \emph {et~al.},\ }\bibfield  {title} {\bibinfo {title} {Probing
  topological spin liquids on a programmable quantum simulator},\ }\href
  {https://doi.org/10.1126/science.abi8794} {\bibfield  {journal} {\bibinfo
  {journal} {Science}\ }\textbf {\bibinfo {volume} {374}},\ \bibinfo {pages}
  {1242} (\bibinfo {year} {2021})}\BibitemShut {NoStop}%
\bibitem [{\citenamefont {Joshi}\ \emph {et~al.}(2023)\citenamefont {Joshi}
  \emph {et~al.}}]{Joshi_2023}%
  \BibitemOpen
  \bibfield  {author} {\bibinfo {author} {\bibfnamefont {M.~K.}\ \bibnamefont
  {Joshi}} \emph {et~al.},\ }\bibfield  {title} {\bibinfo {title} {Exploring
  large-scale entanglement in quantum simulation},\ }\href
  {https://doi.org/10.1038/s41586-023-06768-0} {\bibfield  {journal} {\bibinfo
  {journal} {Nature}\ }\textbf {\bibinfo {volume} {624}},\ \bibinfo {pages}
  {539} (\bibinfo {year} {2023})}\BibitemShut {NoStop}%
\bibitem [{\citenamefont {Karamlou}\ \emph {et~al.}(2024)\citenamefont
  {Karamlou} \emph {et~al.}}]{Karamlou_2024}%
  \BibitemOpen
  \bibfield  {author} {\bibinfo {author} {\bibfnamefont {A.~H.}\ \bibnamefont
  {Karamlou}} \emph {et~al.},\ }\bibfield  {title} {\bibinfo {title} {Probing
  entanglement in a 2{D} hard-core {B}ose-{H}ubbard lattice},\ }\href
  {https://doi.org/10.1038/s41586-024-07325-z} {\bibfield  {journal} {\bibinfo
  {journal} {Nature}\ }\textbf {\bibinfo {volume} {629}},\ \bibinfo {pages}
  {561} (\bibinfo {year} {2024})}\BibitemShut {NoStop}%
\bibitem [{\citenamefont {Islam}\ \emph {et~al.}(2016)\citenamefont {Islam}
  \emph {et~al.}}]{Kaufman_2016}%
  \BibitemOpen
  \bibfield  {author} {\bibinfo {author} {\bibfnamefont {R.}~\bibnamefont
  {Islam}} \emph {et~al.},\ }\bibfield  {title} {\bibinfo {title} {Quantum
  thermalization through entanglement in an isolated many-body system},\ }\href
  {https://doi.org/10.1126/science.aaf6725} {\bibfield  {journal} {\bibinfo
  {journal} {Science}\ }\textbf {\bibinfo {volume} {353}},\ \bibinfo {pages}
  {794} (\bibinfo {year} {2016})}\BibitemShut {NoStop}%
\bibitem [{\citenamefont {Bernien}\ \emph {et~al.}(2017)\citenamefont {Bernien}
  \emph {et~al.}}]{Bernien_2017}%
  \BibitemOpen
  \bibfield  {author} {\bibinfo {author} {\bibfnamefont {H.}~\bibnamefont
  {Bernien}} \emph {et~al.},\ }\bibfield  {title} {\bibinfo {title} {Probing
  many-body dynamics on a 51-atom quantum simulator},\ }\href
  {https://doi.org/doi.org/10.1038/nature24622} {\bibfield  {journal} {\bibinfo
   {journal} {Nature}\ }\textbf {\bibinfo {volume} {551}},\ \bibinfo {pages}
  {579} (\bibinfo {year} {2017})}\BibitemShut {NoStop}%
\bibitem [{\citenamefont {Turner}\ \emph {et~al.}(2018)\citenamefont {Turner},
  \citenamefont {Michailidis}, \citenamefont {Abanin}, \citenamefont {Serbyn},\
  and\ \citenamefont {Papi\'{c}}}]{Turner_2018}%
  \BibitemOpen
  \bibfield  {author} {\bibinfo {author} {\bibfnamefont {C.~J.}\ \bibnamefont
  {Turner}}, \bibinfo {author} {\bibfnamefont {A.~A.}\ \bibnamefont
  {Michailidis}}, \bibinfo {author} {\bibfnamefont {D.~A.}\ \bibnamefont
  {Abanin}}, \bibinfo {author} {\bibfnamefont {M.}~\bibnamefont {Serbyn}},\
  and\ \bibinfo {author} {\bibfnamefont {Z.}~\bibnamefont {Papi\'{c}}},\
  }\bibfield  {title} {\bibinfo {title} {Weak ergodicity breaking from quantum
  many-body scars},\ }\href {https://doi.org/10.1038/s41567-018-0137-5}
  {\bibfield  {journal} {\bibinfo  {journal} {Nat. Phys.}\ }\textbf {\bibinfo
  {volume} {14}},\ \bibinfo {pages} {745} (\bibinfo {year} {2018})}\BibitemShut
  {NoStop}%
\bibitem [{\citenamefont {Page}(1993)}]{Page_1993}%
  \BibitemOpen
  \bibfield  {author} {\bibinfo {author} {\bibfnamefont {D.~N.}\ \bibnamefont
  {Page}},\ }\bibfield  {title} {\bibinfo {title} {Average entropy of a
  subsystem},\ }\href {https://doi.org/10.1103/PhysRevLett.71.1291} {\bibfield
  {journal} {\bibinfo  {journal} {Phys. Rev. Lett.}\ }\textbf {\bibinfo
  {volume} {71}},\ \bibinfo {pages} {1291} (\bibinfo {year}
  {1993})}\BibitemShut {NoStop}%
\bibitem [{\citenamefont {Popescu}\ \emph {et~al.}(2018)\citenamefont
  {Popescu}, \citenamefont {Short},\ and\ \citenamefont
  {Winters}}]{Popescu_2006}%
  \BibitemOpen
  \bibfield  {author} {\bibinfo {author} {\bibfnamefont {S.}~\bibnamefont
  {Popescu}}, \bibinfo {author} {\bibfnamefont {A.~J.}\ \bibnamefont {Short}},\
  and\ \bibinfo {author} {\bibfnamefont {A.}~\bibnamefont {Winters}},\
  }\bibfield  {title} {\bibinfo {title} {Entanglement and the foundations of
  statistical mechanics},\ }\href {https://doi.org/10.1038/nphys444} {\bibfield
   {journal} {\bibinfo  {journal} {Nat. Phys.}\ }\textbf {\bibinfo {volume}
  {2}},\ \bibinfo {pages} {754} (\bibinfo {year} {2018})}\BibitemShut {NoStop}%
\bibitem [{\citenamefont {Vidmar}\ \emph {et~al.}(2017)\citenamefont {Vidmar},
  \citenamefont {Hackl}, \citenamefont {Bianchi},\ and\ \citenamefont
  {Rigol}}]{Vidmar_2017}%
  \BibitemOpen
  \bibfield  {author} {\bibinfo {author} {\bibfnamefont {L.}~\bibnamefont
  {Vidmar}}, \bibinfo {author} {\bibfnamefont {L.}~\bibnamefont {Hackl}},
  \bibinfo {author} {\bibfnamefont {E.}~\bibnamefont {Bianchi}},\ and\ \bibinfo
  {author} {\bibfnamefont {M.}~\bibnamefont {Rigol}},\ }\bibfield  {title}
  {\bibinfo {title} {Entanglement entropy of eigenstates of quadratic fermionic
  hamiltonians},\ }\href {https://doi.org/10.1103/PhysRevLett.119.020601}
  {\bibfield  {journal} {\bibinfo  {journal} {Phys. Rev. Lett.}\ }\textbf
  {\bibinfo {volume} {119}},\ \bibinfo {pages} {020601} (\bibinfo {year}
  {2017})}\BibitemShut {NoStop}%
\bibitem [{\citenamefont {Nakagawa}\ \emph {et~al.}(2018)\citenamefont
  {Nakagawa}, \citenamefont {Watanabe}, \citenamefont {Fujita},\ and\
  \citenamefont {Sugiura}}]{Nakagawa_NC_2018}%
  \BibitemOpen
  \bibfield  {author} {\bibinfo {author} {\bibfnamefont {Y.~O.}\ \bibnamefont
  {Nakagawa}}, \bibinfo {author} {\bibfnamefont {M.}~\bibnamefont {Watanabe}},
  \bibinfo {author} {\bibfnamefont {H.}~\bibnamefont {Fujita}},\ and\ \bibinfo
  {author} {\bibfnamefont {S.}~\bibnamefont {Sugiura}},\ }\bibfield  {title}
  {\bibinfo {title} {Universality in volume-law entanglement of scrambled pure
  quantum states},\ }\href {https://doi.org/10.1038/s41467-018-03883-9}
  {\bibfield  {journal} {\bibinfo  {journal} {Nat. Commun.}\ }\textbf {\bibinfo
  {volume} {9}},\ \bibinfo {pages} {1635} (\bibinfo {year} {2018})}\BibitemShut
  {NoStop}%
\bibitem [{\citenamefont {Yu}\ \emph {et~al.}(2023)\citenamefont {Yu},
  \citenamefont {Gong},\ and\ \citenamefont {Cirac}}]{Yu_PRR_2023}%
  \BibitemOpen
  \bibfield  {author} {\bibinfo {author} {\bibfnamefont {X.-H.}\ \bibnamefont
  {Yu}}, \bibinfo {author} {\bibfnamefont {Z.}~\bibnamefont {Gong}},\ and\
  \bibinfo {author} {\bibfnamefont {J.~I.}\ \bibnamefont {Cirac}},\ }\bibfield
  {title} {\bibinfo {title} {Free-fermion page curve: Canonical typicality and
  dynamical emergence},\ }\href
  {https://doi.org/10.1103/PhysRevResearch.5.013044} {\bibfield  {journal}
  {\bibinfo  {journal} {Phys. Rev. Res.}\ }\textbf {\bibinfo {volume} {5}},\
  \bibinfo {pages} {013044} (\bibinfo {year} {2023})}\BibitemShut {NoStop}%
\bibitem [{\citenamefont {Bianchi}\ \emph {et~al.}(2022)\citenamefont
  {Bianchi}, \citenamefont {Hackl}, \citenamefont {Kieburg}, \citenamefont
  {Rigol},\ and\ \citenamefont {Vidmar}}]{Bianchi_2022}%
  \BibitemOpen
  \bibfield  {author} {\bibinfo {author} {\bibfnamefont {E.}~\bibnamefont
  {Bianchi}}, \bibinfo {author} {\bibfnamefont {L.}~\bibnamefont {Hackl}},
  \bibinfo {author} {\bibfnamefont {M.}~\bibnamefont {Kieburg}}, \bibinfo
  {author} {\bibfnamefont {M.}~\bibnamefont {Rigol}},\ and\ \bibinfo {author}
  {\bibfnamefont {L.}~\bibnamefont {Vidmar}},\ }\bibfield  {title} {\bibinfo
  {title} {Volume-law entanglement entropy of typical pure quantum states},\
  }\href {https://doi.org/10.1103/PRXQuantum.3.030201} {\bibfield  {journal}
  {\bibinfo  {journal} {PRX Quantum}\ }\textbf {\bibinfo {volume} {3}},\
  \bibinfo {pages} {030201} (\bibinfo {year} {2022})}\BibitemShut {NoStop}%
\bibitem [{\citenamefont {Essler}\ and\ \citenamefont
  {Fagotti}(2016)}]{Essler_JSM_2016}%
  \BibitemOpen
  \bibfield  {author} {\bibinfo {author} {\bibfnamefont {F.~H.~L.}\
  \bibnamefont {Essler}}\ and\ \bibinfo {author} {\bibfnamefont
  {M.}~\bibnamefont {Fagotti}},\ }\bibfield  {title} {\bibinfo {title} {Quench
  dynamics and relaxation in isolated integrable quantum spin chains},\ }\href
  {https://doi.org/10.1088/1742-5468/2016/06/064002} {\bibfield  {journal}
  {\bibinfo  {journal} {J. Stat. Mech.}\ }\textbf {\bibinfo {volume} {2016}},\
  \bibinfo {pages} {064002} (\bibinfo {year} {2016})}\BibitemShut {NoStop}%
\bibitem [{\citenamefont {Calabrese}\ and\ \citenamefont
  {Cardy}(2005)}]{Calabrese_JSM_2005}%
  \BibitemOpen
  \bibfield  {author} {\bibinfo {author} {\bibfnamefont {P.}~\bibnamefont
  {Calabrese}}\ and\ \bibinfo {author} {\bibfnamefont {J.}~\bibnamefont
  {Cardy}},\ }\bibfield  {title} {\bibinfo {title} {Evolution of entanglement
  entropy in one-dimensional systems},\ }\href
  {https://doi.org/10.1088/1742-5468/2005/04/P04010} {\bibfield  {journal}
  {\bibinfo  {journal} {J. Stat. Mech.}\ }\textbf {\bibinfo {volume} {2005}},\
  \bibinfo {pages} {P04010} (\bibinfo {year} {2005})}\BibitemShut {NoStop}%
\bibitem [{\citenamefont {Gogolin}\ and\ \citenamefont
  {Eisert}(2016)}]{Gogolin16}%
  \BibitemOpen
  \bibfield  {author} {\bibinfo {author} {\bibfnamefont {C.}~\bibnamefont
  {Gogolin}}\ and\ \bibinfo {author} {\bibfnamefont {J.}~\bibnamefont
  {Eisert}},\ }\bibfield  {title} {\bibinfo {title} {Equilibration,
  thermalisation, and the emergence of statistical mechanics in closed quantum
  systems},\ }\href {https://doi.org/10.1088/0034-4885/79/5/056001} {\bibfield
  {journal} {\bibinfo  {journal} {Rep. Prog. Phys.}\ }\textbf {\bibinfo
  {volume} {79}},\ \bibinfo {pages} {056001} (\bibinfo {year}
  {2016})}\BibitemShut {NoStop}%
\bibitem [{\citenamefont {L.~D'Alessio}\ and\ \citenamefont
  {Rigol}(2016)}]{Rigol16}%
  \BibitemOpen
  \bibfield  {author} {\bibinfo {author} { \bibnamefont
  {L.~D'Alessio}, \bibfnamefont {Y.~Kafri}},\ \bibfnamefont {A.~Polkovnikov}\ and\ \bibinfo {author}
  {\bibfnamefont {M.}~\bibnamefont {Rigol}},\ }\bibfield  {title} {\bibinfo
  {title} {From quantum chaos and eigenstate thermalization to statistical
  mechanics and thermodynamics},\ }\href
  {https://doi.org/10.1080/00018732.2016.1198134} {\bibfield  {journal}
  {\bibinfo  {journal} {Adv. Phys.}\ }\textbf {\bibinfo {volume} {65}},\
  \bibinfo {pages} {239} (\bibinfo {year} {2016})}\BibitemShut {NoStop}%
\bibitem [{\citenamefont {Borgonovi}\ \emph {et~al.}(2016)\citenamefont
  {Borgonovi}, \citenamefont {Izrailev}, \citenamefont {Santos},\ and\
  \citenamefont {Zelevinsky}}]{Borgonovi16}%
  \BibitemOpen
  \bibfield  {author} {\bibinfo {author} {\bibfnamefont {F.}~\bibnamefont
  {Borgonovi}}, \bibinfo {author} {\bibfnamefont {F.}~\bibnamefont {Izrailev}},
  \bibinfo {author} {\bibfnamefont {L.}~\bibnamefont {Santos}},\ and\ \bibinfo
  {author} {\bibfnamefont {V.}~\bibnamefont {Zelevinsky}},\ }\bibfield  {title}
  {\bibinfo {title} {Quantum chaos and thermalization in isolated systems of
  interacting particles},\ }\href
  {https://doi.org/https://doi.org/10.1016/j.physrep.2016.02.005} {\bibfield
  {journal} {\bibinfo  {journal} {Phys. Rep.}\ }\textbf {\bibinfo {volume}
  {626}},\ \bibinfo {pages} {1} (\bibinfo {year} {2016})}\BibitemShut {NoStop}%
\bibitem [{\citenamefont {Nezhadhaghighi}\ and\ \citenamefont
  {Rajabpour}(2014)}]{Rajabpour14}%
  \BibitemOpen
  \bibfield  {author} {\bibinfo {author} {\bibfnamefont {M.~G.}\ \bibnamefont
  {Nezhadhaghighi}}\ and\ \bibinfo {author} {\bibfnamefont {M.~A.}\
  \bibnamefont {Rajabpour}},\ }\bibfield  {title} {\bibinfo {title}
  {Entanglement dynamics in short- and long-range harmonic oscillators},\
  }\href {https://doi.org/10.1103/PhysRevB.90.205438} {\bibfield  {journal}
  {\bibinfo  {journal} {Phys. Rev. B}\ }\textbf {\bibinfo {volume} {90}},\
  \bibinfo {pages} {205438} (\bibinfo {year} {2014})}\BibitemShut {NoStop}%
\bibitem [{\citenamefont {Faiez}\ \emph {et~al.}(2020)\citenamefont {Faiez},
  \citenamefont {$\check{\textrm{S}}$afr\'{a}nek}, \citenamefont {Deutsch},\
  and\ \citenamefont {Aguirre}}]{Faiez_2020}%
  \BibitemOpen
  \bibfield  {author} {\bibinfo {author} {\bibfnamefont {D.}~\bibnamefont
  {Faiez}}, \bibinfo {author} {\bibfnamefont {D.}~\bibnamefont
  {$\check{\textrm{S}}$afr\'{a}nek}}, \bibinfo {author} {\bibfnamefont {J.~M.}\
  \bibnamefont {Deutsch}},\ and\ \bibinfo {author} {\bibfnamefont
  {A.}~\bibnamefont {Aguirre}},\ }\bibfield  {title} {\bibinfo {title} {Typical
  and extreme entropies of long-lived isolated quantum systems},\ }\href
  {https://doi.org/10.1103/PhysRevA.101.052101} {\bibfield  {journal} {\bibinfo
   {journal} {Phys. Rev. A}\ }\textbf {\bibinfo {volume} {101}},\ \bibinfo
  {pages} {052101} (\bibinfo {year} {2020})}\BibitemShut {NoStop}%
\bibitem [{\citenamefont {Neumann}(1929)}]{von_Neumann29}%
  \BibitemOpen
  \bibfield  {author} {\bibinfo {author} {\bibfnamefont {J.~von}\ \bibnamefont
  {Neumann}},\ }\bibfield  {title} {\bibinfo {title} {Beweis des ergodensatzes
  und desh-theorems in der neuen mechanik},\ }\href
  {https://doi.org/10.1007/BF01339852} {\bibfield  {journal} {\bibinfo
  {journal} {Z. Phys.}\ }\textbf {\bibinfo {volume} {57}},\ \bibinfo {pages}
  {30} (\bibinfo {year} {1929})}\BibitemShut {NoStop}%
\bibitem [{\citenamefont {Neumann}(2010)}]{von_Neumann10}%
  \BibitemOpen
  \bibfield  {author} {\bibinfo {author} {\bibfnamefont {J.~von}\ \bibnamefont
  {Neumann}},\ }\bibfield  {title} {\bibinfo {title} {Proof of the ergodic
  theorem and the h-theorem in quantum mechanics},\ }\href
  {https://doi.org/10.1140/epjh/e2010-00008-5} {\bibfield  {journal} {\bibinfo
  {journal} {Euro. Phys. J. H}\ }\textbf {\bibinfo {volume} {35}},\ \bibinfo
  {pages} {201} (\bibinfo {year} {2010})}\BibitemShut {NoStop}%
\bibitem [{\citenamefont {Goldstein}\ \emph {et~al.}(2010)\citenamefont
  {Goldstein}, \citenamefont {Lebowitz}, \citenamefont {Tumulka},\ and\
  \citenamefont {Zanghì}}]{Lebowitz10}%
  \BibitemOpen
  \bibfield  {author} {\bibinfo {author} {\bibfnamefont {S.}~\bibnamefont
  {Goldstein}}, \bibinfo {author} {\bibfnamefont {J.~L.}\ \bibnamefont
  {Lebowitz}}, \bibinfo {author} {\bibfnamefont {R.}~\bibnamefont {Tumulka}},\
  and\ \bibinfo {author} {\bibfnamefont {N.}~\bibnamefont {Zanghì}},\
  }\bibfield  {title} {\bibinfo {title} {Long-time behavior of macroscopic
  quantum systems},\ }\href {https://doi.org/10.1140/epjh/e2010-00007-7}
  {\bibfield  {journal} {\bibinfo  {journal} {Euro. Phys. J. H}\ }\textbf
  {\bibinfo {volume} {35}},\ \bibinfo {pages} {173} (\bibinfo {year}
  {2010})}\BibitemShut {NoStop}%
\bibitem [{\citenamefont {Modak}\ \emph {et~al.}(2020)\citenamefont {Modak},
  \citenamefont {Alba},\ and\ \citenamefont {Calabrese}}]{Modak_JSM_2020}%
  \BibitemOpen
  \bibfield  {author} {\bibinfo {author} {\bibfnamefont {R.}~\bibnamefont
  {Modak}}, \bibinfo {author} {\bibfnamefont {V.}~\bibnamefont {Alba}},\ and\
  \bibinfo {author} {\bibfnamefont {P.}~\bibnamefont {Calabrese}},\ }\bibfield
  {title} {\bibinfo {title} {Entanglement revivals as a probe of scrambling in
  finite quantum systems},\ }\href {https://doi.org/10.1088/1742-5468/aba9d9}
  {\bibfield  {journal} {\bibinfo  {journal} {J. Stat. Mech.}\ }\textbf
  {\bibinfo {volume} {2020}},\ \bibinfo {pages} {083110} (\bibinfo {year}
  {2020})}\BibitemShut {NoStop}%
\bibitem [{\citenamefont {Plenio}\ \emph {et~al.}(2004)\citenamefont {Plenio},
  \citenamefont {Hartley},\ and\ \citenamefont {Eisert}}]{Plenio04}%
  \BibitemOpen
  \bibfield  {author} {\bibinfo {author} {\bibfnamefont {M.~B.}\ \bibnamefont
  {Plenio}}, \bibinfo {author} {\bibfnamefont {J.}~\bibnamefont {Hartley}},\
  and\ \bibinfo {author} {\bibfnamefont {J.}~\bibnamefont {Eisert}},\
  }\bibfield  {title} {\bibinfo {title} {Dynamics and manipulation of
  entanglement in coupled harmonic systems with many degrees of freedom},\
  }\href {https://doi.org/10.1088/1367-2630/6/1/036} {\bibfield  {journal}
  {\bibinfo  {journal} {New J. Phys.}\ }\textbf {\bibinfo {volume} {6}},\
  \bibinfo {pages} {36} (\bibinfo {year} {2004})}\BibitemShut {NoStop}%
\bibitem [{\citenamefont {Lim}\ \emph {et~al.}(2024)\citenamefont {Lim},
  \citenamefont {Lou},\ and\ \citenamefont {Tian}}]{Lih-King_2023}%
  \BibitemOpen
  \bibfield  {author} {\bibinfo {author} {\bibfnamefont {L.-K.}\ \bibnamefont
  {Lim}}, \bibinfo {author} {\bibfnamefont {C.}~\bibnamefont {Lou}},\ and\
  \bibinfo {author} {\bibfnamefont {C.}~\bibnamefont {Tian}},\ }\bibfield
  {title} {\bibinfo {title} {Mesoscopic fluctuations in entanglement
  dynamics},\ }\href {https://doi.org/10.1038/s41467-024-46078-1} {\bibfield
  {journal} {\bibinfo  {journal} {Nat. Commun.}\ }\textbf {\bibinfo {volume}
  {15}},\ \bibinfo {pages} {1775} (\bibinfo {year} {2024})}\BibitemShut
  {NoStop}%
\bibitem [{\citenamefont {Boucheron}\ \emph {et~al.}(2013)\citenamefont
  {Boucheron}, \citenamefont {Lugosi},\ and\ \citenamefont
  {Massart}}]{Boucheron_2013}%
  \BibitemOpen
  \bibfield  {author} {\bibinfo {author} {\bibfnamefont {S.}~\bibnamefont
  {Boucheron}}, \bibinfo {author} {\bibfnamefont {G.}~\bibnamefont {Lugosi}},\
  and\ \bibinfo {author} {\bibfnamefont {P.}~\bibnamefont {Massart}},\ }\href
  {https://doi.org/10.1093/acprof:oso/9780199535255.001.0001} {\emph {\bibinfo
  {title} {{Concentration Inequalities: A Nonasymptotic Theory of
  Independence}}}}\ (\bibinfo  {publisher} {Oxford University Press},\ \bibinfo
  {year} {2013})\BibitemShut {NoStop}%
\bibitem [{\citenamefont {Ping}(2006)}]{Sheng_2006}%
  \BibitemOpen
  \bibfield  {author} {\bibinfo {author} {\bibfnamefont {S.}~\bibnamefont
  {Ping}},\ }\href {https://doi.org/10.1007/3-540-29156-3} {\emph {\bibinfo
  {title} {{Introduction to Wave Scattering, Localization and Mesoscopic
  Phenomena}}}}\ (\bibinfo  {publisher} {Springer, Germany, 2nd edition},\
  \bibinfo {year} {2006})\BibitemShut {NoStop}%
\bibitem [{\citenamefont {Akkermans}\ and\ \citenamefont
  {Montambaux}(2007)}]{Akkermans_2007}%
  \BibitemOpen
  \bibfield  {author} {\bibinfo {author} {\bibfnamefont {E.}~\bibnamefont
  {Akkermans}}\ and\ \bibinfo {author} {\bibfnamefont {G.}~\bibnamefont
  {Montambaux}},\ }\href {https://doi.org/DOI: 10.1017/CBO9780511618833} {\emph
  {\bibinfo {title} {{Mesoscopic Physics of Electrons and Photons}}}}\
  (\bibinfo  {publisher} {Cambridge University Press},\ \bibinfo {year}
  {2007})\BibitemShut {NoStop}%
\bibitem [{\citenamefont {Kane}\ (2003)}]{Kane03}%
  \BibitemOpen
  \bibfield  {author} {\bibinfo {author} {\bibfnamefont {C.~L.}\ \bibnamefont
  {Kane}},\ }\bibfield
  {title} {\bibinfo {title} {Telegraph noise and fractional statistics in the quantum Hall effect},\ }\href
  {https://doi.org/10.1103/PhysRevLett.90.226902} {\bibfield  {journal}
  {\bibinfo  {journal} {Phys. Rev. Lett.}\ }\textbf {\bibinfo {volume} {90}},\
  \bibinfo {pages} {226902} (\bibinfo {year} {2003})}\BibitemShut {NoStop}%
\bibitem [{\citenamefont {Grosfeld}\ \citenamefont {Simon} and\ \citenamefont
  {Stern}(2006)}]{Stern06}%
  \BibitemOpen
  \bibfield  {author} {\bibinfo {author} {\bibfnamefont {E.}~\bibnamefont
  {Grosfeld}}, \bibfnamefont {S.~H.}\ \bibnamefont
  {Simon}, and\ \bibinfo {author} {\bibfnamefont {A.}\ \bibnamefont
  {Stern}},\ }\bibfield  {title} {\bibinfo {title} {Switching noise as a probe of statistics in the fractional quantum Hall effect},\ }\href {https://doi.org/10.1103/PhysRevLett.96.226803}
  {\bibfield  {journal} {\bibinfo  {journal} {Phys. Rev. Lett.}\ }\textbf
  {\bibinfo {volume} {96}},\ \bibinfo {pages} {226803} (\bibinfo {year}
  {2006})}\BibitemShut {NoStop}%
\bibitem [{\citenamefont {Rosenow} and \ \citenamefont {Simon} (2012)}]{Simon12}%
  \BibitemOpen
  \bibfield  {author} {\bibinfo {author} {\bibfnamefont {B.}~\bibnamefont
  {Rosenow}} and \bibfnamefont {S.~H.}\ \bibnamefont
  {Simon},\ }\bibfield  {title} {\bibinfo {title} {Telegraph noise and the Fabry-Perot quantum Hall interferometer},\ }\href {%https://doi.org/10.1103/PhysRevLett.96.226803
  }
  {\bibfield  {journal} {\bibinfo  {journal} {Phys. Rev. B}\ }\textbf
  {\bibinfo {volume} {85}},\ \bibinfo {pages} {201302} (\bibinfo {year}
  {2012})}\BibitemShut {NoStop}%
\bibitem [{\citenamefont {Peschel}\ and\ \citenamefont
  {Chung}(1999)}]{Peschel99}%
  \BibitemOpen
  \bibfield  {author} {\bibinfo {author} {\bibfnamefont {I.}~\bibnamefont
  {Peschel}}\ and\ \bibinfo {author} {\bibfnamefont {M.-C.}\ \bibnamefont
  {Chung}},\ }\bibfield  {title} {\bibinfo {title} {Density matrices for a
  chain of oscillators},\ }\href {https://doi.org/10.1088/0305-4470/32/48/305}
  {\bibfield  {journal} {\bibinfo  {journal} {J. Phys. A: Math. Gen.}\ }\textbf
  {\bibinfo {volume} {32}},\ \bibinfo {pages} {8419} (\bibinfo {year}
  {1999})}\BibitemShut {NoStop}%
\bibitem [{\citenamefont {Audenaert}\ \emph {et~al.}(2002)\citenamefont
  {Audenaert}, \citenamefont {Eisert}, \citenamefont {Plenio},\ and\
  \citenamefont {Werner}}]{Plenio02}%
  \BibitemOpen
  \bibfield  {author} {\bibinfo {author} {\bibfnamefont {K.}~\bibnamefont
  {Audenaert}}, \bibinfo {author} {\bibfnamefont {J.}~\bibnamefont {Eisert}},
  \bibinfo {author} {\bibfnamefont {M.~B.}\ \bibnamefont {Plenio}},\ and\
  \bibinfo {author} {\bibfnamefont {R.~F.}\ \bibnamefont {Werner}},\ }\bibfield
   {title} {\bibinfo {title} {Entanglement properties of the harmonic chain},\
  }\href {https://doi.org/10.1103/PhysRevA.66.042327} {\bibfield  {journal}
  {\bibinfo  {journal} {Phys. Rev. A}\ }\textbf {\bibinfo {volume} {66}},\
  \bibinfo {pages} {042327} (\bibinfo {year} {2002})}\BibitemShut {NoStop}%
\bibitem [{\citenamefont {Botero}\ and\ \citenamefont
  {Reznik}(2004)}]{Botero04}%
  \BibitemOpen
  \bibfield  {author} {\bibinfo {author} {\bibfnamefont {A.}~\bibnamefont
  {Botero}}\ and\ \bibinfo {author} {\bibfnamefont {B.}~\bibnamefont
  {Reznik}},\ }\bibfield  {title} {\bibinfo {title} {Spatial structures and
  localization of vacuum entanglement in the linear harmonic chain},\ }\href
  {https://doi.org/10.1103/PhysRevA.70.052329} {\bibfield  {journal} {\bibinfo
  {journal} {Phys. Rev. A}\ }\textbf {\bibinfo {volume} {70}},\ \bibinfo
  {pages} {052329} (\bibinfo {year} {2004})}\BibitemShut {NoStop}%
\bibitem [{\citenamefont {Plenio}\ \emph {et~al.}(2005)\citenamefont {Plenio},
  \citenamefont {Eisert}, \citenamefont {Drei\ss{}ig},\ and\ \citenamefont
  {Cramer}}]{Plenio05}%
  \BibitemOpen
  \bibfield  {author} {\bibinfo {author} {\bibfnamefont {M.~B.}\ \bibnamefont
  {Plenio}}, \bibinfo {author} {\bibfnamefont {J.}~\bibnamefont {Eisert}},
  \bibinfo {author} {\bibfnamefont {J.}~\bibnamefont {Drei\ss{}ig}},\ and\
  \bibinfo {author} {\bibfnamefont {M.}~\bibnamefont {Cramer}},\ }\bibfield
  {title} {\bibinfo {title} {Entropy, entanglement, and area: Analytical
  results for harmonic lattice systems},\ }\href
  {https://doi.org/10.1103/PhysRevLett.94.060503} {\bibfield  {journal}
  {\bibinfo  {journal} {Phys. Rev. Lett.}\ }\textbf {\bibinfo {volume} {94}},\
  \bibinfo {pages} {060503} (\bibinfo {year} {2005})}\BibitemShut {NoStop}%
\bibitem [{\citenamefont {Eisert}\ \emph {et~al.}(2010)\citenamefont {Eisert},
  \citenamefont {Cramer},\ and\ \citenamefont {Plenio}}]{Eisert_RMP_2010}%
  \BibitemOpen
  \bibfield  {author} {\bibinfo {author} {\bibfnamefont {J.}~\bibnamefont
  {Eisert}}, \bibinfo {author} {\bibfnamefont {M.}~\bibnamefont {Cramer}},\
  and\ \bibinfo {author} {\bibfnamefont {M.~B.}\ \bibnamefont {Plenio}},\
  }\bibfield  {title} {\bibinfo {title} {Colloquium: Area laws for the
  entanglement entropy},\ }\href {https://doi.org/10.1103/RevModPhys.82.277}
  {\bibfield  {journal} {\bibinfo  {journal} {Rev. Mod. Phys.}\ }\textbf
  {\bibinfo {volume} {82}},\ \bibinfo {pages} {277} (\bibinfo {year}
  {2010})}\BibitemShut {NoStop}%
\bibitem [{\citenamefont {Cotler}\ \emph {et~al.}(2016)\citenamefont {Cotler},
  \citenamefont {Hertzberg}, \citenamefont {Mezei},\ and\ \citenamefont
  {Mueller}}]{Cotller16}%
  \BibitemOpen
  \bibfield  {author} {\bibinfo {author} {\bibfnamefont {J.~S.}\ \bibnamefont
  {Cotler}}, \bibinfo {author} {\bibfnamefont {M.~P.}\ \bibnamefont
  {Hertzberg}}, \bibinfo {author} {\bibfnamefont {M.}~\bibnamefont {Mezei}},\
  and\ \bibinfo {author} {\bibfnamefont {M.~T.}\ \bibnamefont {Mueller}},\
  }\bibfield  {title} {\bibinfo {title} {Entanglement growth after a global
  quench in free scalar field theory},\ }\href
  {https://doi.org/10.1007/JHEP11(2016)166} {\bibfield  {journal} {\bibinfo
  {journal} {J. High Energ. Phys.}\ }\textbf {\bibinfo {volume} {2016}},\
  \bibinfo {pages} {166}}\BibitemShut {NoStop}%
\bibitem [{\citenamefont {Hackl}\ \emph {et~al.}(2018)\citenamefont {Hackl},
  \citenamefont {Bianchi}, \citenamefont {Modak},\ and\ \citenamefont
  {Rigol}}]{Hackl_PRA_2018}%
  \BibitemOpen
  \bibfield  {author} {\bibinfo {author} {\bibfnamefont {L.}~\bibnamefont
  {Hackl}}, \bibinfo {author} {\bibfnamefont {E.}~\bibnamefont {Bianchi}},
  \bibinfo {author} {\bibfnamefont {R.}~\bibnamefont {Modak}},\ and\ \bibinfo
  {author} {\bibfnamefont {M.}~\bibnamefont {Rigol}},\ }\bibfield  {title}
  {\bibinfo {title} {Entanglement production in bosonic systems: Linear and
  logarithmic growth},\ }\href {https://doi.org/10.1103/PhysRevA.97.032321}
  {\bibfield  {journal} {\bibinfo  {journal} {Phys. Rev. A}\ }\textbf {\bibinfo
  {volume} {97}},\ \bibinfo {pages} {032321} (\bibinfo {year}
  {2018})}\BibitemShut {NoStop}%
\bibitem [{\citenamefont {Alba}\ and\ \citenamefont
  {Calabrese}(2018)}]{Alba_SPP_2018}%
  \BibitemOpen
  \bibfield  {author} {\bibinfo {author} {\bibfnamefont {V.}~\bibnamefont
  {Alba}}\ and\ \bibinfo {author} {\bibfnamefont {P.}~\bibnamefont
  {Calabrese}},\ }\bibfield  {title} {\bibinfo {title} {{Entanglement dynamics
  after quantum quenches in generic integrable systems}},\ }\href
  {https://doi.org/10.21468/SciPostPhys.4.3.017} {\bibfield  {journal}
  {\bibinfo  {journal} {SciPost Phys.}\ }\textbf {\bibinfo {volume} {4}},\
  \bibinfo {pages} {017} (\bibinfo {year} {2018})}\BibitemShut {NoStop}%
\bibitem [{\citenamefont {Coser}\ \emph {et~al.}(2014)\citenamefont {Coser},
  \citenamefont {Tonni},\ and\ \citenamefont {Calabrese}}]{Calabrese17}%
  \BibitemOpen
  \bibfield  {author} {\bibinfo {author} {\bibfnamefont {A.}~\bibnamefont
  {Coser}}, \bibinfo {author} {\bibfnamefont {E.}~\bibnamefont {Tonni}},\ and\
  \bibinfo {author} {\bibfnamefont {P.}~\bibnamefont {Calabrese}},\ }\bibfield
  {title} {\bibinfo {title} {Entanglement negativity after a global quantum
  quench},\ }\href {https://doi.org/10.1088/1742-5468/2014/12/P12017}
  {\bibfield  {journal} {\bibinfo  {journal} {J. Stat. Mech.}\ }\textbf
  {\bibinfo {volume} {2014}},\ \bibinfo {pages} {P12017} (\bibinfo {year}
  {2014})}\BibitemShut {NoStop}%
\bibitem [{\citenamefont {Adesso}\ and\ \citenamefont
  {Illuminati}(2007)}]{Adesso07}%
  \BibitemOpen
  \bibfield  {author} {\bibinfo {author} {\bibfnamefont {G.}~\bibnamefont
  {Adesso}}\ and\ \bibinfo {author} {\bibfnamefont {F.}~\bibnamefont
  {Illuminati}},\ }\bibfield  {title} {\bibinfo {title} {Entanglement in
  continuous-variable systems: recent advances and current perspectives},\
  }\href {https://doi.org/10.1088/1751-8113/40/28/S01} {\bibfield  {journal}
  {\bibinfo  {journal} {J. Phys. A: Math. Theor.}\ }\textbf {\bibinfo {volume}
  {40}},\ \bibinfo {pages} {7821} (\bibinfo {year} {2007})}\BibitemShut
  {NoStop}%
\bibitem [{\citenamefont {Williamson}(1936)}]{Williamson36}%
  \BibitemOpen
  \bibfield  {author} {\bibinfo {author} {\bibfnamefont {J.}~\bibnamefont
  {Williamson}},\ }\bibfield  {title} {\bibinfo {title} {On the algebraic
  problem concerning the normal forms of linear dynamical systems},\ }\href
  {https://api.semanticscholar.org/CorpusID:124759285} {\bibfield  {journal}
  {\bibinfo  {journal} {Amer. J. Math.}\ }\textbf {\bibinfo {volume} {58}},\
  \bibinfo {pages} {141} (\bibinfo {year} {1936})}\BibitemShut {NoStop}%
\bibitem [{\citenamefont {Peschel}(2003)}]{Peschel_JPA_2003}%
  \BibitemOpen
  \bibfield  {author} {\bibinfo {author} {\bibfnamefont {I.}~\bibnamefont
  {Peschel}},\ }\bibfield  {title} {\bibinfo {title} {Calculation of reduced
  density matrices from correlation functions},\ }\href
  {https://doi.org/10.1088/0305-4470/36/14/101} {\bibfield  {journal} {\bibinfo
   {journal} {J. Phys. A: Math. Gen.}\ }\textbf {\bibinfo {volume} {36}},\
  \bibinfo {pages} {L205} (\bibinfo {year} {2003})}\BibitemShut {NoStop}%
\bibitem [{\citenamefont {Vidal}\ \emph {et~al.}(2003)\citenamefont {Vidal},
  \citenamefont {Latorre}, \citenamefont {Rico},\ and\ \citenamefont
  {Kitaev}}]{Vidal_PRL_2003}%
  \BibitemOpen
  \bibfield  {author} {\bibinfo {author} {\bibfnamefont {G.}~\bibnamefont
  {Vidal}}, \bibinfo {author} {\bibfnamefont {J.~I.}\ \bibnamefont {Latorre}},
  \bibinfo {author} {\bibfnamefont {E.}~\bibnamefont {Rico}},\ and\ \bibinfo
  {author} {\bibfnamefont {A.}~\bibnamefont {Kitaev}},\ }\bibfield  {title}
  {\bibinfo {title} {Entanglement in quantum critical phenomena},\ }\href
  {https://doi.org/10.1103/PhysRevLett.90.227902} {\bibfield  {journal}
  {\bibinfo  {journal} {Phys. Rev. Lett.}\ }\textbf {\bibinfo {volume} {90}},\
  \bibinfo {pages} {227902} (\bibinfo {year} {2003})}\BibitemShut {NoStop}%
\bibitem [{\citenamefont {Peschel}\ and\ \citenamefont
  {Eisler}(2009)}]{Peschel_JPA_2009}%
  \BibitemOpen
  \bibfield  {author} {\bibinfo {author} {\bibfnamefont {I.}~\bibnamefont
  {Peschel}}\ and\ \bibinfo {author} {\bibfnamefont {V.}~\bibnamefont
  {Eisler}},\ }\bibfield  {title} {\bibinfo {title} {Reduced density matrices
  and entanglement entropy in free lattice models},\ }\href
  {https://doi.org/10.1088/1751-8113/42/50/504003} {\bibfield  {journal}
  {\bibinfo  {journal} {J. Phys. A: Math. Theor.}\ }\textbf {\bibinfo {volume}
  {42}},\ \bibinfo {pages} {504003} (\bibinfo {year} {2009})}\BibitemShut
  {NoStop}%
\bibitem [{\citenamefont {T.~Shi}\ and\ \citenamefont
  {Cirac}(2016)}]{Shi18}%
  \BibitemOpen
  \bibfield  {author} {\bibinfo {author} { \bibnamefont
  {T.~Shi}},\ \bibfnamefont {E.~Demler}\ and\ \bibinfo {author}
  {\bibfnamefont {J.~I.}~\bibnamefont {Cirac}},\ }\bibfield  {title} {\bibinfo
  {title} {Variational study of fermionic and bosonic systems with non-Gaussian states: Theory and applications},\ }\href
  {https://doi.org/10.1016/j.aop.2017.11.014} {\bibfield  {journal}
  {\bibinfo  {journal} {Ann. Phys.}\ }\textbf {\bibinfo {volume} {390}},\
  \bibinfo {pages} {245} (\bibinfo {year} {2018})}\BibitemShut {NoStop}%
\bibitem [{\citenamefont {T.~Shi}\ and\ \citenamefont
  {Cirac}(2020)}]{Shi20}%
  \BibitemOpen
  \bibfield  {author} {\bibinfo {author} { \bibnamefont
  {T.~Shi}},\ \bibfnamefont {E.~Demler}\ and\ \bibinfo {author}
  {\bibfnamefont {J.~I.}~\bibnamefont {Cirac}},\ }\bibfield  {title} {\bibinfo
  {title} {Variational approach for many-body systems at finite temperature},\ }\href
  {https://doi.org/10.1103/PhysRevLett.125.180602} {\bibfield  {journal}
  {\bibinfo  {journal} {Phys. Rev. Lett.}\ }\textbf {\bibinfo {volume} {125}},\
  \bibinfo {pages} {180602} (\bibinfo {year} {2020})}\BibitemShut {NoStop}%
\bibitem [{\citenamefont {Higham}(2008)}]{Higham08}%
  \BibitemOpen
  \bibfield  {author} {\bibinfo {author} {\bibfnamefont {N.~J.}\ \bibnamefont
  {Higham}},\ }\href {https://doi.org/10.1137/1.9780898717778} {\emph {\bibinfo
  {title} {{Functions of Matrices}}}}\ (\bibinfo  {publisher} {Society for
  Industrial and Applied Mathematics (SIAM)},\ \bibinfo {year}
  {2008})\BibitemShut {NoStop}%
\bibitem [{\citenamefont {Schachenmayer}\ \emph {et~al.}(2013)\citenamefont
  {Schachenmayer}, \citenamefont {Lanyon}, \citenamefont {Roos},\ and\
  \citenamefont {Daley}}]{Schachenmayer_PRX_2013}%
  \BibitemOpen
  \bibfield  {author} {\bibinfo {author} {\bibfnamefont {J.}~\bibnamefont
  {Schachenmayer}}, \bibinfo {author} {\bibfnamefont {B.~P.}\ \bibnamefont
  {Lanyon}}, \bibinfo {author} {\bibfnamefont {C.~F.}\ \bibnamefont {Roos}},\
  and\ \bibinfo {author} {\bibfnamefont {A.~J.}\ \bibnamefont {Daley}},\
  }\bibfield  {title} {\bibinfo {title} {Entanglement growth in quench dynamics
  with variable range interactions},\ }\href
  {https://doi.org/10.1103/PhysRevX.3.031015} {\bibfield  {journal} {\bibinfo
  {journal} {Phys. Rev. X}\ }\textbf {\bibinfo {volume} {3}},\ \bibinfo {pages}
  {031015} (\bibinfo {year} {2013})}\BibitemShut {NoStop}%
\bibitem [{\citenamefont {Samoilenko}(2007)}]{Samoilenko07}%
  \BibitemOpen
  \bibfield  {author} {\bibinfo {author} {\bibfnamefont {A.~M.}\ \bibnamefont
  {Samoilenko}},\ }\bibfield  {title} {\bibinfo {title} {Quasiperiodic
  oscillations},\ }\href {https://api.semanticscholar.org/CorpusID:5918069}
  {\bibfield  {journal} {\bibinfo  {journal} {Scholarpedia}\ }\textbf {\bibinfo
  {volume} {2}},\ \bibinfo {pages} {1783} (\bibinfo {year} {2007})}\BibitemShut
  {NoStop}%
\bibitem [{\citenamefont {Arnold}(1989)}]{Arnold_1989}%
  \BibitemOpen
  \bibfield  {author} {\bibinfo {author} {\bibfnamefont {V.~I.}\ \bibnamefont
  {Arnold}},\ }\href {https://doi.org/10.1007/978-1-4757-2063-1} {\emph
  {\bibinfo {title} {{Mathematical Methods of Classical Mechanics}}}}\
  (\bibinfo  {publisher} {Springer New York, NY},\ \bibinfo {year}
  {1989})\BibitemShut {NoStop}%
\bibitem [{\citenamefont {Kac}(1959)}]{Kac}%
  \BibitemOpen
  \bibfield  {author} {\bibinfo {author} {\bibfnamefont {M.}~\bibnamefont
  {Kac}},\ }\href {http://www.jstor.org/stable/10.4169/j.ctt5hh8tq} {\emph
  {\bibinfo {title} {{Statistical Independence in Probability, Analysis and
  Number Theory}}}}\ (\bibinfo  {publisher} {Mathematical Association of
  America},\ \bibinfo {year} {1959})\BibitemShut {NoStop}%
\bibitem [{\citenamefont {Ledoux}(2001)}]{Ledoux_2001}%
  \BibitemOpen
  \bibfield  {author} {\bibinfo {author} {\bibfnamefont {M.}~\bibnamefont
  {Ledoux}},\ }\href {https://api.semanticscholar.org/CorpusID:117088211}
  {\emph {\bibinfo {title} {The concentration of measure phenomenon}}}\
  (\bibinfo  {publisher} {AMS, Providence},\ \bibinfo {year}
  {2001})\BibitemShut {NoStop}%
\bibitem [{\citenamefont {Tao}(2012)}]{Tao2012}%
  \BibitemOpen
  \bibfield  {author} {\bibinfo {author} {\bibfnamefont {T.}~\bibnamefont
  {Tao}},\ }\href {https://doi.org/10.1090/gsm/132} {\emph {\bibinfo {title}
  {Topics in Random Matrix Theory}}}\ (\bibinfo  {publisher} {Graduate Studies
  in Mathematics v. 132, American Mathematical Society},\ \bibinfo {year}
  {2012})\BibitemShut {NoStop}%
\bibitem [{\citenamefont {Milman}\ and\ \citenamefont
  {Schechtman}(1982)}]{Milman_1982}%
  \BibitemOpen
  \bibfield  {author} {\bibinfo {author} {\bibfnamefont {V.~D.}\ \bibnamefont
  {Milman}}\ and\ \bibinfo {author} {\bibfnamefont {G.}~\bibnamefont
  {Schechtman}},\ }\href {https://doi.org/10.1007/978-3-540-38822-7} {\emph
  {\bibinfo {title} {Asymptotic Theory of Finite Dimensional Normed Spaces}}}\
  (\bibinfo  {publisher} {Springer Berlin, Heidelberg},\ \bibinfo {year}
  {1982})\BibitemShut {NoStop}%
\bibitem [{\citenamefont {Fang}\ \emph {et~al.}(2018)\citenamefont {Fang},
  \citenamefont {Zhao},\ and\ \citenamefont {Tian}}]{Fang_2018}%
  \BibitemOpen
  \bibfield  {author} {\bibinfo {author} {\bibfnamefont {P.}~\bibnamefont
  {Fang}}, \bibinfo {author} {\bibfnamefont {L.}~\bibnamefont {Zhao}},\ and\
  \bibinfo {author} {\bibfnamefont {C.}~\bibnamefont {Tian}},\ }\bibfield
  {title} {\bibinfo {title} {Concentration-of-measure theory for structures and
  fluctuations of waves},\ }\href
  {https://doi.org/10.1103/PhysRevLett.121.140603} {\bibfield  {journal}
  {\bibinfo  {journal} {Phys. Rev. Lett.}\ }\textbf {\bibinfo {volume} {121}},\
  \bibinfo {pages} {140603} (\bibinfo {year} {2018})}\BibitemShut {NoStop}%
\bibitem [{\citenamefont {Isoard}\ \emph {et~al.}(2021)\citenamefont {Isoard},
  \citenamefont {Milazzo}, \citenamefont {Pavloff},\ and\ \citenamefont
  {Giraud}}]{Isoard_2021}%
  \BibitemOpen
  \bibfield  {author} {\bibinfo {author} {\bibfnamefont {M.}~\bibnamefont
  {Isoard}}, \bibinfo {author} {\bibfnamefont {N.}~\bibnamefont {Milazzo}},
  \bibinfo {author} {\bibfnamefont {N.}~\bibnamefont {Pavloff}},\ and\ \bibinfo
  {author} {\bibfnamefont {O.}~\bibnamefont {Giraud}},\ }\bibfield  {title}
  {\bibinfo {title} {Bipartite and tripartite entanglement in a
  {B}ose-{E}instein acoustic black hole},\ }\href
  {https://doi.org/10.1103/PhysRevA.104.063302} {\bibfield  {journal} {\bibinfo
   {journal} {Phys. Rev. A}\ }\textbf {\bibinfo {volume} {104}},\ \bibinfo
  {pages} {063302} (\bibinfo {year} {2021})}\BibitemShut {NoStop}%
\bibitem [{\citenamefont {Steinhauer}(2016)}]{Steinhauer_2016}%
  \BibitemOpen
  \bibfield  {author} {\bibinfo {author} {\bibfnamefont {J.}~\bibnamefont
  {Steinhauer}},\ }\bibfield  {title} {\bibinfo {title} {Observation of quantum
  {H}awking radiation and its entanglement in an analogue black hole},\ }\href
  {https://doi.org/10.1038/nphys3863} {\bibfield  {journal} {\bibinfo
  {journal} {Nat. Phys.}\ }\textbf {\bibinfo {volume} {12}},\ \bibinfo {pages}
  {959} (\bibinfo {year} {2016})}\BibitemShut {NoStop}%
\bibitem [{\citenamefont {Shi}\ \emph {et~al.}(2023)\citenamefont {Shi} \emph
  {et~al.}}]{Shi_2023}%
  \BibitemOpen
  \bibfield  {author} {\bibinfo {author} {\bibfnamefont {Y.}~\bibnamefont
  {Shi}} \emph {et~al.},\ }\bibfield  {title} {\bibinfo {title} {Quantum
  simulation of {H}awking radiation and curved spacetime with a superconducting
  on-chip black hole},\ }\href {https://doi.org/10.1038/s41467-023-39064-6}
  {\bibfield  {journal} {\bibinfo  {journal} {Nat. Commun.}\ }\textbf {\bibinfo
  {volume} {14}},\ \bibinfo {pages} {3263} (\bibinfo {year}
  {2023})}\BibitemShut {NoStop}%
\bibitem [{\citenamefont {Calabrese}\ and\ \citenamefont
  {Cardy}(2007)}]{Calabrese07}%
  \BibitemOpen
  \bibfield  {author} {\bibinfo {author} {\bibfnamefont {P.}~\bibnamefont
  {Calabrese}}\ and\ \bibinfo {author} {\bibfnamefont {J.}~\bibnamefont
  {Cardy}},\ }\bibfield  {title} {\bibinfo {title} {Quantum quenches in
  extended systems},\ }\href {https://doi.org/10.1088/1742-5468/2007/06/P06008}
  {\bibfield  {journal} {\bibinfo  {journal} {J. Stat. Mech.}\ }\textbf
  {\bibinfo {volume} {2007}},\ \bibinfo {pages} {P06008} (\bibinfo {year}
  {2007})}\BibitemShut {NoStop}%
\end{thebibliography}
\end{document}